\documentclass{elsart}

\usepackage{times}
\usepackage{natbib}
\usepackage{graphics}
\usepackage{amssymb}
\usepackage{ifthen}
\usepackage{verbatim}
\usepackage{url}

\begin{document}

\begin{frontmatter}

\title{Typed Generic Traversal\\With Term Rewriting Strategies\\
{\footnotesize --- \emph{Accepted for publication in
The Journal of Logic and Algebraic Programming} ---}}
\author{Ralf L{\"a}mmel}
\ead{ralf@cs.vu.nl}
\ead[url]{http://www.cs.vu.nl/\~{}ralf}
\address{Free University \& CWI, Amsterdam\\The Netherlands}

\begin{abstract}
A typed model of strategic term rewriting is developed. The key
innovation is that generic traversal is covered. To this end, we
define a typed rewriting calculus $S'_{\gamma}$. The calculus employs
a many-sorted type system extended by designated generic strategy
types $\gamma$. We consider two generic strategy types, namely the
types of type-preserving and type-unifying strategies. $S'_{\gamma}$
offers traversal combinators to construct traversals or schemes
thereof from many-sorted and generic strategies. The traversal
combinators model different forms of one-step traversal, that is, they
process the immediate subterms of a given term without anticipating
any scheme of recursion into terms. To inhabit generic types, we need
to add a fundamental combinator to lift a many-sorted strategy $s$ to
a generic type $\gamma$. This step is called strategy extension. The
semantics of the corresponding combinator states that $s$ is only
applied if the type of the term at hand fits, otherwise the extended
strategy fails. This approach dictates that the semantics of strategy
application must be type-dependent to a certain extent. Typed
strategic term rewriting with coverage of generic term traversal is a
simple but expressive model of generic programming. It has
applications in program transformation and program analysis.

\end{abstract}


\end{frontmatter}

\newcommand{\myparagraph}[1]{\paragraph{#1}\ }
\newtheorem{example}{Example}
\newtheorem{theorem}{Theorem}
\newtheorem{proof}{Proof}
\newcommand{\smallsize}{\footnotesize}
\newcommand{\minisize}{\scriptsize}

\newpage
{\footnotesize\tableofcontents}

%
%
\newpage
%
%

\newcommand{\implies}{\ensuremath{\Rightarrow}}
\newcommand{\acequiv}{\ensuremath{{\cong}}}
\newcommand{\lift}[1]{\ensuremath{#1{\Uparrow}}}
\newcommand{\assign}[2]{{#1}\,{\mapsto}\,{#2}}
\newcommand{\assigns}[1]{{\{}{#1}{\}}}
\newcommand{\negbf}[1]{\ensuremath{\neg\,#1}}
\newcommand{\m}[1]{\esnuremath{}}
\newcommand{\emptytuple}{\ensuremath{\langle\rangle}}
\newcommand{\pair}[2]{\ensuremath{\langle #1, #2 \rangle}}
\newcommand{\ltr}{\ensuremath{\rightarrow}}
\newcommand{\rtl}{\ensuremath{\leftarrow}}
\newcommand{\assoc}{\ensuremath{\stackrel{\leftarrow}{\rightarrow}}}
\newcommand{\s}[1]{\textsc{{#1}}}
\newcommand{\noskip}{\topsep0pt \parskip0pt \partopsep0pt}
\newcommand{\mycalc}{\ensuremath{S'_{\gamma}}}
\newcommand{\mycalclight}{\ensuremath{S'_0}}
\newcommand{\vunit}{\ensuremath{\nu_{u}}}
\newcommand{\vplus}{\ensuremath{\nu_{\circ}}}   
\newcommand{\sunit}{\ensuremath{s_{u}}}
\newcommand{\splus}{\ensuremath{s_{\circ}}}   
\newcommand{\snil}{\ensuremath{s_0}}
\newcommand{\scons}{\ensuremath{s_c}}   
\newcommand{\reduce}[3]{\ensuremath{{\bigcirc}_{#1}^{#2}(#3)}}
\newcommand{\select}[2]{\ensuremath{{\sharp}_{#1}(#2)}}
\newcommand{\fold}[2]{\ensuremath{{(}\!{|}#1,#2{|}\!{)}}}
\newcommand{\allprod}[2]{\ensuremath{#1 \otimes #2}} 
\newcommand{\where}[2]{\ensuremath{#1\ \textrm{where}\ #2}} 
\newcommand{\for}[2]{\ensuremath{#1\ \textsc{For}\ #2}}
\newcommand{\fst}{\ensuremath{\textsc{Fst}}}
\newcommand{\snd}{\ensuremath{\textsc{Snd}}}
\newcommand{\spawn}[2]{\ensuremath{#1\ {\|}\ #2}}
\newcommand{\void}{\ensuremath{\stackrel{\,{\bot}}{\_\!\_\!\_\!\_}}}
\newcommand{\alltu}[1]{\ensuremath{\mathsf{TU}(#1)}}
\newcommand{\allta}[1]{\ensuremath{\mathsf{TA}(#1)}}
\newcommand{\allte}[1]{\ensuremath{\mathsf{TE}(#1)}}
\newcommand{\allts}[1]{\ensuremath{\mathsf{TS}(#1)}}
\newcommand{\annotate}[2]{\ensuremath{#1\,{:}\,#2}}
\newcommand{\extend}[2]{\ensuremath{#1\,{\lhd}\,#2}}
\newcommand{\restrict}[2]{\ensuremath{#1\,{\rhd}\,#2}}
\newcommand{\dom}[1]{\ensuremath{\textsc{Dom}(#1)}}	
\newcommand{\glb}[3]{\ensuremath{\Gamma\,\vdash#1\ \sqcap\ #2\ \leadsto\ #3}}
\newcommand{\comp}[3]{\ensuremath{\Gamma\,\vdash\,#1;#2\ \leadsto\ #3}}
\newcommand{\negbft}[2]{\ensuremath{\Gamma\,\vdash\neg\,{#1}\ \leadsto\ #2}}
\newcommand{\plus}[2]{\ensuremath{#1\,\&\,#2}}
\newcommand{\lplus}[2]{\ensuremath{#1\,\superimpose{$\longleftarrow$}{$\&$}\,#2}}
\newcommand{\rplus}[2]{\ensuremath{#1\,\superimpose{$\longrightarrow$}{$\&$}\,#2}}
\newcommand{\typeless}{\ensuremath{\prec_{\Gamma}}}
\newcommand{\typeleq}{\ensuremath{\preceq_{\Gamma}}}
\newcommand{\alltp}{\ensuremath{\mathsf{TP}}}
\newcommand{\sps}[1]{\begin{array}{cl}&#1\end{array}}
\newcommand{\dnp}{\ \wedge\ }
\newcommand{\np}{\\\wedge&}
\newcommand{\all}[1]{\ensuremath{\Box(#1)}}
\newcommand{\one}[2]{\ensuremath{\Diamond_{#1}(#2)}}
\newcommand{\some}[1]{\ensuremath{\superimpose{$\Box$}{$\Diamond$}^{#1}}}
\newcommand{\somestar}{\some{*}}
\newcommand{\someplus}{\some{+}}
\newcommand{\someS}{\some{}}
\newcommand{\w}[1]{\mbox{\textsl{#1}}}
\newcommand{\ioj}[2]{\ensuremath{#1 \leadsto #2}}
\newcommand{\cioj}[3]{\ensuremath{#1\,\vdash #2 \leadsto #3}}
\newcommand{\ir}[3]{\ensuremath{%
\begin{array}[b]{c}
#2
\\ \hline
#3
\end{array}\hfill
\begin{tabular}[b]{r}\vspace{.5em}\tg{#1}
\end{tabular}\medskip
}}
\newcommand{\tg}[1]{\ensuremath{{[}\mbox{{\minisize\ensuremath{\mathsf{#1}}}}{]}}}
\newcommand{\neutral}[1]{#1^{+/-}}
\newcommand{\ax}[2]{\ensuremath{%
\begin{array}[b]{c}
#2
\end{array}\hfill
\begin{tabular}[c]{r}\vspace{.5em}{\tg{#1}}
\end{tabular}\medskip
}}
\newcommand{\decj}[2]{\medskip\emph{#1}\hfill\framebox{#2}\medskip}
\newcommand{\dech}[1]{\medskip\emph{#1}}
\newcommand{\wfj}[2]{\ensuremath{#1\ \vdash #2}}
\newcommand{\wtj}[3]{\ensuremath{#1\ \vdash #2\,\mbox{\boldmath{$:$}}\, #3}}
\newcommand{\id}{\ensuremath{\epsilon}}
\newcommand{\fail}{\ensuremath{\delta}}
\newcommand{\failure}{\ensuremath{{\uparrow}}}
\newcommand{\seq}[2]{\ensuremath{#1;#2}}
\newcommand{\choice}[2]{\ensuremath{#1 + #2}}
\newcommand{\lchoice}[2]{\ensuremath{#1\,\superimpose{$\leftarrow$}{$+$}\,#2}}
\newcommand{\rchoice}[2]{\ensuremath{#1\,\superimpose{$\rightarrow$}{$+$}\,#2}}
\newcommand{\rec}[2]{\ensuremath{\mu\,#1.\ #2}}
\newcommand{\apply}[2]{\ensuremath{#1\,@\,#2}}
\newcommand{\myend}{\hfill$\diamond$}



\newlength{\basewidth}
\newlength{\topwidth}
\newcommand{\superimpose}[2]{%
  \settowidth{\basewidth}{#1}%
  \settowidth{\topwidth}{#2}%
  \ifthenelse{\lengthtest{\basewidth > \topwidth}}%
    {\makebox[0pt][l]{#1}\makebox[\basewidth]{#2}}%
    {\makebox[0pt][l]{#2}\makebox[\topwidth]{#1}}%
}

\section{Preface}

\paragraph*{Strategic programming}

Term rewriting strategies are of prime importance for the
\emph{implementation} of term rewriting systems. In the present
paper, we focus on another application of strategies, namely on their
utility for \emph{programming}.  Strategies can be used to describe
evaluation and normalisation strategies, e.g., to explicitly control
rewriting for a system that is not confluent or terminating.
Moreover, strategies can be used to perform traversal, and to describe
reusable traversal schemes. In fact, the \emph{typeful treatment of
generic traversal} is the primary subject of the present paper. To
perform traversal in standard rewriting without extra support for
traversal, one has to resort to auxiliary function symbols, and
rewrite rules have to be used to encode the actual traversal for the
signature at hand. This usually implies one rewrite rule per term
constructor, per traversal. This problem has been identified in
\cite{BSV97,VBT98,LVK00,BSV00,BKV01,Visser01-WRS} from different
points of view. In a framework, where traversal strategies are
supported, the programmer can focus on the term patterns which require
problem-specific treatment. All the other patterns can be covered once
and for all by the generic part of a suitable strategy.

\paragraph*{Application potential}

Language concepts for generic term traversal support an important
dimension of generic programming which is useful, for example, for the
implementation of program transformations and program analyses. Such
functionality is usually very uniform for most patterns in the
traversed syntax. In \cite{Visser00}, untyped, suitably parameterised
traversal strategies are used to capture algorithms for free variable
collection, substitution, unification in a generic, that is,
language-independent manner. In~\cite{LV01}, typed traversal
strategies are employed for the specification of refactorings for
object-oriented programs in a concise manner. There are further
ongoing efforts to apply term rewriting strategies to the modular
development of interpreters, to language-independent refactoring, to
grammar engineering, and others.

\paragraph*{\mycalc\ and relatives}

In the present paper, the rewriting calculus \mycalc\ is
developed. The calculus corresponds to a simple but expressive
language for generic programming. The design of
\mycalc\ was influenced by existing rewriting frameworks with support
for strategies as opposed to frameworks which assume a fixed built-in
strategy for normalisation / evaluation. Strategies are supported, for
example, by the specification formalisms Maude~\cite{CELM96,Maude99}
and ELAN~\cite{BKKMR98,BKKR01}. The $\rho$-calculus~\cite{CK99}
provides an abstract model for rewriting including the definition of
strategies. The programming language Stratego~\cite{VBT98} based on
system $S$~\cite{VB98} is entirely devoted to strategic
programming. In fact, the ``$S$'' in \mycalc\ refers to system $S$
which was most influential in the design of \mycalc. The ``$'$'' in
\mycalc\ indicates that even the untyped part of \mycalc\ does not
coincide with system $S$. The ``$\gamma$'' in \mycalc\ stands for the
syntactical domain $\gamma$ of generic strategy types. The idea of
rewriting strategies goes back to Paulson's work on higher-order
implementation of rewriting strategies~\cite{Paulson83} in the context
of the implementation of tactics and tacticals for theorem
proving. The original contribution of \mycalc\ is the \emph{typeful}
approach to generic \emph{traversal} strategies in a
\emph{many-sorted} setting of term rewriting.

\begin{figure}
\vspace{-30\in}
\resizebox{\textwidth}{.53\textheight}{\includegraphics{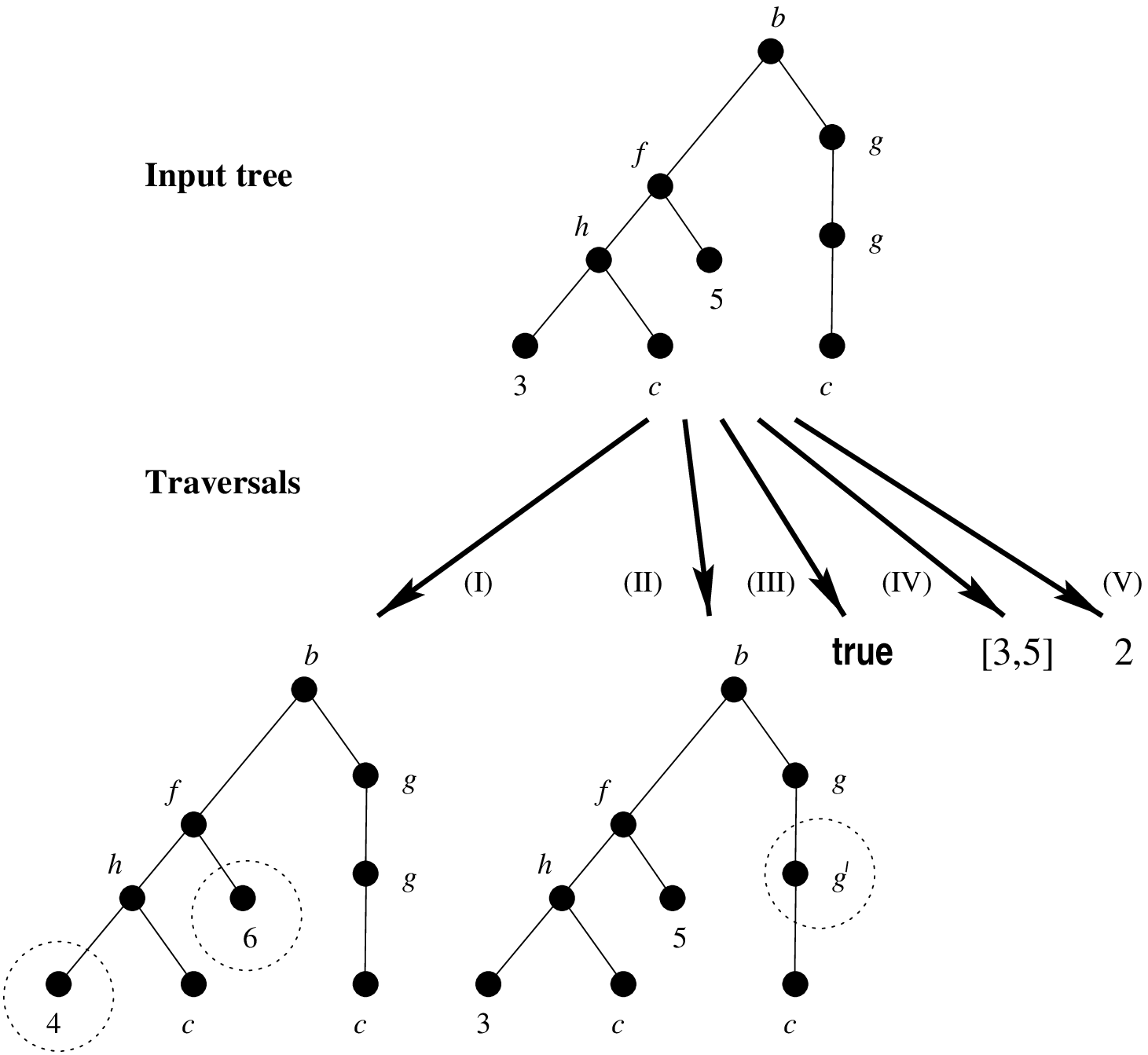}}
\vspace{-20\in}
\caption{Illustration of generic traversal}
\label{F:illustrate}
\bigskip
\end{figure}

\paragraph*{Examples of generic traversal}

In Figure~\ref{F:illustrate}, five examples \s{(I)}--\s{(V)} of
intentionally generic traversal are illustrated. In \s{(I)}, all
naturals in the given term (say, tree) are incremented as modelled by
the rewrite rule $N \to \w{succ}(N)$. We need to turn this rule into a
traversal strategy because the rule on its own is not terminating when
considered as a rewrite system. The strategy should be generic, that
is, it should be applicable to terms of any sort. In \s{(II)}, a
particular pattern is rewritten according to the rewrite rule $g(P)
\to g'(P)$. Assume that we want to control this replacement so that it
is performed in bottom-up manner, and the first (i.e., bottom-most)
matching term is rewritten only. Indeed, in Figure~\ref{F:illustrate},
only one $g$ is turned into a $g'$, namely the deeper one. The
strategy to locate the desired node in the term is completely
generic. While
\s{(I)}--\s{(II)} require type-preserving traversal,
\s{(III)}--\s{(V)}\ require type-changing traversal. 
We say that type-unifying traversal~\cite{LVK00} is needed because the
results of \s{(III)}--\s{(V)}\ are of a fixed, say a unified type. In
\s{(III)}, we test some property of the term, namely if naturals occur
at all. The result is of type Boolean. In \s{(IV)}, we collect all the
naturals in the term using a left-to-right traversal. That is, the
result is a list of integers. Finally, in \s{(V)}, we count all the
occurrences of the function symbol $g$.

\paragraph*{The tension between genericity and specificity}

In addition to a purely many-sorted type system, the rewriting
calculus \mycalc\ offers two designated generic strategy types, namely
the type \alltp\ denoting generic type-preserving strategies, and the
type \alltu{\tau} denoting generic type-unifying strategies with the
unified result type $\tau$. Generic traversal strategies typically
employ many-sorted rewrite rules. Hence, we need to cope with both
many-sorted and generic types, and we somehow need to mediate between
the two levels. Since a traversal strategy must be applicable to terms
of any sort, many-sorted ingredients must be lifted in some way to a
generic type before they can be used in a generic context. As a matter
of fact, a traversal strategy might attempt to apply lifted
many-sorted ingredients to subterms of different sorts. For the sake
of type safety, we have to ensure that many-sorted ingredients are
only applied to terms of the appropriate sort. \mycalc\ offers a
corresponding type-safe combinator for so-called strategy
extension. The many-sorted strategy $s$ is lifted to the generic
strategy type $\gamma$ using the form \extend{s}{\gamma}.  The
extended strategy will immediately fail when applied to a term of a
sort that is different from the domain of $s$. Generic strategies are
composed in a manner that they recover from failure of extended
many-sorted ingredients by applying appropriate generic defaults or by
recursing into the given term.

\paragraph*{Value of typing}

The common arguments in favour of compile-time as opposed to run-time
type checking remain valid for strategic term rewriting. Let us
reiterate some of these arguments in our specific setting to justify
our contribution of a typed rewriting calculus. To start with, the
type system of \mycalc\ and the corresponding reduction semantics
should obviously prevent us from constructing ill-typed
terms. Consider, for example, the rewrite rule $\s{Inc} = N \to
\w{succ}(N)$ of type $\w{Nat} \to \w{Nat}$\ for incrementing naturals
in the context of example \s{(I)}\ above. The left-hand side of
rewrite rule \s{Inc} would actually match with all terms of all sorts,
but it only produces well-typed terms when applied to naturals. A
typed calculus\ prevents us from applying a rewrite rule to a term of
an inappropriate sort.  Admittedly, most rewrite rules use pattern
matching to destruct the input term. In this case, ill-typed terms
cannot be produced. Still, a typed calculus prevents us from
\emph{even attempting} the application of a rewrite rule to a term of
an inappropriate sort. This is very valuable because such attempts
are likely to represent a design flaw in a strategy. Furthermore, a
typed calculus should also prevent the programmer from combining
specific and generic strategies in certain undesirable ways. Consider,
for example, an asymmetric choice $\lchoice{\ell}{\id}$ where a rewrite
rule $\ell$ is strictly preferred over the identity strategy \id, and
only if $\ell$ fails, the identity strategy \id\ triggers.  This
choice is controlled by success and failure of $\ell$. One could argue
that this strategy is generic because the identity strategy \id\ is
applicable to any term. Actually, we favour two other possible
interpretations. One option is to refuse this composition altogether
because we would insist on the types of the branches in a choice to be
the same. Another option is to favour the many-sorted argument type
for the type of the compound strategy. In fact, strategies should not
get generic too easily since we otherwise lose the valuable precision
of a many-sorted type system. Untyped strategic programming suffers
from symptoms such that strategies fail in unexpected manner, or
generic defaults apply to easily. This is basically the same problem
as for untyped programming in Prolog. \mycalc\ addresses all the
aforementioned issues, and it provides static typing for many-sorted
and generic strategies.

\paragraph*{Beyond parametric polymorphism}

Some strategy combinators are easier to type than others. Combinators
for sequential composition, signature-specific congruence operators
and others are easy to type in a many-sorted setting.  By contrast,
generic traversal primitives, e.g., a combinator to apply a strategy
$s$ to all immediate subterms of a given term, are more challenging
since standard many-sorted types are not applicable, and the
well-established concept of parametric polymorphism is insufficient to
model the required kind of genericity. Let us consider the type
schemes underlying the two different forms of generic traversal:
\begin{itemize}
\item $\alltp\ \equiv\ \forall \alpha.\ \alpha \to \alpha$
\hfill \mbox{(i.e., type-preserving traversal)}
\item $\alltu{\tau}\ \equiv\ \forall \alpha.\ \alpha \to \tau$
\hfill \mbox{(i.e., type-unifying traversal)}
\end{itemize}
In the schemes, we point out that $\alpha$ is a universally quantified
type variable. It is easy to see that these schemes are appropriate. A
type-preserving traversal processes terms of any sort (i.e.,
$\alpha$), and returns terms of the same sort (i.e., $\alpha$);
similarly for the type-unifying case. In fact, \mycalc\ does not
enable us to inhabit somewhat arbitrary type schemes. The above two
schemes are the only schemes which can be inhabited with the traversal
combinators of \mycalc. This is also the reason that we do not favour
type schemes to represent types of generic strategies in the first
place, but we rather employ the designated constants \alltp\ and
\alltu{\tau}. If we read the above type schemes in the sense of
parametric polymorphism~\cite{Reynolds83,Wadler89}, we can only
inhabit them in a trivial way. The first scheme can only be inhabited
by the identity function. The second scheme can only be inhabited by a
constant function returning some fixed value of type $\tau$. Generic
traversal goes beyond parametric polymorphism for two
reasons. Firstly, traversal strategies can observe the structure of
terms, that is, they can descend into terms of arbitrary sorts, test
for leafs and compound terms, count the number of immediate subterms,
and others. Secondly, traversal strategies usually exhibit non-uniform
behaviour, that is, there are different branches for certain
distinguished sorts in a traversal. Although strategies are statically
typed in \mycalc, the latter property implies that the reduction
semantics of strategies is type-dependent.

\paragraph*{Structure of the paper}

In Section~\ref{S:rationale}, we provide a gentle introduction to the
subject of strategic programming, and to the rewriting calculus
\mycalc. Examples of traversal strategies are given. The design of
the type system is motivated. As an aside, we use the term ``type''
for types of variables, constant symbols, function symbols, terms,
strategies, and combinators. We also use the term ``sort'' in the
many-sorted sense if it is more suggestive. In
Section~\ref{S:many-sorted}, we start the formal definition of
\mycalc\ with its many-sorted core. In this phase, we cannot
yet cover the traversal primitives. A minor contribution is here that
we show in detail how to cope with type-changing rewrite rules. In
Section~\ref{S:generic}, we provide a type system for generic
strategies. The two aforementioned schemes of type preservation and
type unification are covered. A few supplementary issues to complement
\mycalc\ are addressed in Section~\ref{S:sophi}. Implementation issues
and related work are discussed in Section~\ref{S:impl} and
Section~\ref{S:related}. The paper is concluded in
Section~\ref{S:concl}.

\paragraph*{Objective}

An important meta-goal of the present paper is to develop a simple and
self-contained model of typeful generic programming in the sense of
generic traversal of many-sorted terms. To this end, we basically
resort to a first-order setting of term rewriting. We want to clearly
identify the necessary machinery to accomplish generic traversal in
such a simple setting. We also want to enable a simple implementation
of the intriguing concept of typed generic traversal. The \mycalc\
expressiveness is developed in a stepwise manner. In the course of the
paper, we show that our type system is sensible from a strategic
programmer's point of view. We contend that the type system of
\mycalc\ disciplines strategic programs in a useful and not too
restrictive manner.

\paragraph*{Acknowledgement}

The work of the author was supported, in part, by the Netherlands
Organisation for Scientific Research (NWO), in the project
``\emph{Generation of Program Transformation Systems}''. Some core
ideas spelled out in the paper took shape when I visited the
\emph{Protheo} group at LORIA Nancy in October 2000. I am
particularly grateful for the interaction with my colleague Joost
Visser---in Nancy and in general. Many thanks to Chris\-tophe
Ringeissen who shared an intensive and insightful ELAN session with
Joost and me. I want to thank David Basin, Johan Jeuring, Claude
Kirchner, Paul Klint, Pierre-{\'E}tienne Moreau, Christophe
Ringeissen, Ulf Sch{\"u}nemann, Jurgen Vinju, Eelco Visser, and
Stephanie Weirich for discussions on the subject of the paper. Many
thanks to the anonymous WRS'01 workshop referees for their comments on
an early fragment of this paper (cf.~\cite{Laemmel01-WRS}). This
fragment was later invited for a special issue in the Journal of
Symbolic Computation prior to acceptance of the full paper by the
Journal of Logic and Algebraic Programming. I am grateful for the
detailed remarks and suggestions by the JALP referees. Finally, I am
very grateful to Jan Kort and Jurgen Vinju for their help with
proof-reading the final version of this paper.


\section{Rationale}
\label{S:rationale}

We set up a rewriting calculus \mycalc\ inspired by
ELAN~\cite{BKKMR98,BKKR01} and system $S$~\cite{VB98}. Some basic
knowledge of strategic rewriting is a helpful background for the
present paper (cf.~\cite{BKK96,VB98,CK99}). First, we give an overview
on the primitive strategy combinators of \mycalc. Then, we illustrate
how to define new combinators by means of strategy
definitions. Afterwards, we pay special attention to generic
traversal, that is, we explain the meaning of the traversal primitives,
and we illustrate their expressiveness. In the last part of the
section, we sketch the type system of \mycalc. The subsequent
sections~\ref{S:many-sorted}--\ref{S:sophi} provide a formal
definition of \mycalc.

\subsection{Primitive combinators}
\label{SS:grammar}

In an abstract sense, a term rewriting strategy is a partial
mapping from a term to a term, or to a set of terms. In an extreme
case, a strategy performs normalisation, that is, it maps a term to a
normal form. We use $s$ and $t$, possibly subscripted or primed, to
range over strategy expressions, or terms, respectively. The
application of a strategy $s$ to a term $t$ is denoted by
\apply{s}{t}. The result $r$ of strategy application is called a
\emph{reduct}. It is either a term or ``\failure'' to denote failure.
The primitive combinators of the rewriting calculus \mycalc\ are shown
in Figure~\ref{F:grammar}. Note that we use the term ``combinator'' for
all kinds of operators on strategies, even for constant strategies
like \id\ and \fail\ in Figure~\ref{F:grammar}.

\begin{figure}
{\smallsize

\[\begin{array}{lcll}
s & ::= & t \to t &
\mbox{(Rewrite rule)}
\\
  & |   & \id  &
\mbox{(Identity)}
\\
  & |   & \fail &
\mbox{(Failure)}
\\
  & |   & \seq{s}{s} &
\mbox{(Sequential composition)}
\\
  & |   & \choice{s}{s} &
\mbox{(Non-deterministic choice)}
\\
  & |   & \negbf{s} &
\mbox{(Negation by failure)}
\\
  & |   & c &
\mbox{(Congruence for constant symbol)}
\\
  & |   & f(s,\ldots,s)\ \ \ \ &
\mbox{(Congruence for function symbol)}
\\
  & |   & \all{s} &
\mbox{(Apply strategy to all children)}
\\
  & |   & \one{}{s} &
\mbox{(Apply strategy to one child)}
\\
  & |   & \reduce{}{s}{s} &
\mbox{(Reduce all children)}
\\
  & |   & \select{}{s} &
\mbox{(Select one child)}
\\
 & |   & \void &
\mbox{(Build empty tuple, i.e., \emptytuple)}
\\
 & |   & \spawn{s}{s} &
\mbox{(Apply two strategies to input)}
\\
  & |   & \extend{s}{\gamma} &
\mbox{(Extend many-sorted strategy)}
\end{array}\]

}
\caption{Primitives of \mycalc}
\label{F:grammar}
\bigskip
\end{figure}

\paragraph*{Rewrite rules as strategies}

There is a form of strategy $t_l \to t_r$ for first-order, one-step
rules to be applied at the top of the term. The idea is that if the
given term matches the left-hand side $t_l$, then the input is
rewritten to the right-hand side $t_r$ with the variables in $t_r$
bound according to the match. Otherwise, the rewrite rule
considered as a strategy fails. We adopt some common restrictions for
rewrite rules. The left-hand side $t_l$ determines the bound
variables. (Free) variables on the right-hand $t_r$ side also occur in
$t_l$.

\paragraph*{Basic combinators}

Besides rule formation, there are standard primitives for the identity
strategy (\id), the failure strategy (\fail), sequential composition
(\seq{\cdot}{\cdot}), non-deterministic choice
(\choice{\cdot}{\cdot}), and negation by failure
($\negbf{\cdot}$). Non-deterministic choice means that there is no
prescribed order in which the two argument strategies are considered.
Negation by failure means that \negbf{s}\ fails if and only if $s$
succeeds. In case of success of \negbf{s}, the input term is simply
preserved.  In addition to non-deterministic choice, we should also
allow for asymmetric choice, namely left- vs.\ right-biased choice. We
assume the following syntactic sugar:
\begin{eqnarray*}
\lchoice{s_1}{s_2} & \equiv & \choice{s_1}{(\seq{\negbf{s_1}}{s_2})}\\
\rchoice{s_1}{s_2} & \equiv & \lchoice{s_2}{s_1}
\end{eqnarray*}
That is, in \lchoice{s_1}{s_2}, the left argument has higher priority
than the right one. $s_2$ will only be applied if $s_1$ fails. From an
operational perspective, it would very well make sense to consider
asymmetric choice as a primitive since the above reconstruction
suggests the repeated attempt to perform the preferred strategy. We do
not include asymmetric choice as a primitive because we want to keep
the calculus \mycalc\ as simple as possible.

\paragraph*{Congruences}

Recall that rewrite rules when considered as strategies are applied
at the top of a term. From here on, we use the term ``child'' to
denote an immediate subterm of a term, i.e., one of the $t_i$ in a
term of the form $f(t_1,\ldots,t_n)$. The congruence strategy $f(s_1,
\ldots, s_n)$ provides a convenient way to apply strategies to the
children of a term with $f$ as outermost symbol. More precisely, the
argument strategies $s_1$, \ldots, $s_n$ are applied to the parameters
$t_1$, \ldots, $t_n$ of a term of the form $f(t_1,\ldots,t_n)$. If all
these strategy applications deliver proper term reducts $t'_1$,
\ldots, $t'_n$, then the term $f(t'_1,\ldots,t'_n)$ is constructed,
i.e., the outermost function symbol is preserved. If any child cannot
be processed successfully, or if the outermost function symbol of the
input term is different from $f$, then the strategy fails. The
congruence $c$ for a constant $c$ can be regarded as a test for the
constant $c$. One might consider congruences as syntactic sugar for
rewrite rules which apply strategies to subterms based on
\emph{where}-clauses as introduced later. We treat congruences as
primitive combinators because this is helpful for our presentation:
the generalisation of congruences ultimately leads to the notion of a
generic traversal combinator.

\paragraph*{Notational conventions}

\w{Slanted} type style is used for constant symbols, function symbols,
and sorts. The former start in lower case, the latter in upper case.
\s{Small Caps} type style is used for names of strategies. Variables
in term patterns are potentially subscripted letters in upper case.
We use some common notation to declare constant and function symbols
such as $\w{fork}\,:\,\w{Tree} \times \w{Tree} \to \w{Tree}$. Here,
``$\times$'' denotes the Cartesian product construction for the
parameters of a function symbol.

\begin{example}
\label{X:fliptop}
We can already illustrate a bit of strategic rewriting with the
combinators that we have explained so far. Let us consider the
following problem. We want to flip the top-level subtrees in a binary
tree with naturals at the leafs.  We assume the following symbols to
construct such trees:
\begin{eqnarray*}
\w{zero} & : & \w{Nat}\\
\w{succ} & : & \w{Nat} \to \w{Nat}\\
\w{leaf} & : & \w{Nat} \to \w{Tree}\\
\w{fork} & : & \w{Tree} \times \w{Tree} \to \w{Tree}
\end{eqnarray*}
$N$ and $T$ optionally subscripted or primed are used as variables
of sort \w{Nat} and \w{Tree}, respectively. We can specify the problem
of flipping top-level subtrees with a standard rewrite system. We need
to employ an auxiliary function symbol \w{fliptop} in order to operate
at the top-level.
\[\w{fliptop}(\w{fork}(T_1,T_2)) \to \w{fork}(T_2,T_1)\]
Note that there is no rewrite rule which eliminates \w{fliptop}\ when
applied to a leaf. We could favour the invention of an error tree for
that purpose. Now, let us consider a strategy \s{FlipTop} to flip
top-level subtrees:
\[
\s{FlipTop} = \w{fork}(T_1,T_2) \to \w{fork}(T_2,T_1)\]
That is, we define a strategy, in fact, a rewrite rule \s{FlipTop}\
which rewrites a fork tree by flipping the subtrees. Note that this
rule is non-terminating when considered as a standard rewrite
system. However, when considered as strategy, the rewrite rule is only
applied at the top of the input term, and application is not iterated
in any way. Note also that an application of the strategy \s{FlipTop}\
to a leaf will simply fail. There is no need to invent an error
element. If we want \s{FlipTop} to succeed on a leaf, we can define
the following variant of \s{FlipTop}. We show two equivalent
definitions:
\[\begin{array}{lcl}
\s{FlipTop}' & = & \lchoice{\s{FlipTop}}{\id}\\
& = & \choice{\s{FlipTop}}{\w{leaf}(\id)}
\end{array}\]
In the first formulation, we employ left-biased choice and the identity
\id\ to recover from failure if \s{FlipTop}\ is not applicable. In the
second formulation, we use a case discrimination such that \s{FlipTop}
handles the constructor \w{fork}, and the constructor \w{leaf}\ is
covered by a separate congruence for \w{leaf}.
\end{example}

\paragraph*{Generic traversal combinators}

Congruences can be used for type-specific traversal. Generic traversal
is supported by designated \mycalc\ combinators \all{\cdot}, \one{}{\cdot},
\reduce{}{\cdot}{\cdot}, and \select{}{\cdot}. These traversal
combinators have with congruences in common that they operate on the
children of a term. Since traversal combinators have to cope with any
number of children, one might view them as list-processing functions.
The strategy \all{s} applies the argument strategy $s$ to all children
of the given term. The strategy \one{}{s} applies the argument
strategy $s$ to exactly one child of the given term. The selection of
the child is non-deterministic but constrained by the
success-and-failure behaviour of $s$. The strategies \all{s}\ and
\one{}{s} are meant to be type-preserving since they preserve the
outermost function symbol. The remaining traversal combinators deal
with type-unifying traversal. The strategy \reduce{}{\splus}{s}\
reduces all children. Here $s$ is used to process the children, and
$\splus$\ is used for the pairwise composition of the intermediate
results. We will later discuss the utility of different orders for
processing children. The strategy \select{}{s}\ processes one child
via $s$. The selection of the child is non-deterministic but
constrained by the success-and-failure behaviour of $s$ as in the case
of the type-preserving \one{}{s}.

There are two trivial combinators which are needed for a typeful
treatment of type-unifying strategies. They do not perform traversal
but they are helpers. The strategy \void\ builds the empty tuple
\emptytuple. The strategy \void\ allows us to discard in a sense the
current term of whatever sort, and replace it by the trivial term
\emptytuple. This is useful if we want to migrate to the fixed and
content-free empty tuple type. Such a migration is sometimes needed if
we are not interested in the precise type of the term at hand, e.g.,
if want to encode constant strategies, that is, strategies which
return a fixed term. The strategy \spawn{s_1}{s_2}\ applies the two
strategies $s_1$ and $s_2$ to the input term, and forms a pair from
the results. This is a fundamental form of decomposition relevant for
type-unifying traversal. Obviously, one can nest the application of
the combinator \spawn{\cdot}{\cdot}\ if more than two strategies
should be applied to the input term.

The last combinator \extend{\cdot}{\cdot}\ in Figure~\ref{F:grammar}
serves for lifting a many-sorted strategy such as a rewrite rule to
the generic level. We postpone discussing this combinator until
Section~\ref{SS:typeful} when typed strategic programming is
discussed.

\subsection{Strategy definitions}

New strategy combinators can be defined by means of the abstraction
mechanism for strategy definition. We use $\nu$, possibly subscripted,
for formal strategy parameters in strategy definitions. A strategy
definition $\varphi(\nu_1,\ldots,\nu_n) = s$ introduces an $n$-ary
combinator $\varphi$. Strategy definitions can be recursive. When we
encounter an application $\varphi(s_1,\ldots,s_n)$ of $\varphi$, then
we replace it by the instantiation
$s\assigns{\assign{\nu_1}{s_1},\ldots,\assign{\nu_n}{s_n}}$ of the
body $s$ of the definition of $\varphi$. This leads to a sufficiently
lazy style of unfolding strategy definitions.

\begin{figure}
{\smallsize\begin{center}$\begin{array}{lcll}
\s{Try}(\nu) & = & \lchoice{\nu}{\id}
& \mbox{(Apply $\nu$ if possible, succeed otherwise)}
\\
\s{Repeat}(\nu) & = & \s{Try}(\nu;\s{Repeat}(\nu))\ \ \ \ 
& \mbox{(Apply $\nu$ as often as possible)}
\\
\s{Chi}(\nu,\nu_t,\nu_f) & = & \lchoice{(\seq{\nu}{\nu_t})}{(\seq{\void}{\nu_f)}}
& \mbox{(``Characteristic function'')}
\end{array}$\end{center}}
\caption{Reusable strategy definitions}
\label{F:defined0}
\bigskip
\end{figure}

In Figure~\ref{F:defined0}, three simple strategy definitions are
shown. These definitions embody idioms which are useful in strategic
programming. Firstly, $\s{Try}(s)$\ denotes the idiom to try $s$ but
to succeed via $\id$ if $s$ fails. Secondly, $\s{Repeat}(s)$ denotes
exhaustive iteration in the sense that $s$ is performed as many times
as possible. Thirdly, $\s{Chi}(s,s_t,s_f)$ is intended to map success
and failure of $s$ to ``constants'' $s_t$ and $s_f$, respectively. To
this end, $s$ is supposed to compute \emptytuple\ (if it succeeds),
while $s_t$ and $s_f$ map the ``content-free'' \emptytuple\ to some
term. The helper \void\ is used in the right branch to prepare for
the application of the constant strategy $s_f$.

\begin{example}
\label{X:flipall}
Recall Example~\ref{X:fliptop} where we defined a strategy \s{FlipTop}
for flipping top-level subtrees. Let us define a strategy \s{FlipAll}
which flips subtrees at all levels:
\[\s{FlipAll} = \s{Try}(\seq{\s{FlipTop}}{\w{fork}(\s{FlipAll},\s{FlipAll})})\]
Note how the congruence for fork trees is used to apply \s{FlipAll} to
the subtrees of a fork tree.
\end{example}

\paragraph*{Polyadic strategies}

Many strategies need to operate on several terms.  Consider, for
example, a strategy for addition. It is supposed to take two
naturals. There are several ways to accomplish strategies with
multiple term arguments. Firstly, the programmer could be required to
define function symbols for grouping. Although this is a very simple
approach to deal with polyadic strategies, it is rather inconvenient
for the programmer because (s)he has to invent designated function
symbols. Secondly, we could introduce a special notation to allow a
kind of polyadic strategy application with multiple term
positions. This will not lead to an attractive simple
calculus. Thirdly, we could consider curried strategy
application. This would immediately lead to a higher-order
calculus. Recall that we want stay in a basically first-order
setting. Fourthly, polyadic strategies could be based on polymorphic
tuple types. This is the option we choose. There are distinguished
constructors for tuples. The constant symbol \emptytuple\ represents
the empty tuple, and a pair is represented by \pair{t_1}{t_2}. The
notions of rewrite rules and congruence strategies are immediately
applicable to tuples. For simplicity, we do not consider arbitrary
polymorphic types in \mycalc, but we restrict ourselves to polymorphic
tuples in \mycalc.

\begin{example}
To map a pair of naturals to the first component, the rewrite rule
$\pair{N_1}{N_2} \to N_1$ is appropriate. To flip the top-level
subtrees of a pair of fork trees, the congruence
\pair{\s{FlipTop}}{\s{FlipTop}}\ is appropriate.
\end{example}

\begin{example}
\label{X:add}
The following confluent and terminating rewrite system defines addition of
naturals in the common manner:
\begin{eqnarray*}
& & \w{add} : \w{Nat} \times \w{Nat} \to \w{Nat}\\
& & \w{add}(N,\w{zero}) \to N\\
& & \w{add}(N_1,\w{succ}(N_2)) \to \w{succ}(\w{add}(N_1,N_2))
\end{eqnarray*}
That is, \w{add} is a function symbol to group two naturals to be
added. We rely on a normalisation strategy such as innermost to
actually perform addition. By contrast, we can also define a polyadic
strategy $\s{Add}$ which takes a pair of naturals:
\begin{eqnarray*}
\s{Dec}     & = & \w{succ}(N) \to N\\
\s{Inc}     & = & N \to \w{succ}(N)\\
\s{Add}_{\w{base}} & = & \pair{N}{\w{zero}} \to N\\
\s{Add}_{\w{step}} & = & \seq{\pair{\id}{\s{Dec}}}{\seq{\s{Add}}{\s{Inc}}}\\
\s{Add}            & = & \choice{\s{Add}_{\w{base}}}{\s{Add}_{\w{step}}}
\end{eqnarray*}
For clarity of exposition, we defined a number of auxiliary
strategies.  $\s{Dec}$ attempts to decrement a natural. $\s{Inc}$
increments a natural.  Actual addition is performed according to the
scheme of primitive recursion with the helpers $\s{Add}_{\w{base}}$
and $\s{Add}_{\w{step}}$ for the base and the step case. Both cases
are mutually exclusive. The base case is applicable if the second
natural is a \w{zero}. The step case is applicable if the second
natural is a non-zero value since \s{Dec} will otherwise fail. Notice
how a congruence for pairs is employed in the step case.
\end{example}

\paragraph*{Where-clauses}

For convenience, we generalise the concept of rewrite rules as
follows. A rewrite rule is of the form $t \to b$ where $t$ is the term
of the left-hand side as before, and $b$ is the right-hand side body
of the rule. In the simplest case, a body $b$ is a term $t'$ as
before. However, a body can also involve where-clauses. Then $b$ is of
the following form:
\[\where{b'}{x=\apply{s}{t'}}\]
The meaning of such a body with a where-clause is that the term reduct
which results from the strategy application \apply{s}{t'} is bound to
$x$ for the evaluation of the remaining body $b'$.  For simplicity, we
assume a linear binding discipline, that is, $x$ is not bound
elsewhere in the rule.

\begin{example}
\label{X:add'}
We illustrate the utility of where-clauses by a concise reconstruction of
the strategy \s{Add} from Example~\ref{X:add}:
\[\begin{array}{rcl}
\s{Add}            & = & \pair{N}{\w{zero}} \to N\\
                   & \choice{}{} &
\pair{N_1}{\w{succ}(N_2)} \to
\where{\w{succ}(N_3)}{N_3 = \apply{\s{Add}}{\pair{N_1}{N_2}}}
\end{array}\]
The strategy takes roughly the form of an eager
functional program with pattern-match case \`a la SML.
\end{example}

\subsection{Generic traversal}

Let us discuss the core asset of \mycalc, namely its combinators for
generic traversal in some detail. To prepare the explanation of the
corresponding primitives, we start with a discussion of how to encode
traversal in standard rewriting. By ``standard rewriting'', we mean
many-sorted, first-order rewriting based on a fixed normalisation
strategy. We derive the strategic style from this
encoding. Afterwards, we will define a number of reusable schemes for
generic traversal in terms of the \mycalc\ primitives.  Ultimately, we
will provide the encodings for the traversal problems posed in the
introduction.

\paragraph*{Traversal functions}

Suppose we want to traverse a term of a certain sort.  In the course
of traversing into the term, we need to process the subterms of it at
maybe all levels. In general, these subterms are of different
sorts. If we want to encode traversal in standard rewriting, we
basically need an auxiliary function symbol for each traversed sort to
map it to the corresponding result type. Usually, one has to define
one rewrite rule per constructor in the signature at hand.

\begin{example}
\label{X:count}
Let us define a traversal to count leafs in a tree.  Note that a
function from trees to naturals is obviously type-changing. Consider
the following rewrite rule:
\[\s{Count}_{\w{leaf}} = \w{leaf}(N) \to \w{succ}(\w{zero})\]
This rule directly models the essence of counting leafs, namely it
says that a leaf is mapped to $1$, i.e., $\w{succ}(\w{zero})$. In
standard rewriting, we cannot employ the above rewrite rule since it
is type-changing. Instead, we have to organise a traversal with
rewrite rules for an auxiliary function symbol \w{count}:
\[\begin{array}{l}
\w{count} : \w{Tree} \to \w{Nat}\\
\w{count}(\w{leaf}(N)) \to \w{succ}(\w{zero})\\
\w{count}(\w{fork}(T_1,T_2)) \to \w{add}(\w{count}(T_1),\w{count}(T_2))
\end{array}\]
The first rewrite rule restates $\s{Count}_{\w{leaf}}$ in a
type-preserving manner. The second rewrite rule is only there to
traverse into fork trees. In this manner, we can cope with arbitrarily
nested fork trees, and we will ultimately reach the leafs that have to
be counted. Note that if we needed to traverse terms which involve
other constructors, then designated rewrite rules had to be provided
along the schema used in the second rewrite rule for \w{fork} above.
That is, although we are only interested in leafs, we still have to
skip all other constructors to reach leafs. To be precise, we have to
perform addition all over the place to compute the total number of
leafs from the number of leafs in subterms.
\end{example}

\paragraph*{Traversal strategies}

Traversal based on such auxiliary function symbols and rewrite rules
gets very cumbersome when larger signatures, that is, more
constructors, are considered. This problem has been clearly
articulated in~\cite{BSV97,VBT98,LVK00,BSV00,BKV01,Visser01-WRS}.  An
application domain which deals with large signatures is program
transformation. Signatures correspond here to language syntaxes. The
aforementioned papers clearly illustrate the inappropriateness of the
manual encoding of traversal functions for non-trivial program
transformation systems.  The generic traversal facet of strategic
programming solves this problem in the most general way. In strategic
rewriting, we do not employ auxiliary function symbols and rewrite
rules to encode traversal, but we rely on expressiveness to process the
children of a term in a uniform manner. In fact, traversal combinators
allow us to process children regardless of the outermost constructor
and the type of the term at hand.

\begin{example}
\label{X:count1}
Let us attempt to rephrase Example~\ref{X:count} in strategic style.
We do not want to employ auxiliary function symbols, but we want to
employ the type-changing rewrite rule $\s{Count}_{\w{leaf}}$\ for
handling the terms of interest. In our first attempt, we do not yet
employ traversal combinators. We define a strategy \s{Count} as
follows:
\begin{eqnarray*}
\s{Unwrap}_{\w{fork}}   & = & \w{fork}(T_1,T_2) \to \pair{T_1}{T_2}\\
\s{Count}_{\w{leaf}} & = & \w{leaf}(N) \to \w{succ}(\w{zero})\\
\s{Count}_{\w{fork}} & = & \seq{\s{Unwrap}_{\w{fork}}}{\seq{\pair{\s{Count}}{\s{Count}}}{\s{Add}}}\\
\s{Count}            & = & \choice{\s{Count}_{\w{leaf}}}{\s{Count}_{\w{fork}}}
\end{eqnarray*}
The helper $\s{Count}_{\w{fork}}$ specifies how to count the leafs of
a proper fork tree. That is, we first turn the fork tree into a pair
of its subtrees via $\s{Unwrap}_{\w{fork}}$, then we perform counting
for the subtrees by means of a type-changing congruence on pairs, and
finally the resulting pair is fed to the strategy for addition. Note
that the recursive formulation of \s{Count}\ allows us to traverse
into arbitrarily nested fork trees. In order to obtain a more generic
version of \s{Count}, we can use a traversal combinator to abstract
from the concrete constructor in $\s{Count}_{\w{fork}}$. Here is a
variant of \s{Count} which can cope with any constructor with one or
more children:
\begin{eqnarray*}
\s{Count}_{\w{leaf}} & = & \w{leaf}(N) \to \w{succ}(\w{zero})\\
\s{Count}_{\w{any}}  & = & \reduce{}{\s{Add}}{\s{Count}}\\
\s{Count}            & = & \lchoice{\s{Count}_{\w{leaf}}}{\s{Count}_{\w{any}}}
\end{eqnarray*}
That is, we use the combinator \reduce{}{\cdot}{\cdot}\ to reduce
children accordingly. Note that left-biased choice is needed in the
new definition of \s{Count} to make sure that $\s{Count}_{\w{leaf}}$
is applied whenever possible, and we only descend into the term for
non-leaf trees. We should finally mention that the strategy \s{Count}
is not yet fully faithful regarding typing because we pass the
many-sorted strategy \s{Count} to \reduce{}{\cdot}{\cdot} whereas the
argument for processing the children is intentionally generic.
\end{example}

\begin{figure}
{\smallsize

\noindent
$\begin{array}{lclr}
\s{Con} & = & \all{\fail}
&
\mbox{(Test for a constant)}
\\
\s{Fun} & = & \one{}{\id}
&
\mbox{(Test for a compound term)}
\\
\somestar(\nu) & = & \all{\s{Try}(\nu)}
&
\mbox{(Process several children)}
\\
\someplus(\nu) & = & \seq{\negbf{\all{\negbf{\nu}}}}{\somestar(\nu)}
&
\mbox{(Process at least one child)}
\\
\s{TD}(\nu)   & = & \seq{\nu}{\all{\s{TD}(\nu)}}
&
\mbox{(Top-down traversal)}
\\
\s{BU}(\nu)  & = & \seq{\all{\s{BU}(\nu)}}{\nu}
&
\mbox{(Bottom-up traversal)}
\\
\s{OnceTD}(\nu)    & = & \lchoice{\nu}{\one{}{\s{OnceTD}(\nu)}}
&
\mbox{(Process one node in top-down manner)}
\\
\s{OnceBU}(\nu)    & = & \rchoice{\nu}{\one{}{\s{OnceBU}(\nu)}}
&
\mbox{(Process one node in bottom-up manner)}
\\
\s{Innermost}(\nu) & = & \s{Repeat}(\s{OnceBU}(\nu))
&
\mbox{(Innermost evaluation strategy)}
\\
\s{StopTD}(\nu) & = & \lchoice{\nu}{\all{\s{StopTD}(\nu)}}
&
\mbox{(Top-down traversal with ``cut'')}
\end{array}$
}
\caption{Definitions of type-preserving combinators}
\label{F:defined1TP}
\bigskip
\end{figure}

\begin{figure}
{\smallsize

\noindent
$\begin{array}{lclr}

\s{Any}(\nu) & = & \choice{\nu}{\select{}{\s{Any}(\nu)}}
\\
\s{Tm}(\nu) & = & \lchoice{\nu}{\select{}{\s{Tm}(\nu)}}
\\
\s{Bm}(\nu) & = & \rchoice{\nu}{\select{}{\s{Bm}(\nu)}}
\\
\s{CF}(\nu,\vunit,\vplus) & = &
\choice{(\seq{\s{Con}}{\seq{\void}{\vunit}})}{(\seq{\s{Fun}}{\reduce{}{\vplus}{\nu}})}
\\
\s{Crush}(\nu,\vunit,\vplus) & = &
\seq{(\spawn{%
\nu
}{%
\s{CF}(\s{Crush}(\nu,\vunit,\vplus),\vunit,\vplus)
})}{\vplus}
\\
\s{StopCrush}(\nu,\vunit,\vplus) & = & \lchoice{\nu}{%
\s{CF}(\s{StopCrush}(\nu,\vunit,\vplus),\vunit,\vplus)}
\end{array}$

}
\caption{Definitions of type-unifying combinators}
\label{F:defined1TU}
\bigskip
\end{figure}

\paragraph*{Traversal schemes}

In Figure~\ref{F:defined1TP} and Figure~\ref{F:defined1TU}, we derive
some combinators for generic traversal. Most of the combinators should
actually be regarded as reusable definitions of traversal schemes. The
definitions immediately illustrate the potential of the generic
traversal combinators. Several of the definitions from the
type-preserving group are adopted from~\cite{VB98}. We postpone
discussing typing issues for a minute. Let us read a few of the given
definitions. The strategy $\s{TD}(s)$ applies $s$ to each node in
top-down manner. This is expressed by sequential composition such that
$s$ is first applied to the current node, and then we recurse into the
children. It is easy to see that if $s$ fails for any node, the
traversal fails entirely. A similar derived combinator is
\s{StopTD}. However, left-biased choice instead of sequential
composition is used to transfer control to the recursive part.  Thus,
if the strategy succeeds for the node at hand, the children will not
be processed anymore. Another insightful, intentionally
type-preserving example is \s{Innermost} which directly models the
innermost normalisation strategy known from standard rewriting. The
first three type-unifying combinators \s{Any},
\s{Tm}, and \s{Bm} deal with the selection of a subterm. They all have
in common that they resort to the selection combinator
\select{}{\cdot} to determine a suitable child. They differ in the
sense that they perform search either non-deterministically, or in
top-down manner, or in bottom-up manner. One might wonder whether it
is sensible to vary the horizontal order as well. We will discuss this
issue later. The combinator \s{CF} complements \reduce{}{\cdot}{\cdot}
to also cope with a constant. To this end, there is an additional
parameter \vunit\ for the ``neutral element'' to be applied when a
constant is present. The combinators \s{Crush}\ and
\s{StopCrush}\ model deep reduction based on the same kind of
monoid-like argument strategies as \s{CF}. As an aside, the term
crushing has been coined in the related context of polytypic
programming~\cite{Meertens96}. The combinator \s{Crush} evaluates each
node in the tree, and hence, it needs to succeed for each node. The
reduction of the current node and the recursion into the children is
done in parallel based on \spawn{\cdot}{\cdot}. The corresponding pair
of intermediate results is reduced with the binary monoid
operation. \s{StopCrush} is similar to \s{StopTD} in the sense that
the current node is first evaluated, and only if evaluation fails,
then we recurse into the children.

\begin{figure}
{\smallsize

$\begin{array}{lclr}
& & \mbox{Combinators on Booleans}
\\
\s{False} & = & \emptytuple \to \w{false}
& \mbox{(Build ``false'')}
\\
\s{True}  & = & \emptytuple \to \w{true}
& \mbox{(Build ``true'')}
\\
\\
& & \mbox{Combinators on naturals}
\\
\s{Nat}   & = & \choice{\w{zero}}{\w{succ}(\id)}
& \mbox{(Test by congruences)}
\\
\s{Zero}  & = & \emptytuple \to \w{zero}
& \mbox{(Build ``0'')}
\\
\s{One}   & = & \emptytuple \to \w{succ}(\w{zero})
& \mbox{(Build ``1'')}
\\
\s{Inc}   & = & N \to \w{succ}(N)
& \mbox{(Increment)}
\\
\s{Add}   & = & ...
& \mbox{(Addition; see Example~\ref{X:add})}
\\
\\
& & \mbox{Combinators on lists of naturals}
\\
\s{Nil}   & = & \emptytuple \to \w{nil}
& \mbox{(Build ``nil'')}
\\
\s{Singleton} & = & N \to \w{cons}(N,\w{nil})
& \mbox{(Construct a singleton list)}
\\
\s{Append} & = & ...
& \mbox{(Append two lists; definition postponed)}
\\
\\
& & \mbox{Actual encodings of \s{(I)}--\s{(IV)}}
\end{array}$

$\begin{array}{lcl}
\s{(I)}   & = & \s{StopTD}(\seq{\s{Nat}}{\s{Inc}})\\
\s{(II)}  & = & \s{OnceBU}(g(P) \to g'(P))\\
\s{(III)} & = & \s{Chi}(\s{Any}(\seq{\s{Nat}}{\void}),\s{True},\s{False})\\
\s{(IV)}  & = & \s{StopCrush}(\seq{\s{Nat}}{\s{Singleton}},\s{Nil},\s{Append})\\
\s{(V)}  & = & \s{Crush}(\s{Chi}(\seq{g(\id)}{\void},\s{One},\s{Zero}),\s{Zero},\s{Add})
\end{array}$

}
\caption{Untyped encodings for traversal problems from Figure~\ref{F:illustrate}}
\label{F:untyped}
\bigskip
\end{figure}

\begin{example}
\label{X:untyped}
Let us solve the problems (I)--(V) illustrated in
Figure~\ref{F:illustrate} in the introduction of the paper.  In
Figure~\ref{F:untyped}, we first define some auxiliary strategies on
naturals, Booleans, and lists of naturals, and then, the ultimate
traversals (I)--(V) are defined in terms of the combinators from
Figure~\ref{F:defined0}--Figure~\ref{F:defined1TU}. Note that the
encodings are not yet fully faithful regarding typing. We will later
revise these encodings accordingly.  Let us explain the strategies in
detail.

\begin{description}
\item[\s{(I)}] We are supposed to increment all naturals. The combinator
\s{StopTD} is employed to descend into the given term as long as we do not
find a natural recognised via \s{Nat}. When we encounter a natural
in top-down manner, we apply the rule \s{Inc}\ for incrementing
naturals.  Note that we must not further descend into the term. In
fact, if we used \s{TD} instead of \s{StopTD}, we describe a
non-terminating strategy.  Also note that a bottom-up traversal is not
an option either. If we used \s{BU} instead of \s{StopTD}, we
model the replacement of a natural $N$ by $2N+1$.
\item[\s{(II)}] We want to replace terms of the form $g(P)$ by $g'(P)$.
As we explained in the introduction, the replacement must not be done
exhaustively. We only want to perform one replacement where the
corresponding redex should be identified in bottom-up manner. These
requirements are met by the combinator \s{OnceBU}.
\item[\s{(III)}] We want to find out if naturals occur in the term. The
result should be encoded as a Boolean; hence, the two branches
\s{True}\ and \s{False}\ in \s{Chi}. We look for naturals again via the
auxiliary strategy \s{Nat}. The kind of deep matching we need is
provided by the combinator \s{Any} which non-deterministically looks
for a child where \s{Nat} succeeds. \s{Nat} is followed by \void\ to
express that we are not looking for actual naturals but only for the
property if there are naturals at all. The application of \s{Chi}
turns success and failure into a Boolean.
\item[\s{(IV)}] To collect all naturals in a term, we need to perform a
kind of deep reduction. Here, it is important that reduction with cut
(say, \s{StopCrush}) is used because a term representing a non-zero
natural $N$ ``hosts'' the naturals $N-1, \ldots, 0$ due to the
representation of naturals via the constructors \w{succ} and
\w{zero}. These hosted naturals should not be collected. Recall that
crushing uses monoid-like arguments.  In this example, \s{Append} is
the associative operation of the monoid, and the strategy \s{Nil}\ to
build the empty list represents the unit of \s{Append}.
\item[\s{(V)}] Finally, we want to count all occurrences of $g$. In order
to locate these occurrences, we use the congruence $g(\id)$. In this
example, it is important that we perform crushing exhaustively, i.e.,
without cut, since terms rooted by $g$ might indeed host further
occurrences of $g$. We assume that all occurrences of $g$ have to be
counted.
\end{description}
\end{example}

Note the genericity of the defined strategies (I)--(V). They can be
applied to any term. Of course, the strategies are somewhat specific
because they refer to some concrete constant or function symbols,
namely \w{true}, \w{false}, \w{zero}, \w{succ}, $g$, and $g'$.

\subsection{Typed strategies}
\label{SS:typeful}

Let us now motivate the typeful model of strategic programming
underlying \mycalc. The ultimate challenge is to assign types to
generic traversal strategies like \s{TD}, \s{StopTD}, or
\s{Crush}. Recall our objective for \mycalc\ to stay in a
basically first-order many-sorted term rewriting setting.  The type
system we envisage should be easy to define and implement.

\paragraph*{Many-sorted types}

Let us start with a basic, many-sorted fragment of \mycalc\ without
support for generic traversal. We use $\tau$ and $\pi$, possibly
subscripted or primed, to range over term types or strategy types,
respectively. Term types are sorts and tuple types. We use
\pair{\tau_1}{\tau_2} to denote the product type for pairs
\pair{t_1}{t_2}. The type of the empty tuple \emptytuple\ is simply
denoted as \emptytuple. A strategy type $\pi$ is a first-order
function type, that is, $\pi$ is of the form $\tau \to \tau'$. Here,
$\tau$ is the type of the input term, and $\tau'$ is the type of the
term reduct. We also use the terms domain and co-domain for $\tau$ or
$\tau'$, respectively. The type declaration for a strategy combinator
$\varphi$ which does not take any strategy arguments is of the form
$\varphi: \pi$.  The type declaration for a strategy combinator
$\varphi$\ with $n \geq 1$ arguments is represented in the following
format:
\[\varphi: \pi_1 \times \cdots \times \pi_n \to \pi_0\]
Here, $\pi_1$, \ldots, $\pi_n$ denote the strategy types for the
argument strategies, and $\pi_0$ denotes the strategy type of an
application of $\varphi$. All the $\pi_i$ are again of the form
$\tau_i \to \tau'_i$.  Consequently, strategy combinators correspond
to second-order functions on terms. This can be checked by counting
the level of nesting of arrows ``$\to$'' in a combinator type.

\begin{example}
We show the type of \s{FlipAll}\ from Example~\ref{X:flipall}, the
type of the congruence combinator $\w{fork}(\cdot,\cdot)$ for the
function symbol \w{fork} used in Example~\ref{X:flipall}, and the type
of \s{Add}\ from Example~\ref{X:add}.
\[\begin{array}{lcl}
\s{FlipAll} & : & \w{Tree} \to \w{Tree}\\
\w{fork}    & : & (\w{Tree} \to \w{Tree}) \times (\w{Tree} \to \w{Tree})
\to (\w{Tree} \to \w{Tree})\\
\s{Add}     & : & \pair{\w{Nat}}{\w{Nat}} \to \w{Nat}
\end{array}\]
\end{example}

\paragraph*{Type inference vs.\ type checking}

For simplicity, we assume that the types of all function and constant
symbols, variables, and strategy combinators are explicitly
declared. This is well in line with standard practice in term
rewriting and algebraic specification.  Declarations for variables,
rewriting functions and strategies are common in several frameworks
for rewriting, e.g., in CASL, ASF+SDF, and ELAN. Note however that
this assumption is not essential. Inference of types for all symbols
is feasible. In fact, type inference is simple because the special
generic types of \mycalc\ are basically like constant types, and their
inhabitation is explicitly marked by the combinator
\extend{\cdot}{\cdot}. We will eventually add a bit of parametric
polymorphism to \mycalc\ but since we restrict ourselves to top-level
quantification, type inference will still be feasible.

\begin{example}
\label{X:append}
To illustrate type declarations, we define a strategy \s{Append}\ to
append two lists. For simplicity, we do not consider a polymorphic
\s{Append}, but one that appends lists of naturals.  We declare all
the constant and function symbols (namely \w{nil} and \w{cons}), and
variables for lists (namely $L_1$, $L_2$, $L_3$) and naturals as list
elements (namely $N$).

{\smallsize
\[\begin{array}{lcl}
\w{nil}       & : & \w{NatList}\\
\w{cons}      & : & \w{Nat} \times \w{NatList} \to \w{NatList}\\
L_1, L_2, L_3 & : & \w{NatList}\\
N             & : & \w{Nat}\\
\s{Append}    & : & \pair{\w{NatList}}{\w{NatList}} \to \w{NatList}\\
\s{Append}    & = & \pair{\w{nil}}{L} \to L\\
              & + & \pair{\w{cons}(N,L_1)}{L_2} \to
\where{\w{cons}(N,L_3)}{L_3=\apply{\s{Append}}{\pair{L_1}{L_2}}}
\end{array}\]}
\end{example}

\paragraph*{Generic types}

In order to provide types for generic strategies, we need to extend
our basically many-sorted type system. To this end, we identify
distinguished generic types for strategies which are applicable to all
sorts. We use $\gamma$ to range over generic strategy types. There are
two generic strategy types. The type \alltp\ models generic
type-preserving strategies. The type \alltu{\tau}\ models
type-unifying strategies where all types are mapped to $\tau$. These
two forms correspond to the main characteristics of \mycalc. The types
\alltp\ and \alltu{\cdot}\ can be integrated into an initially
many-sorted system in a simple manner.

\begin{example} The following types are the intended ones for the
illustrative strategies defined in Figure~\ref{F:untyped}.
\[\begin{array}{lcl}
\s{(I)}, \s{(II)}  & : & \alltp\\
\s{(III)} & : & \alltu{\w{Boolean}}\\
\s{(IV)}  & : & \alltu{\w{NatList}}\\
\s{(V)}   & : & \alltu{\w{Nat}}
\end{array}\]
\end{example}

\paragraph*{Parametric polymorphism}

Generic strategy types capture the kind of genericity needed for
generic traversal while being able to mix uniform and sort-specific
behaviour. In order to turn \mycalc\ in a somewhat complete
programming language, we also need to enable parametric polymorphism.
Consider, for example, the combinator \s{Crush}\ for deep reduction in
Figure~\ref{F:defined1TU}. The result type of reduction should be a
parameter. The overall scheme of crushing is in fact not dependent on
the actual unified type. The arguments passed to \s{Crush}\ are the
only strategies to operate on the parametric type for unification. We
employ a very simple form of parametric polymorphism. Types of
strategy combinators may contain type variables which are explicitly
quantified at the top level~\cite{Milner78,CW85}. We use $\alpha$,
possibly subscripted, for term-type variables. Thus, in general, a
type of a strategy combinator is of the following form:
\[\varphi: \forall \alpha_1.\ \ldots.\ \forall \alpha_m.\ \pi_1 \times \cdots \times \pi_n \to \pi_0\]
We assume that any type variable in $\pi_0$, \ldots, $\pi_n$ is
contained in the set $\{\alpha_1,\ldots,\alpha_m\}$.  Furthermore, we
assume explicit type application, that is, the application of a
type-parameterised strategy combinator $\varphi$\ involves type
application using the following form:
\[\varphi[\tau_1,\ldots,\tau_m](s_1,\ldots,s_n)\]
For convenience, an actual implementation of \mycalc\ is likely to
support implicit type application. Also, a more complete language
design would include support for parameterised datatypes such as
parameterised lists as opposed to lists of naturals in
Example~\ref{X:append}. For brevity, we omit parameterised datatypes
in the present paper since they are not strictly needed to develop a
typeful model of generic traversal, and a corresponding extension is
routine. Parameterised algebraic datatypes are well-understood in the
context of algebraic specification and rewriting. The instantiation of
parameterised specifications or modules is typically based on
signature morphisms as supported, e.g., in CASL~\cite{CASL01}\ or
ELAN~\cite{BKKMR98}.  A more appealing approach to support
parameterised datatypes would be based on a language design with full
support for polymorphic functions and parameterised data types as in
the functional languages SML and Haskell.

\begin{example}
\label{X:alpha}
Here are the types for the strategy combinators from
Figure~\ref{F:defined0}--Figure~\ref{F:defined1TU}.  All traversal
schemes which involve a type-unifying facet, need to be parameterised
by the unified type.

{\smallsize\[\begin{array}{lcl}
\s{Try}, \s{Repeat} & : & \alltp \to \alltp\\
\s{Chi} & : & \forall \alpha. \alltu{\emptytuple} \to
(\emptytuple \to\alpha) \to
(\emptytuple \to\alpha) \to
\alltu{\alpha}\\
\s{Con}, \s{Fun} & : & \alltp\\
\somestar, \ldots, \s{StopTd} & : & \alltp \to \alltp\\
\s{Any}, \s{Tm}, \s{Bm}
& : & \forall \alpha. \alltu{\alpha} \to \alltu{\alpha}\\
\s{CF}, \s{Crush}, \s{StopCrush}
& : & \forall \alpha. \alltu{\alpha} \times (\emptytuple \to \alpha)
\times (\pair{\alpha}{\alpha} \to \alpha) \to \alltu{\alpha}
\end{array}\]}

\noindent
We update all definitions which involve type parameters:

{\smallsize\[\begin{array}{lcl}
\s{Chi}[\alpha](\nu,\nu_t,\nu_f) & = &
\lchoice{(\seq{\nu}{\nu_t})}{(\seq{\void}{\nu_f)}}
\\
\s{Any}[\alpha](\nu) & = & \choice{\nu}{\select{}{\s{Any}[\alpha](\nu)}}
\\
\s{Tm}[\alpha](\nu) & = & \lchoice{\nu}{\select{}{\s{Tm}[\alpha](\nu)}}
\\
\s{Bm}[\alpha](\nu) & = & \rchoice{\nu}{\select{}{\s{Bm}[\alpha](\nu)}}
\\
\s{CF}[\alpha](\nu,\vunit,\vplus) & = &
\choice{(\seq{\s{Con}}{\seq{\void}{\vunit}})}{(\seq{\s{Fun}}{\reduce{}{\vplus}{\nu}})}
\\
\s{Crush}[\alpha](\nu,\vunit,\vplus) & = &
\seq{(\spawn{%
\nu
}{%
\s{CF}[\alpha](\s{Crush}[\alpha](\nu,\vunit,\vplus),\vunit,\vplus)
})}{\vplus}
\\
\s{StopCrush}[\alpha](\nu,\vunit,\vplus) & = & \lchoice{\nu}{%
\s{CF}[\alpha](\s{StopCrush}[\alpha](\nu,\vunit,\vplus),\vunit,\vplus)}
\end{array}\]}
\end{example}

\begin{example}
\label{X:Try'}
Let us also give an example of a polymorphic strategy definition which
does not rely on the generic strategy types \alltp\ and \alltu{\cdot}\
at the same time. Consider the declaration $\s{Try}: \alltp \to
\alltp$ from Example~\ref{X:alpha}. This type is motivated by the use
of \s{Try}\ in the definition of traversal strategies, e.g., in the
definition of $\somestar(\cdot)$\ in
Figure~\ref{F:defined1TP}. However, the generic type of \s{Try}\
invalidates the application of \s{Try} in Example~\ref{X:flipall}
where it was used to recover from failure of a many-sorted
strategy. We resolve this conflict of interests by the introduction of
a polymorphic combinator $\s{Try}'$\ for many-sorted strategies, and
we illustrate it by a corresponding revision of
Example~\ref{X:flipall}:
\[\begin{array}{lcl}
\s{Try}' & : & \forall \alpha.\ (\alpha \to \alpha) \to (\alpha \to \alpha)\\
\s{Try}'[\alpha](\nu) & = & \lchoice{\nu}{\id}\\
\s{FlipAll} & = & \s{Try}'[\w{Tree}](\seq{\s{Flip}}{\w{fork}(\s{FlipAll},\s{FlipAll})})
\end{array}\]
Hence, we strictly separate many-sorted vs.\ generic recovery from
failure.
\end{example}

\paragraph*{Strategy extension}

The remaining problem with generic strategies is the mediation between
many-sorted and generic strategy types. If we look back to the
simple-minded definition of \s{(I)}\ in Figure~\ref{F:untyped}, we see
that \seq{\s{Nat}}{\s{Inc}} is used as an argument for \s{StopTD}.
The argument is of the many-sorted type $\w{Nat} \to
\w{Nat}$. However, the combinator \s{StopTD} should presumably insist
on a generic argument because the argument strategy is potentially
applied to nodes of all possible sorts. Obviously,
\seq{\s{Nat}}{\s{Inc}} will fail for all terms other than naturals
because \s{Nat}\ performs a type check via congruences for the
constructors of sort \w{Nat}. It turns out that failure of
\seq{\s{Nat}}{\s{Inc}}\ controls the traversal scheme \s{StopTD} in an
appropriate manner. However, if the programmer would have forgotten
the type guard \s{Nat}, the traversal is not type-safe anymore.  In
general, we argue as follows:

\newpage

\begin{quote}
A programmer has to explicitly turn many-sorted strategies into
generic ones. The reduction semantics is responsible for the type-safe
application of many-sorted ingredients in generic contexts.
\end{quote}
To this end, \mycalc\ offers the combinator \extend{\cdot}{\cdot} to
turn a many-sorted strategy into a generic one. A strategy of the form
$\extend{s}{\gamma}$ models the extension of the strategy $s$ to be
applicable to terms of all sorts. In $\extend{s}{\gamma}$, the
$\gamma$ is a generic type, and the strategy $s$ must be of a
many-sorted type $\tau \to \tau'$.  The type $\tau \to \tau'$ of $s$
and the generic type $\gamma$ must be related in a certain way, namely
the type scheme underlying $\gamma$ has to cover the many-sorted type
$\tau \to \tau'$. Strategy extension is performed in the most basic
way, namely $\extend{s}{\gamma}$ fails for all terms of sorts which
are different from the domain $\tau$ of $s$. The reduction semantics
of $\apply{\extend{s}{\gamma}}{t}$\ is truly type-dependent, that is,
reduction involves a check to see whether the type of $t$ coincides
with the domain of $s$ to enable the application of $s$. One should
not confuse this kind of explicit type test and the potential of
failure with an implicit dynamic type check that might lead to program
abort. In typed strategic rewriting, strategy extension is a
programming idiom to create generic strategies. In a sense, failure is
the initial generic default for an extended strategy. Subsequent
application of \choice{\cdot}{\cdot} and friends can be used to
establish behaviour other than failure. That is, one can recover from
failure caused by \extend{\cdot}{\cdot}, and one can resort to a more
useful generic default, e.g., \id, or the recursive branch of a
generic traversal. Strategy extension is essential for the type-safe
application of many-sorted ingredients in the course of a generic
traversal.

\begin{example}
\label{X:typeful}
We revise Example~\ref{X:untyped} to finally supply the typeful
solutions for the traversal problems (I)--(V) from the
introduction. The following definitions are in full compliance with
the \mycalc\ type system:

{\smallsize\[\begin{array}{rcl}
\s{(I)}   & = & \s{StopTD}(\extend{\s{Inc}}{\alltp})\\
\s{(II)}  & = & \s{OnceBU}(\extend{g(P) \to g'(P)}{\alltp})\\
\s{(III)} & = & \s{Chi}[\w{Boolean}](\s{Any}[\emptytuple](\seq{\extend{\s{Nat}}{\alltp}}{\void}),\s{True},\s{False})\\
\s{(IV)}  & = & \s{StopCrush}[\w{NatList}](\seq{\extend{\s{Nat}}{\alltu{\w{Nat}}}}{\s{Singleton}},\s{Nil},\s{Append})\\
\s{(V)}  & = & \s{Crush}[\w{Nat}](\s{Chi}[\w{Nat}](\seq{\extend{g(\id)}{\alltp}}{\void},\s{One},\s{Zero}),\s{Zero},\s{Add})
\end{array}\]}

\noindent
The changes concern the inserted applications of
\extend{\cdot}{\cdot}, and the actual type parameters for
type-unifying combinators. In the definition of \s{(I)}, the strategy
\s{Inc}\ clearly needs to be lifted to \alltp; similarly for the
rewrite rule in \s{(II)}. Note that the original test for naturals is
gone in the revision of \s{(I)}. The mere type of \s{Inc} sufficiently
restricts its applicability. In the definition of \s{(III)}, the
strategy \s{Nat} is used to check for naturals, and it is lifted to
\alltp. The type-unifying facet of \s{(III)} is enforced by the
subsequent application of \void, and it is also pointed out by the
application of \s{Chi}. In the definition of \s{(IV)}, the strategy
\s{Nat} is used to select naturals, and it is lifted to
\alltu{\w{Nat}}. The subsequent application of \s{Singleton} converts
naturals to singleton lists of naturals. The extension performed in
\s{(V)}\ can be justified by similar arguments as for \s{(III)}. In
both cases, \s{Chi}\ is applied to map the success and failure
behaviour of a strategy to distinguished constants.
\end{example}

\paragraph*{Static type safety}

The resulting typed calculus \mycalc\ obeys a number of convenient
properties. Firstly, \mycalc\ supports statically type-safe strategic
programming. Secondly, each strategy expression is strictly either
many-sorted or generic.  Thirdly, many-sorted strategies cannot become
generic just by accident, say due to the context in which they are
used. Strategies rather become generic via explicit use of
\extend{\cdot}{\cdot}. Fourthly, the type-dependent facet of the
reduction semantics is completely restricted to \extend{\cdot}{\cdot}.
The semantics of all other strategy combinators does not involve type
dependency. There are no implicit dynamic type checks.

Admittedly, any kind of type-dependent reduction is somewhat
non-standard because type systems in the tradition of the
$\lambda$-cube are supposed to meet the type-erasure
property~\cite{Barendregt92,BLRU97}. That is, reduction is supposed to
lead to the same result even if type annotations are removed. An
application of the combinator \extend{\cdot}{\cdot} implies a type
inspection at ``run time'', but this inspection is concerned with the
treatment of different behaviours depending on the actual term
type. Also, the inspection is requested by the programmer as opposed
to an implicit dynamic type check that is performed providently by a
run-time system. Similar expressiveness has also been integrated into
other statically typed languages
(cf.~\cite{ACPP91,ACPR92,HM95,CGL95,DRW95,CWM99,Glew99}).


\section{Many-sorted strategies}
\label{S:many-sorted}

We start the formal definition of \mycalc. As a warm-up, we discuss
many-sorted strategies. For simplicity, we postpone formalising
strategy definitions until Section~\ref{SS:prog}. First, we will
define the reduction semantics of a basic calculus $S'_0$
corresponding to an initial untyped fragment of \mycalc.  The
corresponding piece of syntax is shown in Figure~\ref{F:syntax-basic}.
Then, we develop a simple type system starting with many-sorted
type-preserving strategies. We will discuss some standard properties
of the type system.  Afterwards, we elaborate the type system to cover
type-changing strategies and tuples for polyadic strategies. We use
inference rules, say deduction rules, in the style of Natural
semantics~\cite{Kahn87,Despeyroux88,Pettersson94} for both the
reduction semantics and the type system.

\begin{figure}
{\smallsize

\dech{Syntax}
$$\begin{array}{lclr}
c   & & & \mbox{(Constant symbols)}\\
f,g   & & & \mbox{(Function symbols)}\\
x   & & & \mbox{(Term variables)}\\
t & ::= & c\ |\ f(t, \ldots, t)\ |\ x & \mbox{(Terms)}\\
r & ::= & t\ |\ \failure & \ \ \ \mbox{(Reducts of rewriting)}\\
s & ::= & t \to b\ |\ \id\ |\ \fail\ |\ \seq{s}{s}\ |\ \choice{s}{s}\ |\ \negbf{s}\ |\ c\ |\ f(s,\ldots,s)
& \mbox{(Strategies)}\\
b & ::= & t 
& \mbox{(Rule bodies)}
\end{array}$$

}
\caption{Syntax of the basic calculus $S'_0$}
\label{F:syntax-basic}
\bigskip
\end{figure}

\begin{figure}
{\smallsize

\decj{Reduction of strategy applications}{%
\ioj{\apply{s}{t}}{r}
}

\parbox{.47\textwidth}{%
Positive rules%
}\hfill\parbox{.5\textwidth}{%
Negative rules%
}

\medskip

\parbox[t]{.47\textwidth}{

\ir{rule^+}{%
\exists \theta.\ (\theta(t_l) = t \wedge \theta(t_r) = t')
}{%
\ioj{\apply{t_l \to t_r}{t}}{t'}
}

\ax{id^+}{%
\ioj{\apply{\id}{t}}{t}
}

\ir{neg^+}{%
\ioj{\apply{s}{t}}{\failure}
}{%
\ioj{\apply{\negbf{s}}{t}}{t}
}

\ir{seq^+}{%
\ioj{\apply{s_1}{t}}{t^*}
\dnp
\ioj{\apply{s_2}{t^*}}{t'}
}{%
\ioj{\apply{\seq{s_1}{s_2}}{t}}{t'}
}

\ir{choice^+.1}{%
\ioj{\apply{s_1}{t}}{t'}
}{%
\ioj{\apply{\choice{s_1}{s_2}}{t}}{t'}
}

\ir{choice^+.2}{%
\ioj{\apply{s_2}{t}}{t'}
}{%
\ioj{\apply{\choice{s_1}{s_2}}{t}}{t'}
}

\ax{cong^+.1}{%
\ioj{\apply{c}{c}}{c}
}

\ir{cong^+.2}{%
\sps{
\ioj{\apply{s_1}{t_1}}{t'_1}
\np
\cdots
\np
\ioj{\apply{s_n}{t_n}}{t'_n}
}
}{%
\ioj{\apply{f(s_1,\ldots,s_n)}{\\f(t_1,\ldots,t_n)}}{f(t'_1,\ldots,t'_n)}
}

}\hfill\parbox[t]{.47\textwidth}{

\ir{rule^-}{%
\not\exists \theta.\ \theta(t_l) = t
}{%
\ioj{\apply{t_l \to t_r}{t}}{\failure}
}

\ax{fail^-}{%
\ioj{\apply{\fail}{t}}{\failure}
}

\ir{neg^-}{%
\ioj{\apply{s}{t}}{t'}
}{%
\ioj{\apply{\negbf{s}}{t}}{\failure}
}

\ir{seq^-.1}{%
\ioj{\apply{s_1}{t}}{\failure}
}{%
\ioj{\apply{\seq{s_1}{s_2}}{t}}{\failure}
}

\ir{seq^-.2}{%
\ioj{\apply{s_1}{t}}{t^*}
\dnp
\ioj{\apply{s_2}{t^*}}{\failure}
}{%
\ioj{\apply{\seq{s_1}{s_2}}{t}}{\failure}
}

\ir{choice^-}{%
\sps{
\ioj{\apply{s_1}{t}}{\failure}
\np
\ioj{\apply{s_2}{t}}{\failure}
}
}{%
\ioj{\apply{\choice{s_1}{s_2}}{t}}{\failure}
}

\ir{cong^-.1}{%
c \not= t
}{%
\ioj{\apply{c}{t}}{\failure}
}

\ir{cong^-.2}{%
f \not=g
}{%
\ioj{\apply{f(s_1,\ldots,s_n)}{\\g(t_1,\ldots,t_m)}}{\failure}
}

\ir{cong^-.3}{%
\exists i \in \{1,\ldots,n\}.\ \ioj{\apply{s_i}{t_i}}{\failure}
}{%
\ioj{\apply{f(s_1,\ldots,s_n)}{\\f(t_1,\ldots,t_n)}}{\failure}
}
}
}
\caption{Reduction semantics for the basic calculus $S'_0$}
\label{F:apply-basic}
\bigskip
\end{figure}

\subsection{The basic calculus $S'_0$}

\paragraph*{Reduction of strategy applications}

As for the dynamic semantics of strategies, say the \emph{reduction
semantics}, we employ the judgement $\ioj{\apply{s}{t}}{r}$ for the
reduction of strategy applications. Here, $r$ is the reduct that
results from the application of the strategy $s$ to the term
$t$. Recall that a reduct is either a term $t$ or ``$\failure$''
denoting failure (cf.\ Figure~\ref{F:syntax-basic}). We assume that
strategies are only applied to ground terms, and then also yield
ground terms. The latter assumption is not essential but it is well in
line with standard rewriting. In Figure~\ref{F:apply-basic}, we define
the reduction semantics of strategy application for the initial
calculus $S'_0$. The inference rules formalise our informal
explanations from Section~\ref{SS:grammar}. The reduction semantics of
\mycalc\ is a big-step semantics, that is, $r$ in
$\ioj{\apply{s}{t}}{r}$ models the final result of the execution of
the strategy $s$.

\paragraph*{Notational conventions}

We use the common mix-fix notation for judgements in Natural
semantics, that is, a judgement basically amounts to a mathematical
relation over the ingredients such as $s$, $t$, and $r$ in the example
$\ioj{\apply{s}{t}}{r}$. The remaining symbols ``@'' and
``$\leadsto$'' only hint at the intended meaning of the judgement. As
for $\ioj{\apply{s}{t}}{r}$, we say that the reduct $r$ is
``computed'' from the application of $s$ to $t$. The direction of
computation is indicated by ``$\leadsto$''. Deduction rules are tagged
so that we can refer to them. Deduction rules define, as usual, how to
derive valid judgements from given valid judgements. Hence, semantic
evaluation or type inference amounts to a proof starting from the
axioms in a Natural semantics specification. As for the reduction
semantics, we use rule tags that contain ``$+$'' whenever the reduct
is known to be a proper term whereas ``$-$'' is used for remaining
cases with failure as the reduct. We also use the terms ``positive''
vs.\ ``negative'' rules. To avoid confusion, we should point out that
the term ``reduction'' has two meanings in the present paper, namely
reduction in the sense of the reduction semantics for strategies, and
reduction in the sense of traversal where the children of a term are
reduced by monoid-like combinators (recall \reduce{}{\cdot}{\cdot}).

\paragraph*{Deduction rules}

The axioms for \id\ and \fail\ are trivial. Let us read, for example,
the rules for negation. The application \apply{\negbf{s}}{t} returns
$t$ if the application \apply{s}{t}\ returns ``\failure'' (cf.\
\tg{neg^+}). If \apply{s}{t}\ results in a proper term reduct, then
\apply{\negbf{s}}{t}\ evaluates to ``\failure'' (cf.\ \tg{neg^-}).
These rules also illustrate why we need to include failure as reduct.
Otherwise, a judgement could not query whether a certain strategy
application did \emph{not} succeed. Recall that asymmetric choice also
depends on this ability.  Let us also look at the rules for the other
combinators. The rule \tg{seq^+} directly encodes the idea of
sequential composition where the intermediate term $t^*$ that is
obtained via $s_1$\ is then further reduced via $s_2$. Sequential
composition fails if one of the two ingredients $s_1$ or $s_2$ fails
(cf.\ \tg{seq^-.1}, \tg{seq^-.2}).  As for choice, there is one
positive rule for each operand of the choice (cf.\ \tg{choice^+.1},
\tg{choice^+.2}). Choice allows recovery from failure because if
one branch of the choice evaluates to ``$\failure$'', the other branch
can still succeed. Choice fails if both options do not admit success
(cf.\ \tg{choice^-}). The congruences for constants are trivially
defined (cf.\ \tg{cong^+.1}, \tg{cong^-.1}). The congruences for
function symbols are defined in a schematic manner to cover arbitrary
arities (cf.\ \tg{cong^+.2}, \tg{cong^-.2}, \tg{cong^-.3}).

\begin{figure}
{\smallsize

\dech{Syntax}
\begin{eqnarray*}
b & ::= & \cdots\ |\ \where{b}{x = \apply{s}{t}}
\end{eqnarray*}

\parbox[t]{.47\textwidth}{

\decj{Evaluation\\of rule bodies}{\ioj{b}{r}}

Positive rules

\medskip

\ax{body^+.1}{%
\ioj{t}{t}
}

\ir{body^+.2}{%
\ioj{\apply{s}{t}}{t'}
\dnp
\ioj{b\assigns{\assign{x}{t'}}}{t''}
}{%
\ioj{\where{b}{x = \apply{s}{t}}}{t''}
}

Negative rules

\medskip

\ir{body^-.1}{%
\ioj{\apply{s}{t}}{\failure}
}{%
\ioj{\where{b}{x = \apply{s}{t}}}{\failure}
}

\ir{body^-.2}{%
\ioj{\apply{s}{t}}{t'}
\dnp
\ioj{b\assigns{\assign{x}{t'}}}{\failure}
}{%
\ioj{\where{b}{x = \apply{s}{t}}}{\failure}
}

}\hfill\parbox[t]{.47\textwidth}{

\decj{Reduction\\of strategy applications}{\ioj{\apply{s}{t}}{r}}

Positive rule

\medskip

\ir{rule^+}{%
\exists \theta.\ (\theta(t_l) = t \dnp \ioj{\theta(b)}{t'})
}{%
\ioj{\apply{t_l \to b}{t}}{t'}
}

Negative rules

\medskip

\ir{rule^-.1}{%
\nexists \theta.\ \theta(t_l) = t
}{%
\ioj{\apply{t_l \to b}{t}}{\failure}
}

\ir{rule^-.2}{%
\exists \theta.\ (\theta(t_l) = t \dnp \ioj{\theta(b)}{\failure})
}{%
\ioj{\apply{t_l \to b}{t}}{\failure}
}
}
}
\caption{Extension for where-clauses}
\label{F:apply-where}
\bigskip
\end{figure}

\paragraph*{Where-clauses}

In Figure~\ref{F:apply-basic}, rule bodies were assumed to be terms.
In Figure~\ref{F:apply-where}, an extension is supplied to cope with
where-clauses as motivated earlier. The semantics of rewrite rules as
covered in Figure~\ref{F:apply-basic} is surpassed by the new rules in
Figure~\ref{F:apply-where}. Essentially, we resort to a new judgement
for the evaluation of rule bodies. A rule body which consists of a
term, evaluates trivially to this term (cf.\ \tg{body^+.1}). A rule
body of the form \where{b}{x=\apply{s}{t}} is evaluated by first
performing the strategy application \apply{s}{t}, and then binding the
intermediate term reduct $t'$ (if any) to $x$ in the remaining body
$b$ (cf.\ \tg{body^+.2}). Obviously, a rewrite rule can now fail for
two reasons, either because of an infeasible match (cf.\
\tg{rule^-.1}), or due to a failing subcomputation in a where-clause
(cf.\ \tg{rule^-.2}, \tg{body^-.1}, and \tg{body^-.2}). For brevity,
we will abstract from where-clauses in the formalisation of the type
system for \mycalc. As the reduction semantics indicates, where-clause
do not pose any challenge for formalisation.

\subsection{Type-preserving strategies}
\label{SS:tp}

We want to provide a type system for the basic calculus $S'_0$. We
first focus on type-preserving strategies. We use $S'_{tp}$ to denote
the resulting calculus. In fact, type-changing strategies are not
standard in rewriting.  So we will consider them in a separate step
in Section~\ref{SS:tc}. In general, the typed calculus \mycalc\ is
developed in a stepwise and modular fashion.

\paragraph*{Type expressions}

We already sketched the type syntax in Section~\ref{SS:typeful}. As
for purely many-sorted strategies, the forms of term and strategy
types are trivially defined by the following grammar:
\[\begin{array}{lclr}
\sigma & & & \ \ \ \mbox{(Sorts)}\\
\tau   & ::= & \sigma & \ \ \ \mbox{(Term types)}\\
\pi    & ::= & \tau \to \tau & \ \ \ \mbox{(Strategy types)}
\end{array}\]

\paragraph*{Contexts}

In the upcoming type judgements, we use a context parameter $\Gamma$
to keep track of sorts $\sigma$, and to map constant symbols $c$,
function symbols $f$ and term variables $x$ to types. Initially, we
use the following grammar for contexts:
\[\begin{array}{lclr}
\Gamma & ::= &
\emptyset\ |\ \Gamma, \Gamma
& \mbox{(Contexts as sets)}\\
& | & \sigma\ |\ c : \sigma\ |\ f: \sigma \times \cdots \times \sigma \to \sigma\ \ \ %
& \mbox{(Signature part)}\\
& | & x : \tau
& \mbox{(Term variables)}
\end{array}\]
We will have to consider richer contexts when we formalise strategy
definitions in Section~\ref{SS:prog}.  Let us state the requirements
for a well-formed context $\Gamma$. We assume that there are different
name spaces for the various kinds of symbols and variables. Also, we
assume that constant symbols, function symbols and variables are not
associated with different types in $\Gamma$. That is, we do not
consider overloading. All sorts used in some type declaration in
$\Gamma$ also have to be declared themselves in $\Gamma$. Finally,
when contexts are composed via $\Gamma_1,\Gamma_2$ we require that the
sets of symbols and variables in $\Gamma_1$ and $\Gamma_2$ are
disjoint. Note that disjoint union of contexts will not be used before
Section~\ref{SS:prog}. In fact, our contexts are completely static
until then.

\begin{figure}
{\smallsize

\parbox{.46\textwidth}{%

\decj{Well-formedness\\
of term types}{%
\wfj{\Gamma}{\tau}
}

\ir{tau.1}{%
\sigma \in \Gamma
}{%
\wfj{\Gamma}{\sigma}
}

\decj{Well-formedness\\
of strategy types}{%
\wfj{\Gamma}{\pi}
}

\ir{pi.1}{%
\wfj{\Gamma}{\tau}
}{%
\wfj{\Gamma}{\tau \to \tau}
}

\decj{Well-typedness of terms}{%
\wtj{\Gamma}{t}{\tau}
}

\ir{con}{%
c: \sigma \in \Gamma
}{%
\wtj{\Gamma}{c}{\sigma}
}

\ir{fun}{%
\sps{
f: \sigma_1 \times \cdots \times \sigma_n \to \sigma_0 \in \Gamma
\np
\wtj{\Gamma}{t_1}{\sigma_1}
\np
\cdots
\np
\wtj{\Gamma}{t_n}{\sigma_n}
}
}{%
\wtj{\Gamma}{f(t_1,\ldots,t_n)}{\sigma_0}
}

\ir{var}{%
x: \tau \in \Gamma
}{%
\wtj{\Gamma}{x}{\tau}
}

\decj{Negatable types}{%
\negbft{\pi}{\pi'}
}

\ir{negt.1}{%
\wfj{\Gamma}{\tau}
}{%
\negbft{\tau \to \tau}{\tau \to \tau}
}

\decj{Composable types}{%
\comp{\pi_1}{\pi_2}{\pi}
}

\ir{comp.1}{%
\wfj{\Gamma}{\tau}
}{%
\comp{\tau \to \tau}{\tau \to \tau}{\tau \to \tau}
}

}\hfill\parbox{.49\textwidth}{%

\decj{Well-typedness\\of strategy applications}{%
\wtj{\Gamma}{\apply{s}{t}}{\tau}
}

\ir{apply}{%
\wtj{\Gamma}{s}{\tau \to \tau}
\dnp
\wtj{\Gamma}{t}{\tau}
}{%
\wtj{\Gamma}{\apply{s}{t}}{\tau}
}

\decj{Well-typedness\\of strategies}{%
\wtj{\Gamma}{s}{\pi}
}

\ir{rule}{%
\wtj{\Gamma}{t_l}{\tau}
\dnp
\wtj{\Gamma}{t_r}{\tau}  
}{%
\wtj{\Gamma}{t_l \to t_r}{\tau \to \tau}
}

\ir{id}{%
\wfj{\Gamma}{\tau}
}{%
\wtj{\Gamma}{\id}{\tau \to \tau}
}

\ir{fail}{%
\wfj{\Gamma}{\tau}
}{%
\wtj{\Gamma}{\fail}{\tau \to \tau}
}

\ir{neg}{%
\wtj{\Gamma}{s}{\pi}
\dnp
\negbft{\pi}{\pi'}
}{%
\wtj{\Gamma}{\negbf{s}}{\pi'}
}

\ir{seq}{%
\sps{
\wtj{\Gamma}{s_1}{\pi_1}
\dnp
\wtj{\Gamma}{s_2}{\pi_2}
\np
\comp{\pi_1}{\pi_2}{\pi}
}
}{%
\wtj{\Gamma}{\seq{s_1}{s_2}}{\pi}
}

\ir{choice}{%
\wtj{\Gamma}{s_1}{\pi}
\dnp
\wtj{\Gamma}{s_2}{\pi}
}{%
\wtj{\Gamma}{\choice{s_1}{s_2}}{\pi}
}

\ir{cong.1}{%
c: \sigma \in \Gamma
}{%
\wtj{\Gamma}{c}{\sigma \to \sigma}
}

\ir{cong.2}{%
\sps{
f: \sigma_1 \times \cdots \times \sigma_n \to \sigma_0 \in \Gamma
\np
\wtj{\Gamma}{s_1}{\sigma_1 \to \sigma_1}
\np
\cdots
\np
\wtj{\Gamma}{s_n}{\sigma_n \to \sigma_n}
}
}{%
\wtj{\Gamma}{f(s_1,\ldots,s_n)}{\sigma_0 \to \sigma_0}
}

}

}

\caption{Many-sorted type-preserving strategies}
\label{F:tp}
\bigskip
\end{figure}

\paragraph*{Typing judgements}

The principal judgement of the type system is the type judgement for
strategies. It is of the form \wtj{\Gamma}{s}{\pi}, and it holds if
the strategy $s$ is of strategy type $\pi$ in the context
$\Gamma$. Here is a complete list of all well-formedness and
well-typedness judgements:
{\begin{itemize}\noskip
\item \wfj{\Gamma}{\tau} \hfill (Well-formedness of term types)
\item \wfj{\Gamma}{\pi}\hfill(Well-formedness of strategy types)
\item \wtj{\Gamma}{t}{\tau}\hfill(Well-typedness of terms)
\item \negbft{\pi}{\pi'}\hfill(Negatable types)
\item \comp{\pi_1}{\pi_2}{\pi}\hfill(Composable types)
\item \wtj{\Gamma}{\apply{s}{t}}{\tau}\hfill(Well-typedness of strategy applications)
\item \wtj{\Gamma}{s}{\pi}\hfill(Well-typedness of strategies)
\end{itemize}}

\paragraph*{Typing rules}

The corresponding deduction rules are shown in Figure~\ref{F:tp}. The
present formulation is meant to be very strict regarding type
preservation. For some of the rules, one might feel tempted to
immediately cover type-changing strategies, e.g., for the rules
\tg{apply}\ for strategy application or \tg{comp.1}\ for composable
types in sequential composition. However, we want to enable type
changes in a subsequent step. Let us read some inference rules for
convenience.  Type preservation is postulated by the well-formedness
judgement for strategy types (cf.\ \tg{pi.1}). Rule \tg{apply}\ says
that a strategy application $\apply{s}{t}$ is well-typed if the
strategy $s$ is of type $\tau \to \tau$, and the term $t$ is of type
$\tau$. Obviously, the strategies \id\ and \fail\ have many types,
namely any type $\tau \to \tau$ where $\wfj{\Gamma}{\tau}$ holds (cf.\
\tg{id}\ and \tg{fail}). In turn, compound strategies can also have many
types. The strategy types for compound strategies are regulated by the
rules \tg{neg}, \tg{seq}, \tg{choice}, and \tg{cong.2}. The typing
rules for negation and sequential composition (cf.\ \tg{neg}\ and
\tg{seq}) refer to auxiliary judgements for negatable and composable
types. Their definition is straightforward for the initial case of
many-sorted type-preserving strategies (cf.\ \tg{negt.1}\ and
\tg{comp.1}). The compound strategy \choice{s_1}{s_2}\ for choice is
well-typed if both strategies $s_1$ and $s_2$ are of a common type
$\pi$.  This common type constitutes the type of the choice.

\paragraph*{Properties}

We use $S'_{tp}$ to denote the composition of $S'_0$ defined in
Figure~\ref{F:apply-basic}, and the type system from
Figure~\ref{F:tp}.  The following theorem is concerned with properties
of $S'_{tp}$.  It says that actual strategy types adhere to the scheme
of type preservation, strategy applications are uniquely typed, and
the reduction semantics is properly abstracted in the type system.

\begin{theorem}
\label{T:tp}
The calculus $S'_{tp}$\ for many-sorted type-preserving strategies
obeys the following properties:
\begin{enumerate}
\item Actual strategy types adhere to the scheme of type preservation,
i.e.,\\ for all well-formed contexts $\Gamma$, strategies $s$ and
term types $\tau$, $\tau'$:\\ $\wtj{\Gamma}{s}{\tau \to \tau'}$ implies
$\tau=\tau'$.
\item Strategy applications satisfy unicity of typing (UOT, for short),
i.e.,\\ for all well-formed contexts $\Gamma$, strategies $s$, term
types $\tau$, $\tau'$ and terms $t$:\\
$\wtj{\Gamma}{\apply{s}{t}}{\tau}\ \wedge\ %
\wtj{\Gamma}{\apply{s}{t}}{\tau'}$ implies $\tau = \tau'$.
\item Reduction of strategy applications satisfies subject reduction,
i.e.,\\
for all well-formed contexts $\Gamma$,
strategies $s$,
term types $\tau$
and terms $t$, $t'$:\\
$\wtj{\Gamma}{s}{\tau \to \tau}\ \wedge\ %
\wtj{\Gamma}{t}{\tau}\ \wedge\ %
\ioj{\apply{s}{t}}{t'}$ implies $\wtj{\Gamma}{t'}{\tau}$.
\end{enumerate}
\end{theorem}

In the further development of \mycalc, we will use refinements of
these properties to prove the formal status of the evolving type
system. UOT and subject reduction are basic desirable
properties of type systems
(cf.~\cite{Barendregt92,GeuversPhD,Schmidt94}).  We claim
UOT for strategy applications but not for strategies
themselves because of the typing rules for the constant combinators
\id\ and \fail. UOT for strategy \emph{application} means that the
result type of a strategy application is determined by the type of the
input term. Subject reduction means that if we initiate a reduction of
a well-typed strategy application, then we can be sure that the
resulting term reduct (if any) is of the prescribed type. The
following proof is very verbose to prepare for the elaboration of the
proof in the context of generic types.

\begin{proof}\hfill\mbox{}
\label{P:tp}
(IH abbreviates induction hypothesis in all the upcoming proofs.)
\begin{enumerate}
\item We show adherence to the scheme of type preservation
by induction on $s$ in \wtj{\Gamma}{s}{\pi}. \emph{Base cases}: Type
preservation is directly enforced for rewrite rules, \id, \fail, and
congruences for constants by the corresponding typing rules (cf.\
\tg{rule}, \tg{id}, \tg{fail}, and \tg{cong.1}), that is, the type
position in the conclusion is instantiated according to the
type-preserving form of strategy types. \emph{Induction step}: Type
preservation for \negbf{\cdot}\ (cf.\ \tg{neg}) is implied by the rule
\tg{negt.1} for negatable types. Strictly speaking, we do not need to
employ the IH since the type-preserving shape of the result type is
enforced by \tg{negt.1} regardless of the argument type. As for
\seq{s_1}{s_2}, the auxiliary judgement for composable types enforces
type preservation (cf.\ \comp{\cdots}{\cdots}{\tau\to\tau} in
\tg{comp.1}). Again, the IH does not need to be employed. As for
\choice{s_1}{s_2}, the result type coincides with the argument types,
and hence, type preservation is implied by the IH. Finally, type
preservation for congruences $f(s_1,\ldots,s_n)$ is directly enforced
by the corresponding typing rule (cf.\ the type position in the
conclusion of \tg{cong.2}).
\item Let us first point out that UOT obviously holds for terms
because the inductive definition of \wtj{\Gamma}{t}{\tau}\ enforces a
unique type $\tau$\ for $t$. Here it is essential that we ruled out
overloading of function and constant symbols, and variables. According
to the rule \tg{apply}, the result type of a strategy application is
equal to the type of the input term. Hence, $\apply{s}{t}$ is uniquely
typed.
\item In the type-preserving setting, subject reduction actually means
that the reduction semantics for strategy applications is
type-preserving as prescribed by the type system. That is, if the
reduction of a strategy application \apply{s}{t} with $s:\tau \to
\tau$, $t:\tau$\ yields a proper term reduct $t'$, then $t'$ is also
of type $\tau$. We show this property by induction on $s$ in
\ioj{\apply{s}{t}}{r} while we assume $\wtj{\Gamma}{s}{\tau \to \tau}$
and $\wtj{\Gamma}{t}{\tau}$. To this end, it is crucial to maintain
that the IH can only be employed for a premise
\ioj{\apply{s_i}{t_i}}{t'_i} and a corresponding type $\tau_i$, if we
can prove the following side condition:
\[\begin{array}{l}
\wtj{\Gamma}{s}{\tau \to \tau}\ \wedge\ %
\wtj{\Gamma}{t}{\tau}\
\wedge\ \ldots\\
\ \mbox{implies}\ %
\wtj{\Gamma}{s_i}{\tau_i \to
\tau_i}\ \wedge\ %
\wtj{\Gamma}{t_i}{\tau_i}
\end{array}\]
With the ``\ldots'' we indicate that actual side conditions might
involve additional requirements. The judgements
\wtj{\Gamma}{s_i}{\tau_i \to \tau_i}\ and \wtj{\Gamma}{t_i}{\tau_i}\
have to be approved by consulting the corresponding typing rules that
relate $t$ to $t_i$, and $s$ to $s_i$, and by other means. Note there
are no proof obligations for deduction rules which do not yield a
proper term reduct, namely for negative rules. In particular, there is
no case for \fail\ in the sequel, that is, \fail\ is type-preserving
in a degenerated sense. \emph{Base cases}: As for rewrite rules, we
know that both the left-hand side $t_l$ and the right-hand side $t_r$
are of type $\tau$ as prescribed by \tg{rule}. The substitution
$\theta$ in \tg{rule^+}\ preserves the type of the right-hand side as
implied by basic properties of many-sorted unification and
substitution. Hence, rule application is type-preserving. \id\
preserves the very input term, and hence, it is type-preserving. The
same holds for congruences for constants. \emph{Induction step}:
Negation is type-preserving because the very input term is preserved
as for \id. Thus, we do not need to employ the IH for $s$ in
\negbf{s}. In fact, the IH tells us here that we do not even attempt
to apply $s$ in an ill-typed manner. Let us consider sequential
composition \apply{\seq{s_1}{s_2}}{t}. By the rules \tg{seq} and
\tg{comp.1}, we know that the types of $s_1$, $s_2$ and \seq{s_1}{s_2}
coincide, that is, the common type is $\tau \to
\tau$. We want to show that $t'$ in \ioj{\apply{\seq{s_1}{s_2}}{t}}{t'}\ is
of the same type as $t$.  As $s_1$ must be of the same type as
\seq{s_1}{s_2}, the IH is enabled for \ioj{\apply{s_1}{t}}{t^*}. Thereby,
we know that $t^*$ is of type $\tau$. Since we also know that $s_2$
must be of the same type as \seq{s_1}{s_2}, the IH is enabled for the
second premise \ioj{\apply{s_2}{t^*}}{t'}. Hence, $t'$ is of the same
type as $t$, and sequential composition is type-preserving. As for
choice, reduction of \apply{\choice{s_1}{s_2}}{t}\ directly resorts to
either \apply{s_1}{t}\ or \apply{s_2}{t}\ (cf.\ \tg{choice^+.1} and
\tg{choice^+.2}). We also know that $s_1$, $s_2$\ and
\choice{s_1}{s_2}\ have to be of the same type (cf.\
\tg{choice}). Hence, the IH is enabled for the reduction of the chosen
strategy, be it $s_1$ or $s_2$. Finally, let us consider congruence
strategies where \apply{f(s_1,\ldots,s_n)}{f(t_1,\ldots,t_n)} is
reduced to $f(t'_1,\ldots,t'_n)$ while the $t'_i$ are obtained by the
reduction of the \apply{s_i}{t_i} (cf.\ \tg{cong.2}). Let $f:
\sigma_1 \times \cdots \times \sigma_n \to \sigma_0$ be in
$\Gamma$. Then, we know that for a well-typed term
$f(t_1,\ldots,t_n)$, the $t_i$ must be of type $\sigma_i$ (cf.\
\tg{fun}). We also know that for a well-typed strategy
$f(s_1,\ldots,s_n)$, the $s_i$ must be of type $\sigma_i \to \sigma_i$
(cf.\ \tg{cong.2}). Hence, the IH is enabled for the various
\ioj{\apply{s_i}{t_i}}{t'_i}. Then, the type of $f(t'_1,\ldots,t'_n)$
is the same as $f(t_1,\ldots,t_n)$.
\end{enumerate}
\end{proof}

As an aside, the proof of subject reduction is simplified by the fact
that possibly recursive strategy definitions were omitted. The use of
simple induction on $s$ is enabled by the strong normalisation of the
purely inductive reduction semantics for strategy applications. If
recursive strategy definitions were included, proof by induction on
the depth of derivations is needed. The use of static contexts also
simplifies our proofs.

\subsection{Type-changing strategies}
\label{SS:tc}

In standard rewriting, as a consequence of a fixed normalisation
strategy, rewrite rules are necessarily type-preserving. It does not
make sense to repeatedly look for a redex in a compound term, and then
to apply some type-changing rewrite rule to the redex since this would
potentially lead to an ill-typed compound term. In strategic
rewriting, it is no longer necessary to insist on type-preserving
rewrite rules. One can use strategies to apply type-changing rewrite
rules or other strategies in a disciplined manner making sure that
intermediate results are properly combined as opposed to the type-changing
replacement of a redex in a compound term.

\begin{figure}
{\smallsize

\parbox[t]{.5\textwidth}{

\decj{Well-formedness\\of strategy types}{%
\wfj{\Gamma}{\pi}
}

\ir{pi.1}{%
\wfj{\Gamma}{\tau}
\dnp
\wfj{\Gamma}{\tau'}
}{%
\wfj{\Gamma}{\tau \to \tau'}
}

\decj{Negatable types}{%
\negbft{\pi}{\pi'}
}

\ir{negt.1}{%
\wfj{\Gamma}{\tau}
\dnp
\wfj{\Gamma}{\tau'}
}{%
\negbft{\tau \to \tau'}{\tau \to \tau}
}

\decj{Composable types}{\comp{\pi_1}{\pi_2}{\pi}}

\ir{comp.1}{%
\wfj{\Gamma}{\tau}
\dnp
\wfj{\Gamma}{\tau^*}
\dnp
\wfj{\Gamma}{\tau'}
}{%
\comp{\tau \to \tau^*}{\tau^* \to \tau'}{\tau \to \tau'}
}

}\hfill\parbox[t]{.46\textwidth}{

\decj{Well-typedness\\of strategy applications}{%
\wtj{\Gamma}{\apply{s}{t}}{\tau}
}

\ir{apply}{%
\wtj{\Gamma}{s}{\tau \to \tau'}
\dnp
\wtj{\Gamma}{t}{\tau}
}{%
\wtj{\Gamma}{\apply{s}{t}}{\tau'}
}

\decj{Well-typedness\\ of strategies}{%
\wtj{\Gamma}{s}{\pi}
}

\ir{rule}{%
\wtj{\Gamma}{t_l}{\tau}
\dnp
\wtj{\Gamma}{t_r}{\tau'} 
}{%
\wtj{\Gamma}{t_l \to t_r}{\tau \to \tau'}
}

}
}
\caption{Refinement of $S'_{tp}$
to enable type-changing strategies}
\label{F:tc}
\bigskip
\end{figure}

\paragraph*{Type system update}

In Figure~\ref{F:tc}, the type system for type-preserving strategies
is updated to enable type-changing strategies. We use $S'_{tc}$ to
denote the refinement of $S'_{tp}$ according to the figure. The
refinements amounts to the following adaptations. We replace rule
\tg{pi.1} to characterise potentially type-changing strategies as
well-formed. We also replace the rule \tg{apply} for strategy
application, and the rule \tg{rule} to promote type-changing
strategies. Furthermore, the auxiliary judgements for negatable and
composable strategy types have to be generalised accordingly (cf.\
\tg{negt.1}\ and \tg{comp.1}). The relaxation for composable types
is entirely obvious but we should comment on the typing rule for
negatable types. Negation is said to be type-preserving regardless of
the argument's type. This is appropriate because the only possible
term reduct admitted by negation is the input term itself. The
argument strategy is only tested for failure. Hence, negation by
itself is type-preserving even if the argument strategy would be
type-changing.  All the other typing rules carry over from $S'_{tp}$.
As an aside, we do not generalise the type of \fail\ to go beyond type
preservation.  In fact, one could say that the result type of \fail\
is arbitrary since no term reduct will be returned anyway. However,
such a definition would complicate our claim of UOT.

\begin{theorem}
\label{T:tc}
The calculus $S'_{tc}$\ for potentially type-changing strategies obeys the
following properties:
{\begin{enumerate}\noskip
\item Co-domains of strategies are determined by domains. i.e.,
for all well-formed contexts $\Gamma$,
strategies $s$
and term types $\tau_1$, $\tau'_1$, $\tau_2$, $\tau'_2$:\\
$\wtj{\Gamma}{s}{\tau_1 \to \tau'_1}\ \wedge\ %
\wtj{\Gamma}{s}{\tau_2 \to \tau'_2}\ \wedge\ %
\tau'_1 \not= \tau'_2$
implies\ $\tau_1 \not= \tau_2$.
\item Strategy applications satisfy UOT,
i.e., ... (cf.\ Theorem~\ref{T:tp}).
\item Reduction of strategy applications satisfies subject reduction, i.e.,
for all well-formed contexts $\Gamma$,
strategies $s$,
term types $\tau$, $\tau'$
and terms $t$, $t'$:\\
$\wtj{\Gamma}{s}{\tau \to \tau'}\ \wedge\ %
\wtj{\Gamma}{t}{\tau}\ \wedge\ %
\ioj{\apply{s}{t}}{t'}$ implies $\wtj{\Gamma}{t'}{\tau'}$.
\end{enumerate}}
\end{theorem}

The first property is the necessary generalisation of adherence to the
scheme of type preservation in Theorem~\ref{T:tp}. We require that the
co-domain of a strategy type is uniquely determined by its domain.
That is, there might be different types for a strategy, but once the
type of the input term is fixed, the type of the result is
determined. The second property carries over from Theorem~\ref{T:tp}.
The third property needs to be generalised compared to
Theorem~\ref{T:tp} in order to cover type-changing strategies.

\begin{proof}\hfill\mbox{}
\label{P:tc}
\begin{enumerate}
\item Note that the property trivially holds for type-preserving strategies.
We show the property by induction on $s$ in \wtj{\Gamma}{s}{\pi}.
\emph{Base cases}: The co-domain of a rewrite rule is even uniquely defined
regardless of the domain as an implication of UOT for terms.
The remaining base cases are type-preserving, and hence, they are trivial.
\emph{Induction step}: Negation is trivially covered because it is
type-preserving. As for sequential composition, the domain of
\seq{s_1}{s_2}\ coincides with the domain of $s_1$, the co-domain of
$s_1$ coincides with the domain of $s_2$, and the co-domain of $s_2$
coincides with the co-domain of \seq{s_1}{s_2} (cf.\ \tg{comp.1}). By
applying the IH to $s_1$ and $s_2$, we obtain that the co-domain of
\seq{s_1}{s_2} is transitively determined by its domain. As for
choice, the property follows from the strict coincidence of the types
of $s_1$, $s_2$, and \choice{s_1}{s_2} (cf.\ \tg{choice}) which
immediately enables the IH. Congruences $f(s_1,\ldots,s_n)$ are
trivially covered because they are type-preserving.
\item The simple argument from Proof~\ref{P:tp} regarding the rule
\tg{apply} can be generalised as follows. The domain of the strategy
in \apply{s}{t} needs to coincide with the type of $t$. Since the
co-domain of $s$ is determined by the type of $t$ (cf.\ (1) above), we
know that the type of the reduct is uniquely defined.
\item We need to elaborate our induction proof for Proof~\ref{P:tp}
where we argued that subject reduction for type-preserving strategies
can be proved by showing that the reduction semantics is
type-preserving, too. As for potentially type-changing strategies, we
need to show that reduction obeys the strategy types. Hence, the side
condition for the employment of the IH has to be revised, too. That is,
the IH can be employed for a premise \ioj{\apply{s_i}{t_i}}{t'_i} and
corresponding types $\tau_i$ and $\tau'_i$, if we can prove the following
side condition:
\[\begin{array}{l}
\wtj{\Gamma}{s}{\tau \to \tau'}\ \wedge\ %
\wtj{\Gamma}{t}{\tau}%
\ \wedge\ \ldots\\
\ \mbox{implies}\ %
\wtj{\Gamma}{s_i}{\tau_i \to \tau'_i}\ \wedge\ %
\wtj{\Gamma}{t_i}{\tau_i}
\end{array}\]
\emph{Base cases}: Subject reduction for rewrite rules is implied by
basic properties of many-sorted unification and substitution. The
remaining base cases are type-preserving, and hence, they are covered
by Proof~\ref{P:tp}. \emph{Induction step}: The strategy $s$ in
\negbf{s}\ is not necessarily type-preserving anymore but 
negation by itself adheres to type preservation as prescribed by the
type system (cf.\ \tg{negt.1}). As for sequential composition, we
start from the assumptions \wtj{\Gamma}{\seq{s_1}{s_2}}{\tau \to
\tau'} and \wtj{\Gamma}{t}{\tau}, we want to show that $t'$ in
\ioj{\apply{\seq{s_1}{s_2}}{t}}{t'} is of type $\tau'$.  There must
exist a $\tau^*$ such that \wtj{\Gamma}{s_1}{\tau \to \tau^*}\ and
\wtj{\Gamma}{s_2}{\tau^* \to \tau'}\ (by \tg{comp.1} and \tg{seq}). In
fact, $\tau^*$ is uniquely defined because it is the co-domain of $s_1$
determined by the domain of $s_1$ which coincides with the domain of
$\seq{s_1}{s_2}$. We apply the IH for \apply{s_1}{t}, and hence, we
obtain that the reduction of \apply{s_1}{t} delivers a term $t^*$ of
type $\tau^*$. This enables the IH for the second operand of
sequential composition. Hence, we obtain that the reduction of
\apply{s_2}{t^*} delivers a term $t'$ of type $\tau'$. As for choice,
the arguments from Proof~\ref{P:tp} are still valid since we did not
rely on type preservation. That is, we know that the reduction of the
choice directly resorts to one of the argument strategies, and the
type of the choice has the same type as the two argument
strategies. Hence, subject reduction for choice follows from the
IH. Congruences $f(s_1,\ldots,s_n)$ and the involved arguments are
type-preserving, and hence, subject reduction carries over from
Proof~\ref{P:tp}.
\end{enumerate}
\end{proof}

\begin{figure}
{\smallsize

\dech{Syntax}
\begin{eqnarray*}
\tau & ::= & \cdots\ |\ \emptytuple\ |\ \pair{\tau_1}{\tau_2}\\
t    & ::= & \cdots\ |\ \emptytuple\ |\ \pair{t_1}{t_2}\\
s    & ::= & \cdots\ |\ \emptytuple\ |\ \pair{s_1}{s_2}
\end{eqnarray*}

\parbox[t]{.43\textwidth}{%

\decj{Well-formedness\\of term types}{%
\wfj{\Gamma}{\tau}
}

\ax{tau.2}{%
\wfj{\Gamma}{\emptytuple}
}

\ir{tau.3}{%
\wfj{\Gamma}{\tau_1}
\dnp
\wfj{\Gamma}{\tau_2}
}{%
\wfj{\Gamma}{\pair{\tau_1}{\tau_2}}
}

\decj{Well-typedness\\of terms}{%
\wtj{\Gamma}{t}{\tau}
}

\ax{empty{-}tuple}{\wtj{\Gamma}{\emptytuple}{\emptytuple}}

\ir{pair}{%
\wtj{\Gamma}{t_1}{\tau_1}
\dnp
\wtj{\Gamma}{t_2}{\tau_2}
}{%
\wtj{\Gamma}{\pair{t_1}{t_2}}{\pair{\tau_1}{\tau_2}}
}

}\hfill\parbox[t]{.51\textwidth}{%

\decj{Well-typedness\\of strategies}{\wtj{\Gamma}{s}{\pi}}

\ax{cong.3}{%
\wtj{\Gamma}{\emptytuple}{\emptytuple}
}

\ir{cong.4}{%
\sps{
\wtj{\Gamma}{s_1}{\tau_1\to\tau'_1}
\np
\wtj{\Gamma}{s_2}{\tau_2\to\tau'_2}
}
}{%
\wtj{\Gamma}{\pair{s_1}{s_2}}{\pair{\tau_1}{\tau_2}\to\pair{\tau'_1}{\tau'_2}}
}

\bigskip
\medskip
\begin{quote}
The reduction semantics of tuple congruences is defined precisely as
for ordinary many-sorted constant and function symbols
\end{quote}

}
}
\caption{Tuple types, tuples, and tuple congruences}
\label{F:tuples}
\bigskip
\end{figure}

\subsection{Polyadic strategies}

As we motivated in Section~\ref{S:rationale}, we want to employ tuples to
describe polyadic strategies, that is, strategies which process several
terms. In principle, the following extension for tuples can be composed with
both $S'_{tp}$\ and $S'_{tc}$. However, tuples are only potent in
$S'_{tc}$\ with type-changing strategies enabled.

In Figure~\ref{F:tuples}, we extend the basic calculus $S'_0$\ with
concepts for polyadic strategies in a straightforward manner.  There
are distinguished symbols \emptytuple\ for the empty tuple, and
\pair{\cdot}{\cdot}\ for pairing terms. We use the same symbols
for tuple types, tuples, and congruences on tuples. The judgements for
well-formedness of term types and well-typedness of terms are extended
accordingly (cf.\ \tg{tau.2}, \tg{tau.3}, \tg{empty{-}tuple}, and
\tg{pair}). We also introduce special typing rules for congruences on
tuples (cf.\ \tg{cong.3} and \tg{cong.4}).  Note that the typing rules
for congruences for ordinary symbols relied on a context lookup (cf.\
\tg{cong.1} and \tg{cong.2} in Figure~\ref{F:tp}) while this is not
the case for polymorphic congruences on tuples. Moreover, congruences
on pairs can be type-changing (cf.\ \tg{cong.4}) whereas this is not
an option for many-sorted congruences. We should point out that tuples
are solely intended for argument and result lists while constructor
terms should be purely many-sorted. This intention is enforced because
the argument types of ordinary function symbols are still restricted
to sorts as opposed to tuple types (cf.\ \tg{fun} in
Figure~\ref{F:tp}).

\begin{example}
The strategy \s{Add} from Example~\ref{X:add}\ and
the strategy \s{Count} from Example~\ref{X:count1} are well-typed as
type-changing strategies.
\begin{eqnarray*}
\s{Add}   & : & \pair{\w{Nat}}{\w{Nat}} \to \w{Nat}\\
\s{Count} & : & \w{Tree} \to \w{Nat}
\end{eqnarray*}
As for \s{Add}, we rely on tuple types since addition is encoded as a
strategy which takes a pair of naturals. As for \s{Count}, the
definition from Example~\ref{X:count1} involves a type-changing
congruence on pairs, namely \pair{\s{Count}}{\s{Count}}. This
congruence applies \s{Count}\ to the two subtrees of a fork tree
independently.
\end{example}


\section{Generic strategies}
\label{S:generic}

In the present section, we extend our basic calculus for many-sorted
strategies by types and combinators for generic strategies. First, we
spell out the reduction semantics of type-preserving combinators, and
we formalise the corresponding generic type \alltp. Then, the problem
of mediation between many-sorted and generic strategies is
addressed. There are two directions for mediation. When we qualify a
many-sorted strategy to become generic, then we perform
\emph{extension}. When we instantiate the type of a generic strategy 
for a given sort, then we perform \emph{restriction}. Afterwards, we
define the type-unifying traversal combinators and the corresponding
generic type (constructor) \alltu{\cdot}.

\subsection{Strategies of type \alltp}
\label{SS:alltp}

\paragraph*{Combinators}

In Figure~\ref{F:apply-all-one}, we define the reduction semantics of
the generic traversal primitives \all{\cdot} and \one{}{\cdot} adopted
from system $S$. The rule \tg{all^+.1}\ says that \all{s}\ applied to
a constant immediately succeeds because there are no children which
have to be processed. The rule \tg{all^+.2}\ directly encodes what it
means to apply $s$ to all children of a term $f(t_1, \ldots,
t_n)$. Note that the function symbol $f$ is preserved in the
result. The reduction scheme for \one{}{s}\ is similar.  The rule
\tg{one^+}\ says that $s$ is applied to some subterm $t_i$ of
$f(t_1,\ldots,t_n)$ such that it succeeds for this child. The
semantics is non-deterministic as for choice of the child. One could
also think of a different semantics where the children are tried from
left to right or vice versa until one child is processed
successfully. The negative rule \tg{one^-.1}\ says that \one{}{s}\
applied to a constant fails because there is no child that could be
processed by $s$. The negative rule \tg{one^-.2}\ says that
$\one{}{s}$\ fails if $s$ fails for all children of
$f(t_1,\ldots,t_n)$.  Dually, \all{s}\ fails if $s$ fails for some
child (cf.\ \tg{all^-}).

\begin{figure}
{\smallsize

\dech{Syntax}
\begin{eqnarray*}
s & ::= & \cdots\ |\ \all{s}\ |\ \one{}{s}
\end{eqnarray*}

\decj{Reduction of strategy applications}{%
\ioj{\apply{s}{t}}{t'}
}

\parbox[t]{.51\textwidth}{

\dech{Positive rules}

\medskip

\ax{all^+.1}{%
\ioj{\apply{\all{s}}{c}}{c}
}

\ir{all^+.2}{%
\sps{
\ioj{\apply{s}{t_1}}{t'_1}
\np
\cdots
\np
\ioj{\apply{s}{t_n}}{t'_n}
}
}{%
\ioj{\apply{\all{s}}{f(t_1, \ldots, t_n)}}{\\f(t'_1, \ldots, t'_n)}
}

\ir{one^+}{%
\exists i \in \{1,\ldots,n\}.\ \ioj{\apply{s}{t_i}}{t'_i}
}{%
\ioj{\apply{\one{}{s}}{f(t_1, \ldots, t_n)}}{\\f(t_1, \ldots, t'_i, \ldots, t_n)}
}

}\hfill\parbox[t]{.43\textwidth}{

\dech{Negative rules}

\medskip

\ir{all^-}{%
\exists i \in \{1,\ldots,n\}.\ \ioj{\apply{s}{t_i}}{\failure}
}{%
\ioj{\apply{\all{s}}{f(t_1, \ldots, t_n)}}{\failure}
}

\ax{one^-.1}{%
\ioj{\apply{\one{}{s}}{c}}{\failure}
}

\ir{one^-.2}{%
\sps{
\ioj{\apply{s}{t_1}}{\failure}
\np
\cdots
\np
\ioj{\apply{s}{t_n}}{\failure}
}
}{%
\ioj{\apply{\one{}{s}}{f(t_1, \ldots, t_n)}}{\failure}
}

}
}
\caption{Type-preserving traversal combinators}
\label{F:apply-all-one}
\bigskip
\end{figure}

\paragraph*{The generic type \alltp}

In Figure~\ref{F:alltp}, we extend the typing judgements to formalise
\alltp, and to employ \alltp\ for the relevant combinators. We
establish a syntactical domain $\gamma$ of generic types. We integrate
$\gamma$ into the grammar for types by stating that $\gamma$
corresponds to another form of strategy types $\pi$ complementing
many-sorted strategy types. We start the definition of $\gamma$ with
the generic type \alltp. We use a relation \typeless\ on types to
characterise generic types, say the genericity of types. The relation
\typeleq\ denotes the reflexive closure of \typeless. By $\pi
\typeless \pi'$, we mean that $\pi'$ is more generic than $\pi$. If we
view generic types as type schemes, we can also say that the type
$\pi$ is an \emph{instance} of the type scheme $\pi'$. Rule
\tg{less.1}\ axiomatises \alltp. The rule says that $\tau \to \tau$ is
an instance of \alltp\ for all well-formed $\tau$. This formulation
indeed suggests to consider \alltp\ as the type scheme $\forall
\alpha.\ \alpha \to \alpha$. We urge the reader not to confuse 
\typeleq\ with subtyping. The remaining rules in
Figure~\ref{F:alltp}\ deal with well-typedness of generic
strategies. The constant combinators \id\ and \fail\ are defined to be
generic type-preserving strategies (cf.\ \tg{id}\ and \tg{fail}). As
for negation, we add another rule to the auxiliary judgement
\negbft{\pi}{\pi'} for negatable types. The enabled form of negation
is concerned with generic strategies. The rule \tg{negt.2} states that
any strategy of a generic type $\gamma$ can be negated. As for
sequential composition, we also add a rule to the auxiliary judgement
\comp{\pi_1}{\pi_2}{\pi}\ to cover the case that a generic
type-preserving strategy and another generic strategy are composed
(cf.\ \tg{comp.2}). The typing rules for the traversal combinators
state simply that \all{\cdot}\ and \one{}{\cdot}\ can be used to
derive a strategy of type \alltp\ from an argument strategy of type
\alltp\ (cf.\ \tg{all}\ and \tg{one}).

\begin{figure}
{\smallsize

\dech{Syntax}
\begin{eqnarray*}
\pi    & ::= & \cdots\ |\ \gamma\\
\gamma & ::= & \alltp
\end{eqnarray*}

\parbox[t]{.47\textwidth}{

\decj{Well-formedness\\of strategy types}{%
\wfj{\Gamma}{\pi}
}

\ax{pi.2}{%
\wfj{\Gamma}{\alltp}
}

\decj{Genericity relation}{%
$\pi \typeless \pi'$
}

\ir{less.1}{%
\wfj{\Gamma}{\tau}
}{%
\tau \to \tau \typeless \alltp
}

\decj{Negatable types}{%
\negbft{\pi}{\pi'}
}

\ir{negt.2}{%
\wfj{\Gamma}{\gamma}
}{%
\negbft{\gamma}{\alltp}
}

}\hfill\parbox[t]{.47\textwidth}{

\decj{Composable types}{%
\comp{\pi_1}{\pi_2}{\pi}
}

\ir{comp.2}{%
\wfj{\Gamma}{\gamma}
}{%
\comp{\alltp}{\gamma}{\gamma}
}

\decj{Well-typedness\\of strategies}{%
\wtj{\Gamma}{s}{\pi}
}

\ax{id}{%
\wtj{\Gamma}{\id}{\alltp}
}

\ax{fail}{%
\wtj{\Gamma}{\fail}{\alltp}
}

\ir{all}{%
\wtj{\Gamma}{s}{\alltp}
}{%
\wtj{\Gamma}{\all{s}}{\alltp}
}

\ir{one}{%
\wtj{\Gamma}{s}{\alltp}
}{%
\wtj{\Gamma}{\one{}{s}}{\alltp}
}

}
}
\caption{\alltp---The type of generic type-preserving strategies}
\label{F:alltp}
\bigskip
\end{figure}

\vspace{-50\in}

\paragraph*{Well-defined generic strategy types}

In general, we assume that a well-defined generic type should admit an
instance for every possible term type since a generic strategy should
be applicable to terms of all sorts. To be precise, there should be
exactly one instance per term type. It is an essential property that
there is only one instance per term type.  Otherwise, the type of a
generic strategy application would be ambiguous. The type \alltp\ is
obviously well-defined in this sense.

\paragraph*{Separation of many-sorted and generic strategies}

Note that there are now two levels in our type system, that is, there
are many-sorted types and generic types. The type system strictly
separates many-sorted strategies (such as rewrite rules) and generic
strategies (such as applications of \all{\cdot}). Since there are no
further intermediate levels of genericity, there are only chains of
length $1$ in the partial order \typeleq. Longer chains will be needed
in Section~\ref{SS:overloading} when we consider a possible
sophistication of \mycalc\ to accomplish overloaded strategies.  As
the type system stands, we cannot turn many-sorted strategies into
generic ones, nor the other way around. Also, strategy application, as
it was defined for $S'_{tp}$\ and $S'_{tc}$, only copes with
many-sorted strategies. The type system should allow us to apply a
generic strategy to any term. We will now develop the corresponding
techniques for strategy extension and restriction.

\subsection{Strategy extension}
\label{SS:extend}

Now that we have typed generic traversal combinators at our disposal,
we also want to inhabit the generic type \alltp. So far, we only have
two trivial constants of type \alltp, namely \id\ and \fail. We would
like to construct generic strategies from rewrite rules. We will
formalise the corresponding combinator \extend{\cdot}{\cdot} for
strategy extension. To this end, we also examine other approaches in
order to justify the design of \extend{\cdot}{\cdot}.

\paragraph*{Infeasible approaches}

In the untyped language Stratego, no distinction is made between
rewrite rules and generic strategies. We might attempt to lift this
design to a typed level. We study two such approaches. One infeasible
approach to the inhabitation of generic strategy types like \alltp\ is
that the typing requirements for generic strategy arguments would be
relaxed: whenever we require a generic strategy argument, e.g., for
$s$ in \all{s}, we also would accept a many-sorted $s$. This implicit
approach to the inhabitation of generic types would only be type-safe
if extra dynamic type checks are added. Consider, for example, the
strategy \all{s} where $s$ is of some many-sorted type $\tau \to
\tau$. For the sake of subject reduction (say, type-safe strategy
application), the semantics of
\all{s} had to ensure that every child at hand is of type $\tau$
before it even attempts to apply $s$ to it. If some child is not of
type $\tau$, \all{s} must fail. From the programmer's point of view,
the approach makes it indeed too easy for many-sorted strategies to
get accepted in a generic context. The resulting applicability
failures of many-sorted strategies in generic contexts are not
approved by the strategic programmer. By contrast:
\begin{quote}
We envision a statically type-safe style of strategic programming,
where the employment of many-sorted strategies in a generic context is
approved by the programmer. Moreover, the corresponding calculus
should correspond to a conservative extension of $S'_{tp}$ (or
$S'_{tc}$).
\end{quote}

Another infeasible approach to the inhabitation of generic strategy
types is to resort to a choice combinator (\choice{\cdot}{\cdot}\ or
\lchoice{\cdot}{\cdot}) to compose a many-sorted strategy and a
generic default. We basically end up having the same problem as above.
Let us attempt to turn a rewrite rule $\ell$ into a generic strategy
using the form \lchoice{\ell}{s}\ where $s$ is some generic strategy,
e.g., \id\ or \fail. We could assume that the result type of a choice
corresponds to the least upper bound of the argument types w.r.t.\
\typeleq. One possible argument to refuse such a style arises from the
following simple derivation:
\[
\lchoice{s}{\fail}\ \ \leadsto\ \ %
\choice{s}{\seq{\negbf{s}}{\fail}}\ \ \leadsto\ \ %
\choice{s}{\fail}\ \ \leadsto\ \ %
s
\]
That is, we show that \fail\ is the unit of \lchoice{\cdot}{\cdot}
while assuming that it is the zero of sequential composition. Since
the derivation resembles desirable algebraic identities of choice,
sequential composition and failure that are also met by the
formalisation of \mycalc, we should assume that all strategies in the
derivation are of the same type. This is in conflict with the idea to
use \lchoice{\cdot}{\cdot}\ or \choice{\cdot}{\cdot}\ to inhabit
generic types. Furthermore, the approach would also affect the
reduction semantics in a way that goes beyond a conservative
extension.  We had to redefine the semantics of \choice{\cdot}{\cdot}
to make sure that the argument strategies are applied only if their
type and the type of the term at hand fits. Finally, a too liberal
typing rule for choice makes it easy for a strategic programmer to
confuse two different idioms:
\begin{itemize}
\item recovery from failure, and
\item the inhabitation of generic types.
\end{itemize}
This confusion can lead to unintentionally generic strategies which
then succeed (cf.\ \lchoice{\cdots}{\id}) or fail (cf.\
\lchoice{\cdots}{\fail}) in a surprising manner. The avoidance of
this confusion is among the major advantages that a typed system
offers when compared to the currently untyped Stratego.

\begin{figure}
{\smallsize

\parbox[t]{.45\textwidth}{

\dech{Syntax}
\begin{eqnarray*}
s   & ::= & \cdots\ |\ \extend{s}{\pi}
\end{eqnarray*}

}\hfill\parbox[t]{.45\textwidth}{

\decj{Well-typedness\\of strategies}{%
\wtj{\Gamma}{s}{\pi}
}

\ir{extend}{%
\wtj{\Gamma}{s}{\pi'}
\dnp
\pi' \typeless \pi
}{%
\wtj{\Gamma}{\extend{s}{\pi}}{\pi}
}

}

\decj{Reduction of strategy applications}{\cioj{\Gamma}{\apply{s}{t}}{r}}

\parbox[t]{.45\textwidth}{

\dech{Positive rule}

\medskip

\ir{extend^+}{%
\sps{
\wtj{\Gamma}{s}{\pi'}
\np
\wtj{\Gamma}{t}{\tau}
\np
\exists \tau'.\ \tau \to \tau' \typeleq \pi'
\np
\cioj{\Gamma}{\apply{s}{t}}{t'}
}
}{%
\cioj{\Gamma}{\apply{\extend{s}{\pi}}{t}}{t'}
}

}\hfill\parbox[t]{.45\textwidth}{

\smallskip
\dech{Negative rules}

\medskip

\ir{extend^-.1}{%
\sps{
\wtj{\Gamma}{s}{\pi'}
\np
\wtj{\Gamma}{t}{\tau}
\np
\exists \tau'.\ \tau \to \tau' \typeleq \pi'
\np
\cioj{\Gamma}{\apply{s}{t}}{\failure}
}
}{%
\cioj{\Gamma}{\apply{\extend{s}{\pi}}{t}}{\failure}
}

\ir{extend^-.2}{%
\sps{
\wtj{\Gamma}{s}{\pi'}
\np
\wtj{\Gamma}{t}{\tau}
\np
\nexists \tau'.\ \tau \to \tau' \typeleq \pi'
}
}{%
\cioj{\Gamma}{\apply{\extend{s}{\pi}}{t}}{\failure}
}

}
}
\caption{Turning many-sorted strategies into generic ones}
\label{F:extend}
\bigskip
\end{figure}

\paragraph*{Inhabitation by extension}

The combinator \extend{\cdot}{\cdot}\ serves for the \emph{explicit}
extension of a many-sorted strategy to become applicable to terms of
all sorts. Suppose the type of $s$ is $\tau \to \tau$. Then, of
course, $s$ can only be applied to terms of sort $\tau$ in a type-safe
manner. It is the very meaning of \apply{\extend{s}{\alltp}}{t}\ to
apply $s$ if and only if $t$ is of sort $\tau$. Otherwise,
\apply{\extend{s}{\alltp}}{t}\ fails. Hence, $s$ is extended in a
trivial sense, that is, to behave like \fail\ for all sorts different
from $\tau$. Well-typedness and the reduction semantics of the
combinator \extend{\cdot}{\cdot}\ are defined in
Figure~\ref{F:extend}. We should point out a paradigm shift, namely
that the typing context $\Gamma$ is now also part of the reduction
judgement. That is, the new judgement for the reduction of strategy
applications takes the form $\cioj{\Gamma}{\apply{s}{t}}{r}$. In the
typing rule \tg{extend}, we check if the actual type $\pi'$ of $s$ in
\extend{s}{\pi} is an instance of the type $\pi$ for the planned
extension. In the reduction semantics, in rule \tg{extend^+}, we check
if the type $\tau$ of the term $t$ is covered by the type $\pi'$ of
$s$ in \extend{s}{\pi}. Clearly, this check made the addition of the
typing context $\Gamma$\ necessary. Note that the generic type $\pi$
from \extend{s}{\pi}\ does not play any role during reduction. The
type $\pi$\ is only relevant in the typing rule for strategy extension
to point out the result type of strategy extension.

\paragraph*{Type dependency}

The combinator \extend{\cdot}{\cdot}\ makes it explicit where we want
to become generic. There is no hidden way how many-sorted ingredients
may become generic---accidentally or otherwise. As the reduction
semantics of \extend{\cdot}{\cdot}\ clearly points out, reduction is
truly type-dependent. That is, the reduction of an extended strategy
depends on the run-time comparison of the types of $s$ and $t$ in
\apply{\extend{s}{\pi}}{t}. We assume that all previous rules of
the reduction semantics are lifted to the new form of judgement by
propagating $\Gamma$. Otherwise, all rules stay intact, and hence we
may claim that the incorporation of \extend{\cdot}{\cdot}\ corresponds
to a conservative extension. One should not confuse type-dependent
reduction with dynamic type checks. Type dependency merely means that
a generic strategy admits different behaviours for different sorts. As
an aside, in Section~\ref{SS:elaboration}, we will discuss a
convenient approach to eliminate the typing judgements in the
reduction semantics for strategy extension. The basic idea is to
resort to tagged strategy applications such that types do not need to
be determined at run-time, but a simple tag comparison is sufficient
to perform strategy extension.

\subsection{Restriction}
\label{SS:restriction}

So far, we only considered one direction of mediation between
many-sorted and generic strategy types. We should also refine our type
system so that generic strategies can be easily applied in specific
contexts. Actually, there is not just one way to accommodate
restriction.  Compared to extension, restriction is conceptually much
simpler since restriction is immediately type-safe without further
precautions. In general, we are used to the idea that a generic entity
is used in a specific context, e.g., in the sense of parametric
polymorphism.

\paragraph*{Explicit restriction}

We consider a strategy combinator \restrict{s}{\pi} which has no
semantic effect, but at the level of typing it allows us to consider a
generic strategy $s$ to be of type $\pi$ provided it holds $\pi
\typeless \pi'$ where $\pi'$ is the actual type of $s$. Explicit
restriction is defined in Figure~\ref{F:restrict}. The combinator for
restriction is immediately sufficient if we want to apply a generic
strategy $s$ to a term $t$ of a certain sort $\tau$. If we assume, for
example, that $s$ is of type \alltp, then the well-typed strategy
application \apply{\restrict{s}{\tau\to\tau}}{t}\ can be
employed. Thus, the rule \tg{apply} for strategy applications from
Figure~\ref{F:tp} or the updated rule from Figure~\ref{F:tc} can be
retained without modifications.

\begin{figure}
{\smallsize

\dech{Syntax}
\begin{eqnarray*}
s & ::= & \cdots\ |\ \restrict{s}{\pi}
\end{eqnarray*}

\parbox{.44\textwidth}{

\decj{Well-typedness\\of strategies}{\wtj{\Gamma}{s}{\pi}}

\ir{restrict}{%
\wtj{\Gamma}{s}{\pi'}
\dnp
\pi \typeless \pi'
}{%
\wtj{\Gamma}{\restrict{s}{\pi}}{\pi}
}

}\hfill\parbox{.49\textwidth}{

\decj{Reduction\\of strategy applications}{\cioj{\Gamma}{\apply{s}{t}}{r}}

\ir{\neutral{restrict}}{%
\cioj{\Gamma}{\apply{s}{t}}{r}
}{%
\cioj{\Gamma}{\apply{\restrict{s}{\pi}}{t}}{r}
}

}

}
\caption{Explicit strategy restriction}
\label{F:restrict}
\bigskip
\end{figure}

\paragraph*{Extension and restriction in concert}

For completeness, let us assume that we also can annotate strategies
by their types, say by the form $s:\pi$. This is well-typed if $s$ is
indeed of type $\pi$. The reduction of $s:\pi$ simply resorts to
$s$. Then, in a sense, the three forms \restrict{s}{\pi} (i.e.,
explicit restriction), $s:\pi$ (i.e., type annotation), and
\extend{s}{\pi} (i.e., strategy extension) complement each other as
they deal with the different ways how a strategy $s$ and a type $\pi$
can be related to each other via the partial order
\typeleq. The three forms interact with the type system and the
reduction semantics in the following manner. Type annotations can be
removed without any effect on well-typedness and semantics. By
contrast, a replacement of a restriction \restrict{s}{\pi} by $s$ will
result in an ill-typed program, although $s$ is semantically
equivalent to $\restrict{s}{\pi}$. Finally, a replacement of an
extension \extend{s}{\pi} by $s$ will not just harm well-typedness,
but an ultimate application of $s$ is not even necessarily type-safe.

\paragraph*{Inter-mezzo}

The concepts that we have explained so far are sufficient to assemble
a calculus $S'_{\alltp}$ which covers generic type-preserving
traversal.  Generic type-preserving strategies could be combined with
both $S'_{tp}$\ and $S'_{tc}$. For simplicity, we have chosen the
former as the starting point for $S'_{\alltp}$. For convenience, we
summarise all ingredients of $S'_{\alltp}$:
{\begin{itemize}\noskip
\item The basic calculus $S'_0$ (cf.\ Figure~\ref{F:apply-basic})
\item Many-sorted type-preserving strategies (cf.\ Figure~\ref{F:tp})
\item Generic traversal primitives \all{\cdot} and
\one{}{\cdot}\ (cf.\ Figure~\ref{F:apply-all-one})
\item The generic type \alltp\ (cf.\ Figure~\ref{F:alltp})
\item Strategy extension (cf.\ Figure~\ref{F:extend})
\item Explicit restriction (cf.\ Figure~\ref{F:restrict})
\end{itemize}}%
\noindent
In Theorem~\ref{T:tp}, and Theorem~\ref{T:tc}, we started to address
properties of the many-sorted fragments $S'_{tp}$\ and $S'_{tc}$\ of
\mycalc. Let us update the theorem for $S'_{\alltp}$ accordingly.

\begin{theorem}
\label{T:alltp}
The calculus $S'_{\alltp}$ obeys the following properties:
\begin{enumerate}
\item Strategies satisfy UOT,
and their types adhere to the scheme of type preservation,
i.e.,
for all well-formed contexts $\Gamma$,
strategy types $\pi$, $\pi'$
and strategies $s$:
\begin{enumerate}
\item 
$\wtj{\Gamma}{s}{\pi}\ \wedge\ %
\wtj{\Gamma}{s}{\pi'}$ implies $\pi = \pi'$, and
\item
$\wtj{\Gamma}{s}{\pi}$ implies $\pi \typeleq \alltp$.
\end{enumerate}
\item Strategy applications satisfy UOT,
i.e., ... (cf.\ Theorem~\ref{T:tp}).
\item Reduction of strategy applications satisfies subject reduction,
i.e.,
for all well-formed contexts $\Gamma$,
strategies $s$,
terms $t$, $t'$,
term types $\tau$
and strategy types $\pi$:\\
$\wtj{\Gamma}{s}{\pi}\ \wedge\ %
\wtj{\Gamma}{t}{\tau}\ \wedge\ %
\tau \to \tau \typeleq \pi\ \wedge\ %
\cioj{\Gamma}{\apply{s}{t}}{t'}$ implies $\wtj{\Gamma}{t'}{\tau}$.
\end{enumerate}
\end{theorem}

This theorem is an elaboration of Theorem~\ref{T:tp} for many-sorted
type-preserving strategies. The first property is strengthened since we
now claim UOT for strategies. This becomes possible because
the previously ``overloaded'' combinators \id\ and \fail\ are now of type
\alltp. Further, we need to rephrase what it means that a strategy type
adheres to the scheme of type preservation. Also, the formulation of
subject reduction needs to be upgraded to take \alltp\ into account.

\begin{proof}\hfill\mbox{}
\label{P:alltp}
\begin{enumerate}
\item
\begin{enumerate}
\item 
We prove this property by induction on $s$ in \wtj{\Gamma}{s}{\pi}.
\emph{Base cases}: UOT for rewrite rules is implied by UOT for terms.
Congruences meet UOT because of the requirement for well-formed contexts.
The two base cases for \id\ and \fail\ trivially satisfy UOT by simple
inspection of the type position in the rules \tg{id}\ and \tg{fail}.
\emph{Induction step}: UOT for \negbf{s} is implied by the IH and by the
fact that the auxiliary judgement for negatable types encodes a
function from the argument type to the type of the negated
strategy. UOT for \seq{s_1}{s_2}\ and \choice{s_1}{s_2}\
is implied by the IH.  In both cases, the type of the compound
strategy coincides with the type of the arguments (cf.\ \tg{comp.1},
\tg{comp.2}, \tg{seq}, \tg{choice}). UOT for
$f(s_1,\ldots,s_n)$ follows from well-formedness of contexts. As for,
\all{s}, \one{}{s}, the property can be inferred by inspection of the
type position in the corresponding typing rules. As for
\extend{s}{\pi} and \restrict{s}{\pi}, UOT follows from
the fact that the specified $\pi$ directly constitutes the type of the
extended or restricted strategy.
\item
We prove this property by induction on $s$ in \wtj{\Gamma}{s}{\pi}. We
only need to cover the cases which were updated or newly introduced in
the migration from $S'_{tp}$ to $S'_{\alltp}$. \emph{Base cases}: The
types of \id\ and \fail\ are uniquely defined as \alltp, and hence,
they trivially adhere to the required scheme. \emph{Induction step}:
As an aside, we hardly have to employ the IH for the compound
strategies.  The type of \all{s} and \one{}{s}\ is defined as
\alltp. The forms \negbf{s} and \seq{s_1}{s_2} where type-preserving in
$S'_{tp}$. There are still type-preserving in $S'_{\alltp}$ since the
added rules for negatable and composable types only admit \alltp\ as
an additional possible result type (cf.\ \tg{negt.2} and
\tg{comp.2}). The property holds for choice because the arguments are
type preserving by the IH, and the result type of choice
coincides with the type of the arguments (cf.\ \tg{choice}). The type
$\pi$ in \extend{s}{\pi}, and hence the type of \extend{s}{\pi}
itself, must coincide with \alltp\ since this is the only possible
strategy type admitted as the right argument of \typeless\ in
$S'_{\alltp}$ (cf.\ \tg{extend}\ and \tg{less.1}). Dually, the type
$\pi$ in \restrict{s}{\pi}, and hence the type of \restrict{s}{\pi}
itself, must coincide with a type of the form $\tau \to \tau$ because
this is the only possible form admitted as the left argument of
\typeless\ in $S'_{\alltp}$ (cf.\ \tg{restrict}\ and \tg{less.1}).
\end{enumerate}
\item 
The property follows immediately from
{\begin{itemize}\noskip
\item UOT for terms,
\item UOT for strategies, and
\item the fact that \alltp\ is a ``well-defined generic type'', that is,
fixing the term type $\tau$ processed by a generic strategy, the result
type of strategy application is determined, in fact, it is $\tau$ in the
case of \alltp.
\end{itemize}}
\item
Note that we only deal with type-preserving strategies which allows us to
adopt Proof~\ref{P:tp} to a large extent. Of course, the side condition for
the employment of the IH has to be revised. That is, the IH can be employed
for a premise \ioj{\apply{s_i}{t_i}}{t'_i} and corresponding types $\tau_i$
and $\pi_i$ if we can prove the following side condition:
\[\begin{array}{l}
\wtj{\Gamma}{s}{\pi}\ \wedge\ %
\wtj{\Gamma}{t}{\tau}\ \wedge\ %
\tau \to \tau \typeleq \pi\ \wedge\ \ldots\\
\ \mbox{implies}\ %
\wtj{\Gamma}{s_i}{\pi_i}\ \wedge\ %
\wtj{\Gamma}{t_i}{\tau_i}\ \wedge\ %
\tau_i \to \tau_i \typeleq \pi_i
\end{array}\]
\emph{Base cases}: Proof~\ref{P:tp} is still intact as for 
rewrite rules and congruences for constants since these forms of
strategies were completely preserved in $S'_{\alltp}$. The strategy
\id\ is said to be generically type-preserving according to \tg{id}
while it was ``overloaded'' before. Reduction of \apply{\id}{t}\
yields $t$.  Hence, reduction of \id\ is type-preserving.
\emph{Induction step}: The property holds for negation because
\tg{neg^+} only admits the input term as proper term reduct. The
proof for sequential composition can be precisely repeated as in
Proof~\ref{P:tp} with the only exception that we now have to consider
two cases according to \tg{comp.1}\ and \tg{comp.2}. In both cases,
the type of $s_1$, $s_2$ and \seq{s_1}{s_2} coincide, and the types
adhere to the scheme of type preservation. This is all what the
original proof relied on. Proof~\ref{P:tp} also remains valid as for
choice and congruences. As for the traversal combinators, the property
follows from the IH, and the fact that the shape of the processed term
is preserved. The IH is enabled for any \ioj{\apply{s}{t_i}}{t'_i}
because the strategy $s$ in \all{s}\ and \one{}{s}\ is required to be
of type \alltp, and hence it can cope with any term. The interesting
case is \apply{\extend{s}{\pi}}{t} where we assume that $t$ is of type
$\tau$. We want to employ the IH for the premise
\cioj{\Gamma}{\apply{s}{t}}{t'} in \tg{extend^+}. Hence we are obliged
to show that the type $\pi'$ of $s$ covers $\tau$. This obligation is
precisely captured by the premise $\exists \tau'.\ \tau \to \tau'
\typeleq \pi'$ in \tg{extend^+}.  Thus the IH is enabled, and subject
reduction holds. As for \restrict{s}{\pi}, we directly resort to
$s$. The IH for the reduction of $s$ is trivially implied since $\tau
\to \tau \typeleq \pi$ implies $\tau \to \tau \typeleq \pi'$ for
$\pi \typeless \pi'$ (cf.\ \tg{restrict}) by transitivity of \typeleq.
\end{enumerate}
\end{proof}

\paragraph*{Implicit restriction}

While extension required a dedicated combinator for the reasons we
explained earlier, we do not need to insist on explicit
restriction. Implicit restriction is \emph{desirable} because
otherwise a programmer needs to point out a specific type whenever a
generic strategy is applied in a many-sorted context. Implicit
restriction is \emph{feasible} because restriction has no impact on
the reduction semantics of a strategy. Let us stress that implicit
restriction does not harm type safety in any way. In the worst case,
implicit restriction might lead to accidentally many-sorted
strategies. However, such accidents will not go unnoticed. If we
attempt to apply the intentionally generic strategy in a generic
context or to assign a generic type to it, then the type system will
refuse such attempts.

\begin{example}
Consider the strategy \s{Nat} which was defined in Figure~\ref{F:untyped}.
It involves a congruence $\w{succ}(\id)$, where \id\ is supposed to be
applied to a natural. In $S'_{\alltp}$, we have to rephrase the congruence as
$\w{succ}(\restrict{\id}{\w{Nat}})$. In a calculus with implicit restriction,
$\w{succ}(\id)$ can be retained.
\end{example}

\begin{figure}
{\smallsize

\parbox[t]{.455\textwidth}{

\decj{Composable types}{%
\comp{\pi_1}{\pi_2}{\pi}
}

\ir{comp.3}{%
\wfj{\Gamma}{\tau}
\dnp
\wfj{\Gamma}{\tau'}
}{%
\comp{\tau \to \tau'}{\alltp}{\tau \to \tau'}
}

\ir{comp.4}{%
\wfj{\Gamma}{\tau}
\dnp
\wfj{\Gamma}{\tau'}
}{%
\comp{\alltp}{\tau \to \tau'}{\tau \to \tau'}
}

\decj{Well-typedness of\\strategy applications}{%
\wtj{\Gamma}{\apply{s}{t}}{\tau}
}

\ir{apply}{%
\sps{
\wtj{\Gamma}{s}{\pi}
\dnp
\wtj{\Gamma}{t}{\tau}
\np
\exists \tau'.\ \tau \to \tau' \typeleq \pi
}
}{%
\wtj{\Gamma}{\apply{s}{t}}{\tau'}
}

}\hfill\parbox[t]{.515\textwidth}{

\decj{Well-typedness\\of strategies}{%
\wtj{\Gamma}{s}{\pi}
}

\ir{choice}{%
\sps{
\wtj{\Gamma}{s_1}{\pi_1}
\np
\wtj{\Gamma}{s_2}{\pi_2}
\np
\glb{\pi_1}{\pi_2}{\pi}
}
}{%
\wtj{\Gamma}{\choice{s_1}{s_2}}{\pi}
}

\ir{cong.2}{%
\sps{
f: \sigma_1 \times \cdots \times \sigma_n \to \sigma_0 \in \Gamma
\np
\wtj{\Gamma}{s_1}{\pi_1}
\dnp \sigma_1 \to \sigma_1 \typeleq \pi_1
\np
\cdots
\np
\wtj{\Gamma}{s_n}{\pi_n}
\dnp \sigma_n \to \sigma_n \typeleq \pi_n
}
}{%
\wtj{\Gamma}{f(s_1,\ldots,s_n)}{\sigma_0 \to \sigma_0}
}

}
}
\caption{Refinement of the typing rules for implicit restriction}
\label{F:context}
\bigskip
\end{figure}

\paragraph*{Implicit restriction made explicit}

Let us first consider one form of implicit restriction where we update
all typing rules which have to do with potentially many-sorted
contexts. Basically, we want to state that the type of a compound
strategy \seq{s_1}{s_2}\ or \choice{s_1}{s_2}\ is dictated by a
many-sorted argument (if any). As for congruences, we want to state
that generic strategies can be used as argument strategies. Finally,
we also need to relax strategy application so that we can apply a
generic strategy without further precautions to any term. This
approach to implicit restriction is formalised in
Figure~\ref{F:context}. The updated rule \tg{apply}\ for
well-typedness of strategy applications states that a strategy $s$ of
type $\pi$\ can be applied to a term $t$ of type $\tau$, if the domain
of $\pi$ covers $\tau$. As an aside, note that the definition is
sufficiently general to cope with type-changing strategies. As for
\seq{\cdot}{\cdot}, we relax the definition of composable types to
cover composition of a many-sorted type and the generic type \alltp\
in both possible orders (cf.\ \tg{comp.3}\ and \tg{comp.4}). As for
\choice{\cdot}{\cdot}, we do not insist on equal argument types
anymore, but we employ an auxiliary judgement \glb{\pi_1}{\pi_2}{\pi}
for the greatest lower bound w.r.t.\ \typeleq\ (cf.\
\tg{choice}). Finally, we relax the argument types for congruences via
the \typeleq\ relation.

\paragraph*{Unicity of typing vs.\ principal types}

The value of the refinement in Figure~\ref{F:context} is that we are
very precise about where restriction might be needed. Moreover, we can
maintain UOT for this system. A problem with the above approach is
that several typing rules need to be refined to become aware of
\typeless. There is a simpler approach to implicit restriction. We
can include a typing rule which models that a generic strategy can
also be regarded as a many-sorted strategy. The rule is shown in
Figure~\ref{F:implicit}. This new approach implies that UOT does not
hold anymore for strategies. However, one can easily see that the
multiple types arising from implicit restriction are closed under
\typeleq. Thus, one can safely replace UOT by the existence of a
\emph{principal} type. We simply regard the most generic type of a
strategy as its principal type. Note that UOT still holds for strategy
\emph{application}.

\begin{figure}
{\smallsize

\decj{Well-typedness of strategies}{%
\wtj{\Gamma}{s}{\pi}
}

\ir{implicit}{%
\wtj{\Gamma}{s}{\pi}
\dnp
\pi' \typeless \pi
}{%
\wtj{\Gamma}{s}{\pi'}
}

}

\caption{Implicit restriction relying on principal types}
\label{F:implicit}
\bigskip
\end{figure}

\subsection{Strategies of type \alltu{\cdot}}
\label{SS:alltu}

\begin{figure}
{\smallsize

\dech{Syntax}
\begin{eqnarray*}
s & ::= & \cdots\ |\ \reduce{}{s}{s}\ |\ \select{}{s}\ |\ \void\ |\ \spawn{s}{s}
\end{eqnarray*}

\decj{Reduction of strategy applications}{%
\ioj{\apply{s}{t}}{t'}
}

\parbox[t]{.5\textwidth}{

\dech{Positive rules}

\medskip

\ir{red^+}{%
\sps{
\ioj{\apply{s}{t_1}}{t'_1}
\np
\cdots
\np
\ioj{\apply{s}{t_n}}{t'_n}
\np
\ioj{\apply{\splus}{\pair{t'_1}{t'_2}}}{t'_{n+1}}
\np
\ioj{\apply{\splus}{\pair{t'_{n+1}}{t'_3}}}{t'_{n+2}}
\np
\cdots
\np
\ioj{\apply{\splus}{\pair{t'_{2n-2}}{t'_n}}}{t'_{2n-1}}
}
}{%
\ioj{\apply{\reduce{}{\splus}{s}}{f(t_1, t_2, \ldots, t_n)}}{t'_{2n-1}}
}

\ir{sel^+}{%
\exists i \in \{1,\ldots,n\}.\ \ioj{\apply{s}{t_i}}{t'_i}
}{%
\ioj{\apply{\select{}{s}}{f(t_1, \ldots, t_n)}}{t'_i}
}

\ax{void^+}{%
\ioj{\apply{\void}{t}}{\emptytuple}
}

\ir{spawn^+}{%
\ioj{\apply{s_1}{t}}{t_1}
\dnp
\ioj{\apply{s_2}{t}}{t_2}
}{%
\ioj{\apply{\spawn{s_1}{s_2}}{t}}{\pair{t_1}{t_2}}
}

\dech{Negative rules}

\medskip

\ax{red^-.1}{%
\ioj{\apply{\reduce{}{\splus}{s}}{c}}{\failure}
}

\ir{red^-.2}{%
\exists i \in \{1,\ldots,n\}.\ \ioj{\apply{s}{t_i}}{\failure}
}{%
\ioj{\apply{\reduce{}{\splus}{s}}{f(t_1,\ldots,t_n)}}{\failure}
}

}\hfill\parbox[t]{.47\textwidth}{

\mbox{}

\ir{red^-.3}{%
\sps{
\ioj{\apply{s}{t_1}}{t'_1}
\np
\cdots
\np
\ioj{\apply{s}{t_n}}{t'_n}
\np
\ioj{\apply{\splus}{\pair{t_1}{t_2}}}{\failure}
}
}{%
\ioj{\apply{\reduce{}{\splus}{s}}{f(t_1, t_2, \ldots, t_n)}}{\failure}
}

\ir{red^-.4}{%
\sps{
\!\!\!\!\!\exists i \in \{3,\ldots,n\}.\ (\\
& 
\ioj{\apply{s}{t_1}}{t'_1}
\np
\cdots
\np
\ioj{\apply{s}{t_n}}{t'_n}
\np
\ioj{\apply{\splus}{\pair{t_1}{t_2}}}{t'_{n+1}}
\np
\cdots
\np
\ioj{\apply{\splus}{\pair{t'_{n+i-2}}{t'_i}}}{\failure})
}
}{%
\ioj{\apply{\reduce{}{\splus}{s}}{f(t_1, t_2, \ldots, t_n)}}{\failure}
}

\ax{sel^-.1}{%
\ioj{\apply{\select{}{s}}{c}}{\failure}
}

\ir{sel^-.2}{%
\sps{
\ioj{\apply{s}{t_1}}{\failure}
\np
\cdots
\np
\ioj{\apply{s}{t_n}}{\failure}
}
}{%
\ioj{\apply{\select{}{s}}{f(t_1, \ldots, t_n)}}{\failure}
}

\ir{spawn^-}{%
\ioj{\apply{s_1}{t}}{\failure}%
\ \ \vee\,%
\ioj{\apply{s_2}{t}}{\failure}
}{%
\ioj{\apply{\spawn{s_1}{s_2}}{t}}{\failure}
}

}
}
\caption{Type-unifying combinators}
\label{F:apply-alltu}
\bigskip
\end{figure}

\vspace{-35\in}

\paragraph*{Combinators}

In Figure~\ref{F:apply-alltu}, the reduction semantics of the
combinators for type-unifying traversal is defined.  Thereby, we
complete the combinator suite of \mycalc.  For brevity, we omit the
typing context which might be needed for the type-dependent reduction
of \extend{\cdot}{\cdot}. The combinator
\reduce{}{\cdot}{\cdot}\ is defined in \tg{red^+}. Every child $t_i$ is
processed, and pairwise composition is used to compute a final term
$t'_{2n-1}$ from all the intermediate results. Note that pairwise
composition is performed from left to right. This is a kind of
arbitrary choice at this point, and we will come back to this issue in
Section~\ref{SS:vary}. Note also that the reduction semantics for
\reduce{}{\cdot}{\cdot}\ does not specify a total temporal order on
how pairwise composition is intertwined with processing the
children. There are at least two sensible operational readings of
\tg{red^+}. Either we first process all children, and then we perform
pairwise composition, or we immediately perform pairwise composition
whenever a new child has been processed. The negative rules for
\reduce{}{\cdot}{\cdot}\ are similar to those of
\all{\cdot}\ and \one{}{\cdot}. A constant cannot be reduced as in the
case of \one{}{\cdot} (cf.\ \tg{red^-.1}). Reduction fails if any of
the children cannot be processed as in the case of \all{\cdot} (cf.\
\tg{red^-.2}). There is also the possibility that
pairwise composition fails (cf.\ \tg{red^-.3} and \tg{red^-.4}).

Selection of a child is more easily explained. The overall scheme
regarding both the positive rule and the two negative rules for
\select{}{\cdot}\ is very similar to the combinator \one{}{\cdot}. The
combinator \select{}{\cdot}\ differs from \one{}{\cdot} only in that
the shape of the input term is not preserved. Recall that in the
reduct of an application of \one{}{\cdot}, the outermost function
symbol and all non-processed children carry over from the input
term. Instead, selection simply yields the processed child. As in the
case of \one{}{\cdot}, we cannot process constants (\tg{sel^-.1}), and
we also need to fail if none of the children admits selection (cf.\
\tg{sel^-.2}).

Let us finally consider the auxiliary combinators \void\ and
\spawn{s_1}{s_2}.  The combinator \void\ simply accepts any term, and
reduction yields the empty tuple \emptytuple. The combinator is in a
sense similar to \id\ as it succeeds for every term. However, the term
reduct is of a trivial type, namely \emptytuple\ regardless of the
type of the input term. The strategy \spawn{s_1}{s_2}\ applies both
strategies to the input term, and the intermediate results are paired
(cf.\ \tg{spawn^+}). If either of the strategy applications fails,
\spawn{s_1}{s_2}\ fails, too (cf.\ \tg{spawn^-}).  Note that one could
attempt to describe the behaviour underlying \void\ and
\spawn{\cdot}{\cdot}\ by the following strategies that employ rewrite
rules:

\begin{eqnarray*}
\s{Void} & = & X \to \emptytuple\\
\s{Spawn}(\nu_1,\nu_2) & = &
X \to \where{\where{\pair{Y_1}{Y_2}}{Y_1 = \apply{\nu_1}{X}}}{Y_2 =
\apply{\nu_2}{X}}
\end{eqnarray*}
However, the above rewrite rules and the pattern variables would have
to be generically typed. This is in conflict with the design decisions
that were postulated by us for \mycalc. The types of rewrite rules in
\mycalc\ are required to be many-sorted. All genericity should arise
from distinguished primitive combinators. Recall that these
requirements are meant to support a clean separation of genericity and
specificity, a simple formalisation of \mycalc, and a simple
implementation of the calculus.

\begin{figure}
{\smallsize

\dech{Syntax}
\begin{eqnarray*}
\gamma & ::= & \cdots\ |\ \alltu{\tau}
\end{eqnarray*}

\parbox[t]{.46\textwidth}{

\decj{Well-formedness\\of strategy types}{\wfj{\Gamma}{\pi}}

\ir{pi.3}{%
\wfj{\Gamma}{\tau}
}{%
\wfj{\Gamma}{\alltu{\tau}}
}

\decj{Genericity relation}{%
$\pi \typeless \pi'$
}

\ir{less.2}{%
\wfj{\Gamma}{\tau}
\dnp
\wfj{\Gamma}{\tau'}
}{%
\tau' \to \tau \typeless \alltu{\tau}
}

\decj{Composable types}{\comp{\pi_1}{\pi_2}{\pi}}

\ir{comp.5}{%
\wfj{\Gamma}{\tau}
\dnp
\wfj{\Gamma}{\tau'}
}{%
\comp{\alltu{\tau}}{\tau\to\tau'}{\alltu{\tau'}}
}

}\hfill\parbox[t]{.46\textwidth}{

\decj{Well-typedness\\of strategies}{%
\wtj{\Gamma}{s}{\pi}
}

\ir{red}{%
\sps{
\wtj{\Gamma}{\splus}{\pair{\tau}{\tau} \to \tau}
\np
\wtj{\Gamma}{s}{\alltu{\tau}}
}
}{%
\wtj{\Gamma}{\reduce{}{\splus}{s}}{\alltu{\tau}}
}

\ir{sel}{%
\wtj{\Gamma}{s}{\alltu{\tau}}
}{%
\wtj{\Gamma}{\select{}{s}}{\alltu{\tau}}
}

\ax{void}{%
\wtj{\Gamma}{\void}{\alltu{\emptytuple}}
}

\ir{spawn}{%
\sps{
\wtj{\Gamma}{s_1}{\alltu{\tau_1}}
\np
\wtj{\Gamma}{s_2}{\alltu{\tau_2}}
}
}{
\wtj{\Gamma}{\spawn{s_1}{s_2}}{\alltu{\pair{\tau_1}{\tau_2}}}
}

}
}
\caption{\alltu{\cdot}---The type of generic type-unifying strategies}
\label{F:alltu}
\bigskip
\end{figure}

\paragraph*{The generic type \alltu{\cdot}}

The formalisation of the generic type (constructor) \alltu{\cdot} is
presented in Figure~\ref{F:alltu}. We basically have to perform the
same steps as we discussed for \alltp. Firstly, well-formedness of
\alltu{\cdot}\ is defined (cf.\ \tg{pi.3}). Secondly, the type scheme
underlying \alltu{\tau}\ is defined (cf.\ \tg{less.2}). Thirdly, the
auxiliary judgement for sequential composition is updated (cf.\
\tg{comp.5}) to describe how \alltu{\cdot}\ is promoted. Consider a
sequential composition \seq{s_1}{s_2}\ where $s_1$\ is of type
\alltu{\tau}, and $s_2$ is of type $\tau \to \tau'$. The result is of
type \alltu{\tau'}. Note that a many-sorted strategy followed by a
type-unifying strategy or the sequential composition of two
type-unifying strategies do not amount to a generic strategy.  Still
we could include these constellations in order to facilitate implicit
restriction. The typing rules for the type-unifying strategy
combinators are easily explained. Reduction of all children to a type
$\tau$\ via \reduce{}{\splus}{s}\ requires \splus\ to be able to map a
pair of type $\pair{\tau}{\tau}$\ to a value of type $\tau$ in the
sense of pairwise composition, and the strategy $s$ for processing
the children has to be type-unifying w.r.t.\ the same $\tau$ (cf.\
\tg{red}). The typing rule for the combinator
\select{}{\cdot}\ directly states that the combinator is a transformer
on type-unifying strategies (cf.\ \tg{sel}). The typing rule for
\void\ states that every type is mapped to the most trivial type (cf.\
\tg{void}). Finally, \spawn{\cdot}{\cdot}\ takes two type-unifying
strategies, and produces another type-unifying strategy.  If
\alltu{\tau_1}\ and \alltu{\tau_2}\ are the types of the argument
strategies in \spawn{s_1}{s_2}, then \alltu{\pair{\tau_1}{\tau_2}}\ is
the type of the resulting strategy (cf.\ \tg{spawn}).

\paragraph*{Assembly of \mycalc}

Let us compose the ultimate calculus \mycalc. It accomplishes both
type-preserving and type-changing strategies. Furthermore, tuples are
supported. Ultimately, the generic types \alltp\ and \alltu{\cdot} are
enabled. We favour implicit restriction for \mycalc. For convenience,
we summarise all ingredients for \mycalc:
{\begin{itemize}\noskip
\item The basic calculus $S'_0$ (cf.\ Figure~\ref{F:apply-basic})
\item Many-sorted type-preserving strategies (cf.\ Figure~\ref{F:tp})
\item Many-sorted type-changing strategies (cf.\ Figure~\ref{F:tc})
\item Polyadic strategies (cf.\ Figure~\ref{F:tuples})
\item The combinators \all{\cdot}
and \one{}{\cdot}\ (cf.\ Figure~\ref{F:apply-all-one})
\item The generic type \alltp\ (cf.\ Figure~\ref{F:alltp})
\item The combinators \reduce{}{\cdot}{\cdot}, \select{}{\cdot}, \void, and
\spawn{\cdot}{\cdot} (cf.\ Figure~\ref{F:apply-alltu})
\item The generic type \alltu{\cdot}\ (cf.\ Figure~\ref{F:alltu})
\item Strategy extension (cf.\ Figure~\ref{F:extend})
\item Implicit restriction (cf.\ Figure~\ref{F:context})
\end{itemize}}

\begin{theorem}
\label{T:alltu}
The calculus \mycalc\ obeys the following properties:
\begin{enumerate}
\item Strategies satisfy UOT, i.e., ... (cf.\ Theorem~\ref{T:alltp}).
\item Strategy applications satisfy UOT,  i.e., ... 
(cf.\ Theorem~\ref{T:tp}).
\item Reduction of strategy applications satisfies subject reduction,\\
i.e.,
for all well-formed contexts $\Gamma$,
strategies $s$,
terms $t$, $t'$,
term types $\tau$, $\tau'$,
and strategy types $\pi$:\\
$\wtj{\Gamma}{s}{\pi}\ \wedge\ %
\wtj{\Gamma}{t}{\tau}\ \wedge\ %
\tau \to \tau' \typeleq \pi\ \wedge\ %
\cioj{\Gamma}{\apply{s}{t}}{t'}$ implies $\wtj{\Gamma}{t'}{\tau'}$.
\end{enumerate}
\end{theorem}

We omit the proof because it is a simple combination of the ideas from
Proof~\ref{P:tc} and Proof~\ref{P:alltp}. In the former proof, we
generalised the scheme for many-sorted type-preserving strategies from
Proof~\ref{P:tp} to cope with \emph{type-changing} strategies as
type-unifying strategies are, too.  In the latter, we generalised
Proof~\ref{P:tp} in a different dimension, namely to cope with
\emph{generic} strategies as type-unifying strategies are, too. It is
easy to cope with implicit restriction instead of explicit restriction
in $S'_{\alltp}$, neither does the introduction of tuples pose any
challenge.


\section{Sophistication}
\label{S:sophi}

In the previous two sections we studied the reduction semantics and
the type system for all the \mycalc\ primitives. In this section, we
want to complement this development with a few supplementary
concepts. Firstly, we will consider a straightforward abstraction
mechanism for strategy combinators, that is, strategy
definitions. Secondly, we will refine the model underlying the
formalisation of \mycalc\ to obtain a reduction semantics which does
not employ typing judgements in the reduction semantics anymore. Thirdly,
we describe a form of overloaded strategies, that is, strategies which
are applicable to terms of several types.  Fourthly, we introduce some
syntactic sugar to complement strategy extension by a sometimes more
convenient approach to the inhabitation of generic types, namely
asymmetric type-dependent choice. Finally, we will discuss the
potential for more general or additional traversal combinators.

\subsection{Strategic programs}
\label{SS:prog}

The syntax and semantics of strategic programs is shown in
Figure~\ref{F:sp}. A strategic program is of the form $\Gamma\ \Delta\
s$.  Here $\Gamma$ corresponds to type declarations for the program,
$\Delta$ is a list of strategy definitions, and $s$\ is the main
expression of the program. A strategy definition is of the form
$\varphi(\nu_1,\ldots,\nu_n) = s$ where $\nu_1$, \ldots, $\nu_n$ are
the formal parameters. The parentheses are omitted if $\varphi$ has no
parameters.  We assume that the RHS $s$ does not contain other
strategy variables than $\nu_1$, \ldots, $\nu_n$. Furthermore, we
assume $\alpha$-conversion for the substitution of strategy
variables. In the judgement for the reduction of strategy
applications, we propagate the strategy definitions as context
parameter $\Delta$ (cf.\ \tg{\neutral{prog}}) so that occurrences of
strategy combinators can be expanded accordingly (cf.\
\tg{\neutral{comb}}). Note that the reduction judgement for strategy
applications carries $\Gamma$ in the context in order to enable
strategy extension.

\begin{figure}
{\smallsize

\dech{Syntax}
$$\begin{array}{lclr}
p & ::= & \Gamma\ \Delta\ s
& \mbox{(Programs)}
\\
\nu & & & \mbox{(Strategy variables)}
\\ 
\Gamma & ::= & \cdots\ |\ \varphi:\pi \times \cdots \times \pi \to \pi%
\ |\ \nu:\pi
& \mbox{(Contexts)}
\\
\Delta & ::= & \emptyset\ |\ \Delta, \Delta\ |\ \varphi(\nu,\cdots,\nu) = s
& \mbox{(Definitions)}
\\
s      & ::= & \cdots\ |\ \nu\ |\ \varphi(s,\ldots,s)
& \mbox{(Strategies)}
\end{array}$$

\smallskip

\parbox[t]{.43\textwidth}{

\decj{Reduction\\of programs}{\ioj{\apply{p}{t}}{r}}

\ir{\neutral{prog}}{%
\cioj{\Gamma,\Delta}{\apply{s}{t}}{r}
}{%
\ioj{\apply{\Gamma\ \Delta\ s}{t}}{r}
}

}\hfill\parbox[t]{.52\textwidth}{

\decj{Reduction\\of strategy applications}{\cioj{\Gamma,\Delta}{\apply{s}{t}}{r}}

\medskip

\ir{\neutral{comb}}{%
\sps{
\varphi(\nu_1,\cdots,\nu_n) = s \in \Delta
\np
s' = s\assigns{\assign{\nu_1}{s_1},\ldots,\assign{\nu_n}{s_n}}
\np
\cioj{\Gamma,\Delta}{\apply{s'}{t}}{r}
}
}{%
\cioj{\Gamma,\Delta}{\apply{\varphi(s_1,\ldots,s_n)}{t}}{r}
}

}
}
\caption{Strategic programs: syntax and reduction semantics}
\label{F:sp}
\bigskip
\end{figure}

\begin{figure}
{\smallsize

\parbox[t]{.49\textwidth}{

\decj{Well-typedness\\of programs}{\wtj{\Gamma}{p}{\pi}}

\ir{prog}{%
\wfj{\Gamma}{\Delta} 
\dnp \wtj{\Gamma}{s}{\pi}
}{%
\wtj{}{\Gamma\ \Delta\ s}{\pi}
}

\decj{Well-typedness\\of strategy definitions}{\wfj{\Gamma}{\Delta}}

\ax{def.1}{%
\wfj{\Gamma}{\emptyset}
}

\ir{def.2}{%
\wfj{\Gamma}{\Delta_1}
\dnp
\wfj{\Gamma}{\Delta_2}
}{%
\wfj{\Gamma}{\Delta_1, \Delta_2}
}

\ir{def.3}{%
\sps{
\varphi: \pi_1 \times \cdots \times \pi_n \to \pi_0 \in \Gamma
\np
\wtj{\Gamma, \nu_1:\pi_1, \ldots, \nu_n:\pi_n}{s}{\pi_0}
}
}{%
\wfj{\Gamma}{\varphi(\nu_1,\ldots,\nu_n) = s}
}

}\hfill\parbox[t]{.47\textwidth}{

\decj{Well-typedness\\of strategies}{\wtj{\Gamma}{s}{\pi}}

\ir{arg}{%
\nu:\pi \in \Gamma
}{%
\wtj{\Gamma}{\nu}{\pi}
}

\ir{comb}{%
\sps{
\varphi: \pi_1 \times \cdots \times \pi_n \to \pi_0 \in \Gamma
\np
\wtj{\Gamma}{s_1}{\pi_1}
\np
\cdots
\np
\wtj{\Gamma}{s_n}{\pi_n}
}
}{%
\wtj{\Gamma}{\varphi(s_1,\ldots,s_n)}{\pi_0}
}

}
}
\caption{Well-typedness of strategic programs}
\label{F:sp-wt}
\bigskip
\end{figure}

To consider well-formedness and well-typedness of strategic programs
we need to extend the grammar for contexts $\Gamma$ as it was already
indicated in Figure~\ref{F:sp}.  Contexts may contain type
declarations for strategy combinators and types of strategy
variables. A strategic program is well-formed if the strategy
definitions and the main expression of a program are well-typed (cf.\
\tg{prog}). A strategy definition is well-typed if the body can be
shown to have the declared result type of the combinator while
assuming the appropriate types of the formal parameters in the context
(cf.\ \tg{def.3}). When a strategy variable is encountered by the
well-typedness judgement, its type is determined via the context (cf.\
\tg{arg}). An application of a combinator is well-typed if the types
of the actual parameters are equal to the types of the formal
parameters (cf.\ \tg{comb}). We could also elaborate the latter typing
rule to facilitate implicit restriction. This would allow us to place
generic strategies as actual parameters on many-sorted parameter
positions of strategy combinators.

\begin{figure}
{\smallsize

\dech{Syntax}
$$\begin{array}{lclr}
\alpha &     & & \mbox{(Term-type variables)}\\
\tau   & ::= & \cdots\ |\ \alpha
& \mbox{(Term types)}
\\
\Gamma & ::= & \cdots\ |\ \varphi: \forall \alpha,\ldots,\alpha.\ %
\pi \times \cdots \times \pi \to \pi
& \mbox{(Contexts)}\\
\Delta & ::= & \cdots\ |\ \varphi{[}\alpha,\ldots,\alpha{]}%
(\nu,\ldots,\nu) = s
& \mbox{(Definitions)}\\
s & ::= & \cdots\ |\ \varphi{[}\tau,\ldots,\tau{]}(s,\ldots,s)
& \mbox{(Strategies)}
\end{array}$$
\decj{Well-formedness of term types}{\wfj{\Gamma}{\tau}}

\ir{tau.4}{%
\alpha \in \Gamma
}{%
\wfj{\Gamma}{\alpha}
}

\decj{Well-typedness of strategy definitions}{\wfj{\Gamma}{\Delta}}

\ir{def.4}{%
\sps{
\varphi:
\forall \alpha_1,\ldots,\alpha_m.\ %
\pi_1 \times \cdots \times \pi_n \to \pi_0 \in \Gamma
\np
\wtj{\Gamma,\nu_1:\pi_1,\ldots,\nu_n:\pi_n,
\alpha_1,\ldots,\alpha_m}{s}{\pi_0}
}
}{%
\wfj{\Gamma}{\varphi{[}\alpha_1,\ldots,\alpha_m{]}(\nu_1,\ldots,\nu_n) = s}
}

\decj{Well-typedness of strategies}{\wtj{\Gamma}{s}{\pi}}

\newcommand{\mypi}[1]{%
\pi_{#1}\assigns{\assign{\alpha_1}{\tau_1},\ldots,\assign{\alpha_m}{\tau_m}}
}

\ir{comb{-}forall}{%
\sps{
\varphi:
\forall \alpha_1,\ldots,\alpha_m.\ %
\pi_1 \times \cdots \times \pi_n \to \pi_0 \in \Gamma
\np
\wfj{\Gamma}{\tau_1} \dnp \ldots \dnp \wfj{\Gamma}{\tau_m}
\np
\wtj{\Gamma}{s_1}{\mypi{1}}
\np
\cdots
\np
\wtj{\Gamma}{s_n}{\mypi{n}}
}
}{%
\wtj{\Gamma}{\varphi{[}\tau_1,\ldots,\tau_m{]}(s_1,\ldots,s_n)}{%
\mypi{0}}
}

}

\caption{Type-parametrised strategy definitions}
\label{F:sp'}
\bigskip
\end{figure}

\paragraph*{Type-parameterised strategy definitions}

Let us also enable type-parameterised strategy definitions.  In
Figure~\ref{F:sp'}, we give typing rules to cope with type parameters
in strategy definitions and combinator applications.  The
formalisation is pretty standard. We assume $\alpha$-conversion for
the substitution of type variables.  Term-type variables are regarded
as another form of a term type. The extension of the grammar rule for
$\Gamma$ details that types of strategy combinators might contain type
variables that are quantified at the top level. Type variables are
scoped by the corresponding strategy definition (cf.\ \tg{def.4}). If
the well-formedness judgements for types encounter a term type
variable, it has to be in the context (cf.\ \tg{tau.4}). The
application of a combinator $\varphi$ involves type application,
namely substitution of the type variables by the actual types (cf.\
\tg{comb{-}forall}). For brevity, we do not refine the reduction
semantics from Figure~\ref{F:sp}.

\subsection{$\Gamma$-free strategy extension}
\label{SS:elaboration}

When we introduced strategy extension, we encountered a complication
regarding the reduction semantics. In order to define the
\emph{type-safe} application of a many-sorted strategy $s$ in a
generic context, we have to perform a run-time comparison of the type
of the given term and the type of $s$. To this end, we added the
typing context $\Gamma$ to the judgement for the reduction of strategy
applications, and typing judgements were placed as premises in the
rule for \extend{\cdot}{\cdot}\ (cf.\ Figure~\ref{F:extend}). We would
like to obtain a form of semantics where typing and reduction
judgements are strictly separated. We will employ an intermediary
static elaboration judgement to annotate strategies
accordingly. Furthermore, we assume that terms are tagged by their
types. The resulting reduction semantics is better geared towards
implementation.

\begin{figure}
{\smallsize

\decj{Static elaboration of strategies}{%
\cioj{\Gamma}{s}{s'}
}

\parbox[t]{.42\textwidth}{

\mbox{}

\ax{rule^{\leadsto}}{%
\cioj{\Gamma}{t_l \to t_r}{t_l \to t_r}
}

\ax{id^{\leadsto}}{%
\cioj{\Gamma}{\id}{\id}
}

\ax{fail^{\leadsto}}{%
\cioj{\Gamma}{\fail}{\fail}
}

\ir{neg^{\leadsto}}{%
\cioj{\Gamma}{s}{s'}
}{%
\cioj{\Gamma}{\negbf{s}}{\negbf{s'}}
}

\ir{seq^{\leadsto}}{%
\sps{
\cioj{\Gamma}{s_1}{s'_1}
\np
\cioj{\Gamma}{s_2}{s'_2}
}
}{%
\cioj{\Gamma}{\seq{s_1}{s_2}}{\seq{s'_1}{s'_2}}
}

}\hfill\parbox[t]{.55\textwidth}{

\mbox{}

\ir{choice^{\leadsto}}{%
\sps{
\cioj{\Gamma}{s_1}{s'_1}
\np
\cioj{\Gamma}{s_2}{s'_2}
}
}{%
\cioj{\Gamma}{\choice{s_1}{s_2}}{\choice{s'_1}{s'_2}}
}

\ax{cong^{\leadsto}.1}{%
\cioj{\Gamma}{c}{c}
}

\ir{cong^{\leadsto}.2}{%
\sps{
\cioj{\Gamma}{s_1}{s'_1}
\np
\cdots
\np
\cioj{\Gamma}{s_n}{s'_n}
}
}{%
\cioj{\Gamma}{f(s_1,\ldots,s_n)}{f(s'_1,\ldots,s'_n)}
}

}

}

\caption{General scheme of static elaboration}
\label{F:elaboration}
\bigskip
\end{figure}

\paragraph*{Static elaboration}

So far, we only considered well-typedness and reduction judgements. We
want to refine the model for the formalisation of \mycalc\ to include
a static elaboration judgement of the following form:
\[\cioj{\Gamma}{s}{s'}\]
The general idea of static elaboration is that the input strategy $s$
can be transformed in a semantics-preserving manner. There are several
potential applications of static elaboration. We will emphasise its
application to the problem of eliminating typing judgements in the
reduction semantics of \extend{\cdot}{\cdot}. In addition, one could
employ static elaboration for the definition of syntactic sugar or for
program optimisation~\cite{JV01}. The semantic model for typeful
strategies needs to be updated to consist of three phases:

\newpage

{\begin{enumerate}\noskip
\item The given strategy $s$ is checked to be well-typed.
\item $s$ is elaborated resulting in a strategy $s'$.
\item Given a suitable term $t$, the strategy $s'$ is applied to $t$ 
to derive a reduct.
\end{enumerate}
These phases obviously map nicely to an implementational model where
type checking and elaboration is done once and for all statically,
that is, without insisting on an input term. In general, static
elaboration might be type-dependent, that is, the typing context
$\Gamma$ is part of the elaboration judgement as in the case of the
well-formedness and well-typedness judgements. In
Figure~\ref{F:elaboration}, we initiate the general scheme of static
elaboration. We give trivial rules for all combinators of the basic
calculus $S'_0$ such that we descend into compound strategy
expressions. So far, the judgement encodes the identity function on
strategy expressions. Below, we will provide a special rule for the
elaboration of applications of \extend{\cdot}{\cdot}.

\begin{figure}
{\smallsize

\dech{Syntax}
$$\begin{array}{lclr}
s & ::= & \cdots\ |\ \annotate{s}{\pi} & \mbox{(Strategies)}\\
t & ::= & \cdots\ |\ \annotate{t}{\tau} & \mbox{(Terms)}
\end{array}$$

\parbox[t]{.44\textwidth}{

\decj{Static elaboration\\of strategies}{\cioj{\Gamma}{s}{s'}}

\ir{extend^{\leadsto}}{%
\sps{
\wtj{\Gamma}{s}{\pi'}
\np
\cioj{\Gamma}{s}{s'}
}
}{%
\cioj{\Gamma}{\extend{s}{\pi}}{\extend{s':\pi'}{\pi}}
}

}\hfill\parbox[t]{.49\textwidth}{

\decj{Reduction\\of strategy applications}{\ioj{\apply{s}{t}}{r}}

\dech{Positive rule}

\medskip

\ir{{extend'}^+}{%
\ioj{\apply{s}{t:\tau}}{t'}
}{%
\ioj{\apply{\extend{\annotate{s}{\tau\to\tau'}}}{t:\tau}}{t'}
}

\dech{Negative rules}

\medskip

\ir{{extend'}^-.1}{%
\ioj{\apply{s}{t:\tau}}{\failure}
}{%
\ioj{\apply{\extend{\annotate{s}{\tau\to\tau'}}}{t:\tau}}{\failure}
}

\ir{{extend'}^-.2}{%
\tau \not= \tau''
}{%
\ioj{\apply{\extend{\annotate{s}{\tau\to\tau'}}{\tau}}{t:\tau''}}{\failure}
}

}
}
\caption{Strategy extension relying on type tags}
\label{F:extend'}
\bigskip
\end{figure}

\paragraph*{Type tags}

In order to eliminate the typing premise for the extended strategy in
the reduction semantics of \extend{\cdot}{\cdot} we replace strategy
expressions of the form \extend{s}{\pi}\ by
\extend{\annotate{s}{\pi'}}{\pi}\ where $\pi'$ denotes the actual type
of $s$. Here, we reanimate the notation of type-annotated strategies
that was already proposed earlier.  Since the type is captured in the
elaborated strategy expression, the type of $s$ does not need to be
determined during reduction anymore. Furthermore, we assume that terms
are tagged by their sorts. Obviously, this assumption is useful to
also get rid of the type judgement for the term $t$ in the reduction
semantics for \apply{\extend{\cdots}{\pi}}{t}.  Thus, the original
type dependency reduces to a simple comparison of type tags of the
extended strategy and the term at hand.  The rules for static
elaboration and the new reduction semantics of strategy extension is
shown in Figure~\ref{F:extend'}. The elaboration rule
\tg{extend^{\leadsto}} deviates from the trivial default scheme of
static elaboration by actually adding the inferred type as a tag. The
deduction rule \tg{{extend'}^+} defines the new reduction semantics of
strategy extension.

\begin{figure}
{\smallsize

\decj{Reduction of strategy applications}{%
\ioj{\apply{s}{t}}{r}
}

\ax{cong^+.1}{%
\ioj{\apply{c}{\annotate{c}{\tau}}}{\annotate{c}{\tau}}
}

\ir{cong^+.2}{%
\sps{
\ioj{\apply{s_1}{t_1}}{t'_1}
\np
\cdots
\np
\ioj{\apply{s_n}{t_n}}{t'_n}
}
}{%
\ioj{\apply{f(s_1,\ldots,s_n)}{\annotate{f(t_1,\ldots,t_n)}{\tau}}}{\annotate{f(t'_1,\ldots,t'_n)}{\tau}}}

\ax{all^+.1}{%
\ioj{\apply{\all{s}}{\annotate{c}{\tau}}}{\annotate{c}{\tau}}
}

\ir{all^+.2}{%
\sps{
\ioj{\apply{s}{t_1}}{t'_1}
\np
\cdots
\np
\ioj{\apply{s}{t_n}}{t'_n}
}
}{%
\ioj{\apply{\all{s}}{\annotate{f(t_1, \ldots, t_n)}{\tau}}}{\annotate{f(t'_1, \ldots, t'_n)}{\tau}}
}

}
\caption{Refined reduction semantics to cope with tagged terms}
\label{F:propagate}
\bigskip
\end{figure}

\paragraph*{Tagged terms}

The assumption that terms are tagged by a type has actually two
implications which need to be treated carefully. Firstly, we should
better assume that terms are consistently tagged at all levels. This
means that the terms constituting a rewrite rule have to be tagged,
too. Secondly, we need to make sure that all reduction rules
appropriately deal with tagged terms. In fact, we need to update the
reduction semantics of congruences and generic traversal because they
are not prepared to deal with tags. In Figure~\ref{F:propagate}, we
illustrate the new style of traversal. For brevity, we only show the
positive rules for congruences and for the combinator \all{\cdot}.

\subsection{Overloaded strategies}
\label{SS:overloading}

We want to consider an intermediate form of genericity, namely
overloaded strategies. Overloading means that we can cope with
strategies which are applicable to terms of a number of sorts. We
introduce a designated combinator \plus{\cdot}{\cdot} to gather
strategies of different types in an overloaded strategy. The
combinator \plus{\cdot}{\cdot} is type-dependent in the same way as
strategy extension via \extend{\cdot}{\cdot}. In fact, we say that
\plus{\cdot}{\cdot}\ performs \emph{symmetric type-dependent choice}.
The type of the ultimate term decides which side of the choice is
attempted.  Hence, this choice is not left- or right-biased, nor is it
controlled by success and failure. We use the notation
\plus{\cdot}{\cdot} for the construction of both overloaded strategies
and the corresponding strategy types.

\begin{example}
\label{X:overloading}
Consider the following constructors for naturals and integers:
\begin{eqnarray*}
\w{one}      & : & \w{NatOne}\\
\w{succ}     & : & \w{NatOne} \to \w{NatOne}\\
\w{zero}     & : & \w{NatZero}\\
\w{notzero}  & : & \w{NatOne} \to \w{NatZero}\\
\w{positive} & : & \w{NatZero} \to \w{Int}\\
\w{negative} & : & \w{NatOne} \to \w{Int}\\
\end{eqnarray*}%
\w{NatZero} includes $0$, whereas \w{NatOne} starts with $1$. Integers
are constructed via two branches, one for positive integers including
zero, and another for negative integers. We use \w{NO} as stem of
variables of sort \w{NatOne}. Let us define two overloaded strategies
\s{Inc} and \s{Dec} which are capable of incrementing and decrementing
terms of the three above sorts:
\begin{eqnarray*}
\s{Inc} & : & \plus{\w{NatOne}\to\w{NatOne}}{\plus{\w{NatZero}\to\w{NatZero}}{\w{Int}\to\w{Int}}}\\
\s{Inc} & = & \plus{%
\w{NO} \to \w{succ}(\w{NO})%
\\&}{&\plus{%
\choice{\w{zero} \to \w{notzero}(\w{one})}{\w{notzero}(\s{Inc})}%
\\&}{&%
\choice{\choice{\w{positive}(\s{Inc})}{\w{negative}(\s{Dec})}%
}{\w{negative}(\w{one}) \to \w{positive}(\w{zero})}}}\\
\s{Dec} & : & \plus{\w{NatOne}\to\w{NatOne}}{\plus{\w{NatZero}\to\w{NatZero}}{\w{Int}\to\w{Int}}}\\
\s{Dec} & = & \plus{%
\w{succ}(\w{NO}) \to \w{NO}%
\\&}{&\plus{%
\choice{\w{notzero}(\w{one}) \to \w{zero}}{\w{notzero}(\s{Dec})}%
\\&}{&%
\choice{\choice{\w{positive}(\s{Dec})}{\w{negative}(\s{Inc})}%
}{\w{positive}(\w{zero}) \to \w{negative}(\w{one})}}}
\end{eqnarray*}
The strategies are defined via symmetric type-dependent choice with
three cases, one for each sort. Otherwise, the functionality to
increment and decrement is defined by rewrite rules or in terms of
congruences on the appropriate constructors. As an aside, it is
necessary to assume implicit restriction for overloaded strategies in
order to claim well-typedness for the above definitions. This is
because the using occurrences of \s{Inc} and \s{Dec}\ are used for
specific sorts covered by the overloaded types of \s{Inc} and \s{Dec}.
\end{example}

\begin{figure}
{\smallsize

\parbox[t]{.45\textwidth}{

\smallskip

\dech{Syntax}
\begin{eqnarray*}
\pi  & = & \cdots\ |\ \plus{\pi}{\pi}\\
s    & = & \cdots\ |\ \plus{s}{s}
\end{eqnarray*}

\decj{Well-formedness\\of strategy types}{\wfj{\Gamma}{\pi}}

\ir{pi.4}{%
\sps{
\wfj{\Gamma}{\pi_1}
\dnp
\dom{\pi_1} \leadsto \tau\!{s}_1
\np
\wfj{\Gamma}{\pi_2}
\dnp
\dom{\pi_2} \leadsto \tau\!{s}_2
\np
\tau\!{s}_1 \cap \tau\!{s}_2 = \emptyset
}
}{%
\wfj{\Gamma}{\plus{\pi_1}{\pi_2}}
}

\decj{Domains\\of strategies}{$\dom{\pi} \leadsto \tau\!{s}$}

\ax{dom.1}{%
\dom{\tau \to \tau'} \leadsto \{\tau\}
}

\ir{dom.2}{%
\sps{
\dom{\pi_1} \leadsto \tau\!{s}_1
\np
\dom{\pi_2} \leadsto \tau\!{s}_2
}
}{%
\dom{\plus{\pi_1}{\pi_2}} \leadsto \tau\!{s}_1 \cup \tau\!{s}_2
}

\decj{Genericity relation}{%
$\pi \typeless \pi'$
}

\ir{less.3}{%
\sps{
\wfj{\Gamma}{\pi}
\dnp
\wfj{\Gamma}{\pi'}
\np
\exists \pi''.\ \plus{\pi}{\pi''}\,\acequiv\,\pi'
}
}{%
\pi \typeless \pi'
}

\ir{less.4}{%
\pi_1 \typeless \pi
\dnp
\pi_2 \typeless \pi
}{%
\plus{\pi_1}{\pi_2} \typeless \pi
}

}\hfill\parbox[t]{.48\textwidth}{

\decj{Composable types}{\comp{\pi}{\pi'}{\pi''}}

\ir{comp.6}{%
\sps{
\pi_1\,\acequiv\,\plus{\pi'_1}{\pi''_1}
\np
\pi_2\,\acequiv\,\plus{\pi'_2}{\pi''_2}
\np
\comp{\pi'_1}{\pi'_2}{\pi'_3}
\np
\comp{\pi''_1}{\pi''_2}{\pi''_3}
}
}{%
\comp{\pi_1}{\pi_2}{\plus{\pi'_3}{\pi''_3}}
}

\decj{Well-typedness\\of strategies}{%
\wtj{\Gamma}{s}{\pi}
}

\ir{amp}{%
\sps{
\wtj{\Gamma}{s_1}{\pi_1}
\np
\wtj{\Gamma}{s_2}{\pi_2}
\np
\wfj{\Gamma}{\plus{\pi_1}{\pi_2}}
}
}{%
\wtj{\Gamma}{\plus{s_1}{s_2}}{\plus{\pi_1}{\pi_2}}
}

\decj{Reduction\\of strategy applications}{\cioj{\Gamma}{\apply{s}{t}}{r}}

\ir{\neutral{amp}}{%
\sps{
\!\!\!\!\!\exists i \in \{1,2\}.\ (\\
& 
\wtj{\Gamma}{t}{\tau}
\np
\wtj{\Gamma}{s}{\pi_i}
\np
\dom{\pi_i} \leadsto \tau\!{s}_i
\np
\tau \in \tau\!{s}_i
\np
\cioj{\Gamma}{\apply{s_i}{t}}{r})
}
}{%
\cioj{\Gamma}{\apply{\plus{s_1}{s_2}}{t}}{r}
}

}
}
\caption{Overloaded strategies}
\label{F:overloading}
\bigskip
\end{figure}

\paragraph*{Typing rules}

In Figure~\ref{F:overloading}, the reduction semantics for overloaded
strategies and the corresponding typing rules are defined. The type of
an overloaded strategy is of the form \plus{\tau_1 \to
\tau'_1}{\plus{\cdots}{\tau_n \to \tau'_n}}.  The type models
strategies which are applicable to terms of types $\tau_1$,
\ldots, $\tau_n$. If such a strategy is actually applied to a term of type
$\tau_i$, the result will be of type $\tau'_i$. We use an auxiliary
judgement $\dom{\pi}\,\leadsto\,\tau\!{s}$ to obtain the finite set
$\tau\!{s}$ of term types admitted as domains by a strategy type
$\pi$. We do not attempt to cover generic types in this judgement
because symmetric type-dependent choice cannot involve a generic
strategy. This is because if one branch would be generic, there are no
sorts left to be covered by the other branch. Indeed, we require that
the domains of the types composed by \plus{\cdot}{\cdot}\ must be
disjoint (cf.\ \tg{pi.4}). This requirement enforces immediately UOT
of strategy applications. Furthermore, the requirement also ensures
that type-dependent choice is deterministic, and hence does not
overlap with \choice{\cdot}{\cdot}, i.e., choice controlled by success
and failure.  In \tg{less.3}--\tg{less.4}, we update the relation
\typeless\ on strategy types. To this end, we employ an equivalence
\acequiv\ on strategy types modulo associativity and commutativity of
\plus{\cdot}{\cdot}. Rule \tg{less.3} models that $\pi$ is less
generic than any type $\pi'$ which is equivalent to
$\plus{\pi}{\pi''}$. Clearly, this rule is needed to relate simple
many-sorted and overloaded strategy types to each other. The rule also
relates overloaded strategy types among each other. Rule \tg{less.4}
models that the type of an overloaded strategy is less generic than
another type $\pi$, if both components $\pi_1$ and $\pi_2$ of the
overloaded type are also less generic than $\pi$. This rule relates
overloaded strategy types and generic types to each other. In this
elaboration of \typeless, the simple many-sorted strategies are the
least elements, and the generic types are the greatest elements.

\paragraph*{Reduction semantics}

An overloaded strategy is constructed by symmetric type-dependent
choice \plus{s_1}{s_2} where the types of the arguments $s_1$ and
$s_2$ have to admit the construction of an overloaded strategy type
(cf.\ \tg{amp}). As for the reduction semantics of \plus{s_1}{s_2},
the appropriate $s_i$ is chosen depending on the type of the input
term (cf.\ \tg{amp^+}). The kind of typing premises in the reduction
semantics are similar to the original definition of strategy
extension, and static elaboration could be used again to eliminate
them.  We should note that the refined reduction semantics from
Section~\ref{SS:elaboration} is not prepared to cope with overloaded
strategies. A corresponding generalisation does not pose any
challenge.

\paragraph*{Expressiveness}

Although overloading is convenient in strategic programming, it can
usually be circumvented with some additional coding effort. To
reconstruct Example~\ref{X:overloading} without overloading, we had to
define separate strategies for the different sorts \w{NatZero},
\w{NatOne}, and \w{Int}. Overloading is convenient to describe
many-sorted ingredients of a traversal in the case that the traversal
deals with several term types $\tau_1$, \ldots, $\tau_n$ in a specific
manner. If we use overloading we can compose the many-sorted
ingredients for $\tau_1$, \ldots, $\tau_n$ in one overloaded
strategy. It is then still possible to extend the overloaded strategy
in different ways before we pass it to the ultimate traversal
scheme. Without overloading, we need to immediately represent the
several many-sorted ingredients as a generic strategy by iterating
strategy extension for each $\tau_i$.  Also, while the type system
enforces that the $\tau_1$, \ldots, $\tau_n$ are distinct in the case
of overloading, there is no such guarantee without overloading. In
addition to the convenience added by overloading, it is also worth
mentioning that overloading can be used to reconstruct generic
strategies in some restricted manner. If we consider a fixed
signature, then we can represent the signature-specific instantiations
of generic strategy types as overloaded strategy types. Consider, for
example, the type \alltp. We can reconstruct \alltp\ by overloading
all $\tau \to \tau$ for all well-formed $\tau$ according the given
signature. Note that this construction becomes infinite if we enable
tuple types, but it is finite if we restrict ourselves to traversal of
many-sorted terms. Based on these signature-exhausting overloaded
types, we could represent the generic traversal combinators as
signature-specific overloaded combinators defined in terms of the
many-sorted congruences for all the available function symbols.

\paragraph*{Bibliographical notes}

When we compare symmetric type-dependent choice to other notions of
overloading or ad-hoc polymorphism~\cite{CW85,WB89,Jones95}, we should
note that these other notions are usually based on a form of
declaration as opposed to a combinator. Also, other models of
overloading usually perform overloading resolution at compile time
whereas the dispatch for overloaded strategies happens at
run-time. In~\cite{CGL95}, an extended $\lambda$-calculus $\lambda\&$\
is defined that employs type-dependent reduction in a way very similar
to our approach. Type-dependent reduction is used to model late
binding in the object-oriented sense.  More precisely, type-dependent
reduction is used in $\lambda\&$\ to resort to the most appropriate
``branch'' of a function based on the run-time type of the
argument. This work also discusses the relation of overloading and
intersection types~\cite{CDCV81,BDL95}. This is interesting because,
at a first glance, one could envision that intersection types might be
useful in modelling overloading.  For short, intersection types are
not appropriate to model overloading if type-dependent reduction is
involved. Using intersection types, we say that a function $f$ is of
type $a \cap b$ if $f$ can play the role of both an element of type
$a$ and of type $b$. Overloading in the sense of \mycalc\ and
$\lambda\&$\ relies on type-dependent reduction, and thereby the
selection of the role is crucial for the computation. This facet goes
beyond the common interpretation of intersection types.

\subsection{Asymmetric type-dependent choice}
\label{SS:sugar}

So far, the only way to turn a many-sorted strategy $s$ into a generic
one is based on the form \extend{s}{\pi}. This kind of casting implies
that the lifted strategy will fail at least for all term types
different from the domain of $s$. This is often not desirable, and
hence, an extension usually entails a complementary choice. In the
present section, we want to argue that the separation of lifting (by
\extend{\cdot}{\cdot}) and completion by
\choice{\cdot}{\cdot}\ and friends is problematic. It is however 
possible to support a different style of inhabitation of generic
types. We will define a corresponding combinator for \emph{asymmetric
type-dependent choice}. While the combinator \plus{\cdot}{\cdot}\ for
\emph{symmetric} type-dependent choice from Section~\ref{SS:overloading}
was linked to the notion of overloading, the upcoming asymmetric form
does not rely on overloading.  In fact, the corresponding left-biased
and right-biased forms \lplus{\cdot}{\cdot}\ and \rplus{\cdot}{\cdot}
can be regarded as syntactic sugar defined in terms of strategy
extension.  Asymmetric type-dependent choice means to apply the less
generic strategy if this is type-safe, and to resort to a more
generic strategy otherwise. If we do not consider overloading, then
this form of choice favours the many-sorted operand if this is
type-safe, and it resorts to the generic default otherwise.

\begin{example}
To motivate the idea of asymmetric type-dependent choice, let us
reconsider the traversal scheme \s{StopTD} that was defined
earlier. We repeat its definition for convenience:
\begin{eqnarray*}
\s{StopTD} & : & \alltp \to \alltp\\
\s{StopTD}(\nu) & = & \lchoice{\nu}{\all{\s{StopTD}(\nu)}}
\end{eqnarray*}
Left-biased choice controlled by success and failure is used here to
first try the generic argument $s$ of $\s{StopTD}(s)$ but to descend
into the children if $s$ fails. Let us assume that $s$ was obtained
from a many-sorted strategy $s'$ by strategy extension as in
\extend{s'}{\alltp}. It is important to note that $s$ could fail for
\emph{two} reasons. Firstly, $s$ is faced with a term of a sort
different from the domain of $s'$. Secondly, $s'$ is applicable as for
the typing, but $s'$ is defined in a way to refuse the given term,
e.g., because of unsatisfied preconditions. These two sources of
failure are not separated in the definition of \s{StopTD}. In the
present formulation, \s{StopTD}\ will always recover from failure of
$s$ and descend into the children. In fact, \s{StopTD}\ will always
succeed because \all{\cdot} at least succeeds for leafs.
\end{example}

\paragraph*{Syntactic sugar}

We conclude from the above example that type-mismatch and other
sources of failure are hard to separate in a programming style based
on \extend{\cdot}{\cdot}. We improve the situation as follows. We
introduce asymmetric type-dependent choice. In the left-biased
notation \lplus{s_1}{s_2}, the left operand $s_1$ is regarded as an
update for the default $s_2$. Hence, we call this form left-biased
type-dependent choice. The many-sorted strategy $s_1$ should be
applied if the type of the term at hand fits, and we resort to the
generic default $s_2$ otherwise. For brevity, we do not take
overloaded strategies into account. One essential ingredient of the
definition of asymmetric type-dependent choice is a type guard, that
is, a generic strategy which is supposed to accept terms of a certain
sort and to refuse all other terms. A type guard is constructed from a
many-sorted restriction of \id\ which is then lifted to the generic
type of choice. Here is the syntactic sugar for type guards and
asymmetric type-dependent choice:
\begin{eqnarray*}
\extend{\tau}{\gamma} & \equiv & \extend{(\restrict{\id}{\tau\to\tau})}{\gamma}\\
\lplus{s_1}{s_2} & \equiv & 
\choice{\extend{s_1}{\pi}}{(\seq{\negbf{(\extend{\tau}{\alltp})}}{s_2})}
\ \mbox{where}\ s_1: \tau \to \tau', s_2: \pi\\
\rplus{s_1}{s_2} & \equiv &
\lplus{s_2}{s_1}
\end{eqnarray*}%
A fully formal definition of this syntactic sugar could be given via
the elaboration judgement discussed earlier but we omit this
definition for brevity. The definition of \lplus{s_1}{s_2}\ employs a
negated type guard \negbf{(\extend{\tau}{\alltp})} to block the
application of the generic default $s_2$ in case $s_1$ is applicable
as for typing.

\begin{example}
\label{X:StopTD'}
Let us define a variant of \s{StopTD} which interprets failure of the
argument strategy as global failure. This can be used for some form of
``design by contract''. If the argument strategy ever detects that
some precondition is not met, the corresponding failure will be
properly propagated as opposed to accidental descent.
\begin{eqnarray*}
\s{StopTD}' & : & \forall \alpha.\ (\alpha \to \alpha) \to \alltp\\
\s{StopTD}'[\alpha](\nu) & = & \lplus{\nu}{\all{\s{StopTD}'[\alpha](\nu)}}
\end{eqnarray*}
$\s{StopTD}'$ is different from \s{StopTD}\ in that the argument of
\s{StopTD} is a generic strategy whereas it is many-sorted in the case
of $\s{StopTD}'$. To this end, the type of $\s{StopTD}'$ involves a
type parameter for the sort of the argument. The asymmetric
type-dependent choice to derive a generic strategy from the argument
is part of the definition of $\s{StopTD}'$.
\end{example}

\begin{example}
We should mention that type guards are useful on their own. Recall the
illustrative traversal problem \s{(IV)} to collect all natural numbers
in a tree. The encoding from Example~\ref{X:typeful} relies on a
user-defined strategy \s{Nat} to test for naturals based on the
congruences for the constructors of sort \w{Nat}. The syntactic sugar
for type guards allows us to test for arbitrary sorts without the
cumbersome style of enumerating all constructors. This is illustrated
in the following definition of \s{(IV)} where we use the notation for a
type guard for naturals instead of relying on the user-defined strategy
\s{Nat}:
\[\begin{array}{rcl}
\s{(IV)}  & = & \s{StopCrush}[\w{NatList}](\seq{\extend{\w{Nat}}{\alltu{\w{Nat}}}}{\s{Singleton}},\s{Nil},\s{Append})\\
\end{array}\]
\end{example}

\paragraph*{Complementary forms of choice}

It is instructive to compare the different forms of asymmetric choice
encountered in the present paper. In the case of
\lchoice{s_1}{s_2}, the success of $s_1$ rules out the application of
$s_2$. In the case of \lplus{s_1}{s_2}, the mere type of $s_1$ decides
if the application of $s_2$ will be ever considered. To understand this
twist, consider the following strategy approximating \lplus{s_1}{s_2}:
\[\lchoice{\extend{s_1}{\pi_2}}{s_2}
\ \mbox{where $s_2$ is of type $\pi_2$}\]
This formulation attempts to compensate for the type guard in the
definition of asymmetric type-dependent choice by resorting to
left-biased choice controlled by success and failure. That is, we
attempt to simulate left-biased type-dependent choice by left-biased
choice controlled by success and failure. This attempt is not faithful
since $s_2$ might be applied to a term $t$ even if the types of $s_1$
and $t$ fit, namely if $s_1$ fails on $t$.

To summarise, choice between strategies of the \emph{same type} is
solely modelled by the combinators \choice{\cdot}{\cdot}\ and friends
that are controlled by success and failure. Non-deterministic and
asymmetric choice differ in the sense if there is a preferred order on
the arguments of the choice. For convenience, we might accept
different types for the argument strategies of \choice{\cdot}{\cdot}\
and friends. But then we restrict the type of the choice to the
greatest lower bound of the types of the arguments. By contrast,
type-dependent choice composes strategies of \emph{different} types,
and the type of the choice extends to the least upper bound of the
types of the arguments. The corresponding combinators are not at all
controlled by success and failure. Instead, the type of the term at
hand determines the branch to be taken. The arguments in an asymmetric
type-dependent choice are related via \typeless, whereas the domains
of the arguments in a symmetric type-dependent choice are required to
be disjoint. In conclusion, choice by success and failure and
type-dependent choice complement each other. The division of labour
between the two kinds of choice was also nicely illustrated in
Example~\ref{X:overloading}.

\subsection{Variations on traversal}
\label{SS:vary}

The selection of the traversal primitives of \mycalc\ has been driven
by the requirement not to employ any universal representation type.
For that reason the children are never directly exposed to the
strategic program. Instead, one has to select the appropriate
combinator to process the children. We want to indicate briefly that
there is a potential for generalised or additional traversal
primitives while keeping in mind the aforementioned requirement.

\paragraph*{Order of processing children}

The reduction semantics of the traversal primitives left the order of
processing children largely unspecified. As for \all{\cdot}, the order
does not seem to be an issue since all the children are processed
anyway and independently of each other. Note however that the order
becomes an issue if we anticipate the possibility that processing
fails for one child or several children. Then, different orders will
not just lead to different execution times, but even program
termination might depend on the order. As for \one{}{\cdot}, a
flexible order is desirable for yet another reason. That is, one might
favour the left-most vs.\ the right-most child that can be
processed. The actual choice might be a correctness issue as opposed
to a mere efficiency issue. To cope with such variations, one can
consider refined traversal combinators such as \one{o}{s}\ where we
assume that the order of processing children is constrained by
$o$. There are the following options for such an order constraint $o$:
\begin{itemize}
\item ``\ltr''\ ---\ processing from left to right
\item ``\rtl''\ ---\ processing from right to left
\item unspecified
\end{itemize}
As for the type-unifying traversal combinators, order constraints make
sense as well.  In \reduce{o}{\splus}{s}, the constraint $o$ could be
used to control how the pairwise composition $\splus$ is applied to
the processed children.  A simple investigation of the original
formalisation of \reduce{}{\splus}{s}\ in Figure~\ref{F:apply-alltu}\
makes clear that a left-to-right reduction was specified (although it
was not constrained if pairwise composition is intertwined with
processing the children). A certain order $o$ for reduction might be
relevant to cope with combinators $\splus$ which do not admit
associativity and/or commutativity. As for selection via
\select{o}{s}, basically the same arguments apply as to \one{o}{s}.

\begin{figure}
{\smallsize

\dech{Syntax}
\begin{eqnarray*}
s & ::= & \cdots\ |\ \fold{s}{s}
\end{eqnarray*}

\decj{Reduction of strategy applications}{%
\ioj{\apply{s}{t}}{t'}
}

\parbox[t]{.47\textwidth}{

\dech{Positive rules}

\medskip

\ir{fold^+.1}{%
\ioj{\apply{\snil}{\langle\rangle}}{t'_0}
}{%
\ioj{\apply{\fold{\snil}{\scons}}{c}}{t'_0}
}

\ir{fold^+.2}{%
\sps{
\ioj{\apply{\snil}{\langle\rangle}}{t'_n}
\np
\ioj{\apply{\scons}{\langle{}t_n,t'_n\rangle}}{t'_{n-1}}
\np
\cdots
\np
\ioj{\apply{\scons}{\langle{}t_1,t'_1\rangle}}{t'_0}
}
}{%
\ioj{\apply{\fold{\snil}{\scons}}{f(t_1, \ldots, t_n)}}{t'_0}
}

}\hfill\parbox[t]{.47\textwidth}{

\dech{Negative rules}

\medskip

\ir{fold^-.1}{%
\ioj{\apply{\snil}{\langle\rangle}}{\failure}
}{%
\ioj{\apply{\fold{\snil}{\scons}}{c}}{\failure}
}

\ir{fold^-.2}{%
\ioj{\apply{\snil}{\langle\rangle}}{\failure}
}{%
\ioj{\apply{\fold{\snil}{\scons}}{f(t_1, \ldots, t_n)}}{\failure}
}

\ir{fold^-.3}{%
\sps{
\exists i \in \{1,\ldots,n\}.\\
&
\ioj{\apply{\snil}{\langle\rangle}}{t'_n}
\np
\ioj{\apply{\scons}{\langle{}t_n,t'_n\rangle}}{t'_{n-1}}
\np
\cdots
\np
\ioj{\apply{\scons}{\langle{}t_i,t'_i\rangle}}{\failure}
}
}{%
\ioj{\apply{\fold{\snil}{\scons}}{f(t_1, \ldots, t_n)}}{\failure}
}

}

}
\caption{An intentionally type-unifying traversal combinator for 
folding the children}
\label{F:apply-fold}
\bigskip
\end{figure}

\paragraph*{Pairwise composition vs.\ folding}

It turns out that reduction as modelled by the combinator
\reduce{}{\cdot}{\cdot}\ can be generalised. Instead of separating
the aspects of processing the children and composing intermediate
results, we can also define reduction so that the way how a child is
processed depends on previously processed children. In fact, one can
define a combinator \fold{\cdot}{\cdot}\ which folds directly over the
children of a term very much in the sense of the folklore pattern for
folding a list. Note however that we have to cope with an
intentionally heterogeneous list corresponding to the children of a
given term. That is, in folding over the children of a term, we need a
\emph{generic} ingredient to operate on a given child and the
intermediate result obtained from previous folding steps. The
reduction semantics of the strategy \fold{\snil}{\scons}\ is defined
in Figure~\ref{F:apply-fold}. The first argument $\snil$ encodes the
initial value for folding. In the case of a constant symbol, $\snil$
defines the result of folding (cf.\
\tg{fold^+.1}). For nontrivial terms, we fold over their children by
repeated application of the second argument $\scons$ (cf.\
\tg{fold^+.2}). Without loss of generality, \fold{\cdot}{\cdot}\ is
a right-associative fold.

\begin{example}
Let us attempt a reconstruction of the strategy \s{CF}\ from
Figure~\ref{F:defined1TU}. For convenience, we first show the original
definition in terms of \reduce{}{\cdot}{\cdot}. Then, we show a
reconstruction which employs the combinator \fold{\cdot}{\cdot}.
\begin{eqnarray*}
\s{CF}(\nu,\vunit,\vplus) & = &
\choice{(\seq{\s{Con}}{\seq{\void}{\vunit}})}{(\seq{\s{Fun}}{\reduce{}{\vplus}{\nu}})}\\
& = & \fold{\vunit}{\seq{\pair{\nu}{\id}}{\vplus}}
\end{eqnarray*}
This reconstruction immediately illustrates why the combinator
\fold{\cdot}{\cdot}\ is more powerful than the combinator
\reduce{}{\cdot}{\cdot}. As the second argument in the above 
application of \fold{\cdot}{\cdot}\ points out, a child is processed
independent of the intermediate value of reduction (cf.\ the
congruence \pair{\nu}{\id}), and both values are composed in a subsequent
step by \vplus. This is precisely the scheme underlying
\reduce{}{\cdot}{\cdot}.
\end{example}

We cannot type the combinator \fold{\cdot}{\cdot}\ in a simple way in
our present type system. Consider the intended type of the second
argument. The strategy should process a pair consisting of a term of
any type (corresponding to some child), and a term of the
distinguished type for type unification. This amounts to the type
scheme $\forall \alpha.\ \langle\alpha,\tau\rangle \to \tau$ where
$\tau$ is the unified type for reduction. One could introduce a
designated generic type for that purpose. Unfortunately, more
extensions would be needed to effectively use the additional
generality. It is not obvious how to stay in a many-sorted
setting in this case. Due to these complications we do not attempt to
work out typing rules for \fold{\cdot}{\cdot}.

\paragraph*{Environments and states}

There are other useful type schemes than just \alltp\ and \alltu{\cdot}.
In the following table, we repeat the definition of \alltp\ and
\alltu{\cdot}, and we list three further schemes:
\[\begin{array}{lclr}
\alltp        & \equiv & \forall \alpha.\ \alpha \to \alpha
&\mbox{(Type preservation)}
\\
\alltu{\tau}  & \equiv & \forall \alpha.\ \alpha \to \tau
&\mbox{(Type unification)}
\\
\allta{\tau}  & \equiv & \forall \alpha.\ \pair{\alpha}{\tau} \to \tau
&\mbox{(Accumulation)}
\\
\allte{\tau}  & \equiv & \forall \alpha.\ \pair{\alpha}{\tau} \to \alpha
&\mbox{(\alltp\ with environment passing)}
\\
\allts{\tau}  & \equiv & \forall \alpha. \pair{\alpha}{\tau} \to \pair{\alpha}{\tau}
&\mbox{(\alltp\ with state passing)}
\end{array}\]
A strategy of type \allta{\tau} takes a pair \pair{x}{a} where $x$ can
be of any term type and $a$ is of type $\tau$, and it returns the
resulting value $a'$ of type $\tau$. When thinking of traversal,
\allta{\tau}\ suggest \emph{accumulation} of a value whereas the earlier
\alltu{\tau}\ rather suggests \emph{synthesis} of a value. Both schemes of
traversal are interchangeable, in principle. Then, the type scheme
\allte{\tau}\ denotes all strategies that take a pair \pair{x}{e}\ where
$x$ can be of any term type and $e$ is of type $\tau$, and it returns
a resulting term $x'$. When thinking of traversal, \allte{\tau}\
amounts to a combination of type-preserving traversal and
\emph{environment passing}. Finally, \allts{\tau} can be regarded as a
combination of \alltp\ and \allta{\tau}. In this combination, it is
suggestive to speak of \emph{state passing}.

\begin{figure}
{\smallsize

\decj{Reduction of strategy applications}{%
\ioj{\apply{s}{t}}{t'}
}

\bigskip

\alltp

\ir{one^+}{%
\exists i \in \{1,\ldots,n\}.\ \ioj{\apply{s}{t_i}}{t'_i}
}{%
\ioj{\apply{\one{}{s}}{f(t_1, \ldots, t_n)}}{f(t_1, \ldots, t'_i, \ldots, t_n)}
}

\bigskip

\allte{\cdot}

\ir{one^+}{%
\exists i \in \{1,\ldots,n\}.\ \ioj{\apply{s}{\pair{t_i}{e}}}{t'_i}
}{%
\ioj{\apply{\one{}{s}}{\pair{f(t_1, \ldots, t_n)}{e}}}{f(t_1, \ldots, t'_i, \ldots, t_n)}
}

\bigskip

\allts{\cdot}

\ir{one^+}{%
\exists i \in \{1,\ldots,n\}.\ \ioj{\apply{s}{\pair{t_i}{a}}}{\pair{t'_i}{a'}}
}{%
\ioj{\apply{\one{}{s}}{\pair{f(t_1, \ldots, t_n)}{a}}}{\pair{f(t_1, \ldots, t'_i, \ldots, t_n)}{a'}}
}

}
\caption{Variants of \one{}{\cdot}}
\label{F:one}
\bigskip
\end{figure}

\paragraph*{Designated combinators vs.\ monads}

The ultimate question is how to inhabit the above type
schemes. \mycalc\ is not sufficiently expressive to derive traversal
combinators for the additional generic types from the existing
combinators that cover \alltp\ and \alltu{\cdot}. However, it is not
difficult to define corresponding variations on the existing traversal
primitives. Let us illustrate this idea for the generic type
\allte{\cdot}.  Dedicated traversal combinators should not simply
apply a given strategy to the children, but an environment has to be
pushed through the term, too. Let us characterise a corresponding
variation on \one{}{\cdot}. If \one{}{s} is applied to
\pair{f(t_1,\ldots,t_n)}{e}, then the strategy application to rewrite
a child $t_i$ is of the form \apply{s}{\pair{t_i}{e}}. In
Figure~\ref{F:one}, the positive rules for variants of \one{}{\cdot}\
for \alltp, \allte{\cdot}\ and \allts{\cdot}\ are shown. All other
traversal primitives admit similar variations. In a higher-order
functional programming context, monads~\cite{Spivey90,Wadler92} can be
employed to merge effects like environment or state passing with the
basic scheme of type-preserving or type-unifying traversal. Monads
would also immediately allow us to deal with reducts other than
optional terms, namely lists or sets of terms. In the reduction
semantics of \mycalc, we hardwired the choice of an optional term as
reduct.  This choice corresponds to the maybe monad.


\section{Implementation}
\label{S:impl}

In the sequel, we discuss a Prolog-based implementation of \mycalc,
and we report on an investigation regarding the integration of the
\mycalc\ expressiveness into the rewriting framework ELAN. The Prolog
implementation is convenient to verify our ideas and the
formalisation, but also to prove the simplicity of the approach. We
have chosen Prolog due to its suitability for prototyping language
syntax, typing rules, and dynamic semantics (cf.~\cite{LR01}). The
ELAN-centered investigation backs up our claim that the proposed form
of generic programming can be easily integrated into an existing,
basically first-order, many-sorted rewriting framework.

\begin{figure}
{\smallsize

\parbox[t]{.4\textwidth}{
Library strategies

\bigskip

{\minisize
\begin{verbatim}
try : tp -> tp.
try(S) = S <+ id.

repeat : tp -> tp.
repeat(S) = try(S; repeat(S)).

oncebu : tp -> tp.
oncebu(S) = one(oncebu(S)) <+ S.

stoptd : tp -> tp.
stoptd(S) = S <+ all(stoptd(S)).

\end{verbatim}
}
}\hfill\parbox[t]{.54\textwidth}{
Traversals \s{(I)}\ and \s{(II)}

\bigskip

{\minisize
\begin{verbatim}
data nat   = zero | succ(nat).
data gsort = c
           | g(gsort)
           | gprime(gsort).
data other = b(other,gsort)
           | f(other,nat)
           | h(nat,gsort).

inc : nat -> nat.
inc = N -> succ(N).

traverseI : tp.
traverseI = stoptd(inc < tp).

traverseII : tp.
traverseII = oncebu((g(X) -> gprime(X)) < tp).
\end{verbatim}
}
}
}
\caption{Strategic programs in Prolog}
\label{F:sample.pl}
\bigskip
\end{figure}

\subsection{A Prolog prototype}
It is well-known that deduction rules in the style of Natural
semantics map nicely to Prolog clauses
(cf.~\cite{Despeyroux88}). Prolog's unification and backtracking
enable the straight execution of a large class of deduction
systems. In fact, the Natural semantics definitions from the present
paper are immediately implementable in this manner. The judgements
were mapped to Prolog in the following manner. Well-formedness,
well-typedness, static elaboration and reduction judgements constitute
corresponding predicate definitions. Terms are represented as ground
and basically untyped Prolog terms. Strategic programs are
represented as files of period-terminated Prolog terms encoding type
declarations and strategy definitions. In this manner, Prolog I/O can
be used instead of parsing. Prolog variables are used to encode term
variables in rewrite rules, strategy variables in strategy
definitions, and term-type variables in type declarations.

\paragraph*{Strategies in Prolog}

The encoding of strategies is illustrated in Figure~\ref{F:sample.pl}.
On the left side, the Prolog encoding for some reusable strategies
from Figure~\ref{F:defined0} and Figure~\ref{F:defined1TP} are
shown. On the right side, the strategies for the introductory
traversal problems \s{(I)}\ and \s{(II)}\ from the introduction are
shown. The rewrite rule to increment a natural is for example
represented as \verb|N -> succ(N)|. One can see that the encoding
basically deals with notational conventions of Prolog such as the
period ``.'' to terminate a term to be read from a file. The term
\texttt{tp} denotes the type \alltp. The \texttt{data}\ directive is
used to declare algebraic datatypes contributing to the context
$\Gamma$ of a strategic program. We do not declare types of term
variables since it is very easy to infer their types using the
non-ground representation for rewrite rules.

\paragraph*{Prolog encodings of the judgements}

The implementation of \mycalc\ is illustrated with a few excerpts in
Figure~\ref{F:mycalc.pl}. We show some clauses for the predicates
encoding the reduction of strategy applications and static elaboration
of strategies.  The left-most excerpt shows the very simple
implementation of the combinator \all{\cdot}\ in Prolog. Here we
resort to the Prolog operator ``\verb!=..!'' to access the children as
a list, and we employ a higher-order predicate \verb!map/3! to map the
argument strategy over the children. In the middle, we show the
encoding of the static elaboration rule from
Figure~\ref{F:extend'}. The right-most excerpt implements the
reduction semantics of an annotated application of
\extend{\cdot}{\cdot}. It deviates from the formalisation in
Figure~\ref{F:extend'} in that we do not assume tagged terms but we
rather look up the type of the given term by retrieving the outermost
symbol's result type from a simple context parameter.

\begin{figure}
{\smallsize

\parbox[t]{.3\textwidth}{%
Reduction of \all{\cdot}

\bigskip

{\minisize
\begin{verbatim}
apply(G,all(S),T0,T1)
 :-
    T0 =.. [F|L0],
    map(apply(G,S),L0,L1),
    T1 =.. [F|L1].
\end{verbatim}
}
}\hfill\parbox[t]{.29\textwidth}{%
Elaboration of \extend{\cdot}{\cdot}

\bigskip

{\minisize
\begin{verbatim}
elaborate(G,S<T,(S:Pi)<T)
 :- 
    wtStrategy(G,S,Pi).


\end{verbatim}
}
}\hfill\parbox[t]{.33\textwidth}{%
Reduction of \extend{\cdot}{\cdot}

\bigskip

{\minisize
\begin{verbatim}
apply(G,(S:Tau->_)<_,T0,T1)
 :-
    T0 =.. [F|_],
    fInGamma(F,G,Tau,_),
    apply(G,S,T0,T1).
\end{verbatim}
}
}
}
\caption{Implementation of \mycalc\ in Prolog}
\label{F:mycalc.pl}
\bigskip
\end{figure}

\paragraph*{Prological strategies}

The proposed implementational model is geared towards a direct
implementation of the calculus' formalisation in Prolog, that is,
judgements become predicates. Strategic programming can also be
integrated into Prolog in a more seamless way from the logic
programmer's point of view. Essentially, strategy combinators can be
represented as higher-order predicates. Prolog programmers are used to
this idea which is for example used for list processing. Furthermore,
we abandon rewrite rules altogether, and we assume that many-sorted
functionality is defined in terms of ordinary Prolog predicates. This
approach is not just convenient for logic programmers, but it also
leads to a very compact implementation of strategic programming
expressiveness. In such a Prolog incarnation of strategic programming,
the most complicated issue is typing. In general, all attempts to
impose type systems on Prolog restrict Prolog's expressiveness to a
considerable extent. We cannot expect that all the implementations of
the strategy combinators themselves can be typed-checked. In
particular, the use of the univ operator ``=..'' for generic term
destruction and construction is hardly typeable. Recall that ``=..''
would be needed for the implementation of traversal combinators.
Hence, we need an approach where type-checking is optional, that is,
it can be switched off maybe per Prolog module. We refer
to~\cite{LR01} for a discussion of the Prological incarnation of
strategic programming.

\subsection{Integration into ELAN}
\label{SS:ELAN}

The rewriting framework ELAN supports many-sorted rewriting
strategies.  However, generic traversal combinators are not
offered. ELAN's type system is indeed a many-sorted one. ELAN's module
system offers parameterisation of modules by sorts. One can import the
same parameterised module for different sorts. This leads to a style
of programming where function symbols and strategy combinators are
potentially overloaded. In the sequel, we explain how combinators for
generic traversal and strategy extension can be made available in ELAN
based on the \mycalc\ model of typed strategies.  The simplicity of
the integration model indeed further backs up our claim that \mycalc\
is straightforward to implement. We should point out that there are
ongoing efforts to revise the specification formalism and the system
architecture underlying ELAN.  We base our explanations on ELAN as
of~\cite{BKKMR98,BKKR01}.

\paragraph*{The module \texttt{strat}[X]}

Let us recall some characteristics of many-sorted strategies as
supported by ELAN.  There is a designated library module
\texttt{strat[X]} for strategy combinators parameterised by a sort
\texttt{X}. In fact, certain ELAN strategy combinators are built-in,
but for the sake of a homogeneous situation we assume that all
combinators are provided by the module \texttt{strat[X]}. ELAN offers
a notation for strategy application which can be used in the
\texttt{where}-clauses of a rewrite rule and in the user interface. If
strategies should be composed and applied to terms of a certain sort,
one needs to import the module \texttt{strat[X]} where the formal
parameter \texttt{X} is instantiated by the given sort. By importing
this module for several sorts, the strategy combinators are overloaded
for all the sorts accordingly. This approach implies that parsing
immediately serves for type checking. ELAN also allows one to
\emph{define} new many-sorted strategy combinators. One can also
define combinators for the sort parameter of a module so that the
definitions are reusable for different sorts. As an aside, ELAN's
parameterised modules can be used as a substitute for
type-parameterised strategies in the sense of \mycalc.

\paragraph*{The module \texttt{any}[X]}

In addition to parameterised modules, ELAN offers further means to
define generic functionality, that is, functionality dealing with
terms of arbitrary sorts. We review these techniques to see whether
they are suitable for the implementation of the \mycalc\ combinators
for generic traversal and strategy extension. There is a designated
library module \texttt{any[X]} which supports a form of dynamic typing
and generic term destruction / construction per sort \texttt{X}. The
module uses a universal datatype \texttt{any} in the sense of dynamic
typing, The datatype \texttt{X} and \texttt{any} are mediated via an
injection function defined by the module. Further, the module hosts
\emph{explode} and \emph{implode} functions to destruct and construct
terms of sort \texttt{any}. The children of a term are made accessible
as a list of terms of sort \texttt{any}. Internally, ELAN uses a
pre-processor to generate the rewrite rules for explosion and
implosion.

\paragraph*{Naive encoding of \mycalc}

For brevity, we restrict ourselves to type-preserving strategies in
the sequel. \mycalc\ strategies of type \alltp\ can be encoded as ELAN
strategies of type $\texttt{any} \to \texttt{any}$. One can define
traversal combinators in terms of implosion and explosion based on the
functionality of the module \texttt{any[X]}. The combinator
\all{\cdot}, for example, would be defined in roughly the same manner
as in the above Prolog encoding. First, the given term of sort
\texttt{any} is exploded to access the functor and the children. Then,
the argument strategy is mapped over the children via a dedicated
strategy for list processing. Finally, the original functor and the
processed children are imploded. The combinator for strategy extension
can be encoded in ELAN as follows. Given a strategy $s$ of type
$\texttt{X} \to \texttt{X}$, strategy extension derives a strategy of
type $\texttt{any} \to \texttt{any}$. The application of the extended
$s$ entails the attempt to take away the injection of type $\texttt{X}
\to \texttt{any}$ from the term at hand. If the given term is not of
sort \texttt{X}, the application of the extended strategy fails in
accordance with type safety. The combinator for strategy extension is
overloaded for all possible X, that is, it needs to be placed in a
module parameterised by \texttt{X}. If the strategic programmer wants
to apply a ``generic'' strategy, (s)he has to inject the given term into
\texttt{any} prior to application, and to unwrap the injection from the
result.

\paragraph*{Fully typed encoding of \mycalc}

The above encoding suffers from the following problem. The sort
\texttt{any} is exposed to the strategic programmer in the sense
that generic strategies are known to operate on terms of sort
\texttt{any}. Hence, there is no guarantee that generic strategies
are well-typed in a many-sorted sense. To give an example, an
intentionally type-preserving strategy can map a term of sort 
\texttt{X} to a term of sort \texttt{Y} while this type change
would go unnoticed as long as terms are represented inside the union
type \texttt{any}. Furthermore, the exposition of \texttt{any} allows
a strategic programmer to manipulate compound terms in an inconsistent
manner. Note that explosion and implosion involves lists of
terms. That is, the ELAN type system does not ensure that the
manipulated exploded terms form valid terms in the many-sorted
sense. This implies a potential for implosion failure at run-time.  A
fully typed encoding requires the following elaboration of the naive
approach. In abstract terms, we need to hide the employment of
\texttt{any} for strategic programmers who want to apply generic
strategies, inhabit generic strategy types via strategy extension, or
define new combinators in terms of the basic combinators. Then, a
strategic programmer cannot define ill-typed generic strategies,
neither can (s)he cause implosion failures provided all the
\emph{basic} combinators are implemented in accordance with the
\mycalc\ reduction judgement that is known to be type-safe.  In order
to hide the employment of \texttt{any}, we assume the introduction of
designated sorts for generic strategy types where these sorts are
known to the strategic programmer but not their definition. To give an
example, we assume a sort \texttt{tp} for the \mycalc\ type \alltp\
with the hidden definition $\texttt{any} \to \texttt{any}$ in
ELAN. All the combinators for a generic strategy type are defined in a
module together with the designated sort. Since strategy application
and strategy extension work per sort, we need a parameterised module,
e.g., \texttt{tp[X]} for generic type-preserving strategies which can
be applied to terms of sort \texttt{X}, and which can be derived by
extending many-sorted strategies of type $\texttt{X} \to
\texttt{X}$. Clearly, \texttt{tp[X]} can be regarded as an abstract
datatype (ADT) for generic type-preserving strategies.

To summarise, the described integration model relies on the following
concepts:
\begin{itemize}
\item parameterised modules to overload strategy combinators per sort,
\item type-checking by parsing overloaded many-sorted strategies,
\item dynamic typing to achieve the needed degree of polymorphism, and
\item support for generic term destruction / construction.
\end{itemize}
Because these features are present in ELAN, the support of
\mycalc-like strategies does not require any internal modification of
ELAN. Instead of relying on features like a pre-processor for term
implosion and explosion, we could favour an extension of the rewrite
engine to directly support traversal combinators, and strategy
extension as well. This approach would be, in general, appropriate to
implement \mycalc-like strategies in other frameworks for rewriting or
algebraic specification, e.g., in ASF+SDF~\cite{BHK89,Klint93,B+01}.


\section{Related work}
\label{S:related}

Specific pointers to related work were placed in the technical
sections.  It remains to comment on related work from a more general
point of view.  First, we relate \mycalc\ to existing strategic
rewriting calculi. Then, we discuss other efforts in the rewriting
community to enable some form of generic programming. Finally, we
discuss genericity in functional programming because this paradigm is
very much related to rewriting.

\subsection{Strategic rewriting calculi}

Let us relate the calculus \mycalc\ to those frameworks for strategic
programming which were most influential for its design, namely system
$S$ underlying Stratego~\cite{VB98,VBT98}, and ELAN~\cite{BKKMR98,BKKR01}.

\paragraph*{\mycalc\ vs.\ system $S$ and Stratego}

Our typed rewriting calculus \mycalc\ adopts the untyped system $S$ to
a large extent. We stick to the same semantic model. We also adopt its
traversal combinators \all{\cdot}\ and \one{}{\cdot}. System $S$
suggests a hybrid traversal combinator $\someS(\cdot)$ where the
application of the argument strategy has to succeed for at least one
child but the application is attempted for all children. We leave out
$\someS{\cdot}$\ in \mycalc\ in order to minimise the operator suite
which needs to be covered by the formalisation. The main limitation of
\mycalc\ compared to system $S$ is that we favour standard first-order
rewrite rules with where-clauses as primitive form of strategy. By
contrast, system $S$ provides less standard primitives which are
however sufficient to model rewrite rules as syntactic sugar. These
primitives are matching to bind variables, building terms relying on
previous bindings, and scoping of variables. The additional
flexibility which one gains by this separation is that arbitrary
strategies can be performed between matching and building. One can
simulate this style by using where-clauses in \mycalc. The key
innovation of \mycalc\ when compared to system $S$ is the combinator
\extend{\cdot}{\cdot}\ for strategy extension.  Since the combinator
\extend{\cdot}{\cdot}\ relies on a type-dependent reduction semantics,
one cannot even expect any combinator like this in untyped systems
such as Stratego or system $S$. Furthermore, \mycalc\ also introduces
combinators which are not expressible in system $S$, namely the
combinators \reduce{}{\cdot}{\cdot}\ and \select{}{\cdot}\ for
intentionally type-unifying traversal. Stratego provides a combinator
which can be used to encode type-unifying traversal, namely
$\cdot\#\cdot$. This combinator is meant for generic destruction and
construction of terms very much in the style of the standard univ
operator ``=..'' in Prolog. Interestingly, the combinators
\all{\cdot}\ and \one{}{\cdot} could be encoded in terms of
$\cdot\#\cdot$.  A crucial problem with $\cdot\#\cdot$ is that it
leads to a hopelessly untyped model of traversal since the programmer
accesses the children of a term as a list. While the system $S$
combinators \all{\cdot}\ and \one{}{\cdot} suggest a typeful
treatment, typeful type-unifying strategies cannot be based on
$\cdot\#\cdot$ but other combinators are needed.  This is the reason
that we designed the traversal combinators
\reduce{}{\cdot}{\cdot}\ and \select{}{\cdot}\ for type-unifying
traversal in \mycalc.

\paragraph*{\mycalc\ vs.\ ELAN}

The influence of ELAN is also traceable in \mycalc. We adopt the model
of rewrite rules with where-clauses from ELAN. We also adopt recursive
strategy definitions from ELAN while system $S$ favours a special
recursion operator \rec{\cdot}{\cdot}. In the initial design of a
basically many-sorted type system we also received inspiration from
the ELAN specification language. ELAN and \mycalc\ differ in the
semantic model assumed for reduction. ELAN offers a faithful model of
non-determinism via sets or lists of possible results where the empty
set represents failure. The type system of \mycalc\ does not rely on
the simple model of system $S$.  In fact, our typeful approach to
generic traversal could be integrated with the ELAN-like semantic
model without changing any detail in the type system.

\subsection{Genericity in rewriting}

In \cite{CFM01}, polytypic entities are defined in terms of the
reflection and meta-programming capabilities of Maude. This approach
is hard to compare with \mycalc\ which is based on the idea of static
typing and a designated type system. Furthermore, the mixture of
many-sorted and generic functionality is not considered. Also, the
Maude approach---as any other polytypic approach---does not propose
traversal combinators but traversal is based on polytypic induction.

In~\cite{BKV01}, a fixed set of traversal strategies is supported by
so-called traversal functions extending the algebraic specification
formalism ASF~\cite{BHK89}. The central idea is to declare designated
function symbols for traversal according to predefined strategies for
top-down, bottom-up and accumulating traversal. The programmer refines
a traversal function by rewrite rules for specific sorts. This
approach is less general than the \mycalc\ approach because the
programmer cannot define new traversal schemes. Also, it is more
difficult to separate many-sorted and generic functionality. However,
this approach is sufficient for many common scenarios in program
transformation and analysis~\cite{BKV01,LW01}. In fact, traversal
functions are very convenient to use because of their seamless
integration into ASF+SDF~\cite{BHK89,Klint93,B+01}.

In~\cite{BKKR01}, dynamic typing~\cite{ACPP91,ACPR92} and generic
implosion / explosion \`a la Prolog's ``=..'' are used to traverse
terms. Dynamics tend to spread all over a program which clearly goes
against a many-sorted typing discipline. Also, the use of explosion
and implosion in a program implies a basically untyped manipulation of
exploded terms. Furthermore, basic traversal combinators are not
identified. Their benefits were first identified in the early work on
Stratego~\cite{LV97}. We already explained in Section~\ref{SS:ELAN}
how ELAN's features can be used to implement the \mycalc\ combinators
in a typeful manner. The formalisation of \mycalc\ avoids all kinds of
typing problems in the first place because terms are not converted to
a universal type.

We envision that the design of \mycalc\ could be useful in the further
elaboration of other formal models for rewriting so that typed generic
traversal will be covered. One prime candidate is the
$\rho$-calculus~\cite{CK99} which provides an abstract and very
general formal model for rewriting including strategies.  Generic
traversal combinators have been defined in the $\rho$-calculus (cf.\
$\Psi(s)$ and $\Phi(s)$ in~\cite{CK99} corresponding to
\all{s}\ and \one{}{s}) but these definitions cannot be typed in the
available typed fragments of the $\rho$-calculus~\cite{CKL01}.  There
is ongoing research to organise typed calculi in a so-called
$\rho$-cube, very much in the sense of the
$\lambda$-cube~\cite{Barendregt92}. It is not obvious how certain
typing notions interact with each other if we attempt to cover generic
traversal in this cube, e.g., type-dependent reduction \`a la \mycalc\
vs.\ dependent types.

\subsection{Genericity in functional programming}

The most established notion of genericity or polymorphism is certainly
parametric
polymorphism~\cite{Milner78,GirardPhD,Reynolds74,Reynolds83,CW85,Wadler89}.
It is clear that parametric polymorphism is not sufficient to model
typed generic traversal strategies. Firstly, it does not allow us to
descend into terms. Secondly, parametric polymorphism is also in
conflict with non-uniform behaviour as it can be assembled in terms of
strategy extension. Several forms of polymorphism were proposed that
go beyond parametric polymorphism, namely dynamic
typing~\cite{ACPP91,ACPR92}, extensional polymorphism~\cite{DRW95},
intensional polymorphism~\cite{HM95}, and polytypism~\cite{JJ97}.  A
general observation is that none of the available systems subsumes
\mycalc.  Dynamic typing was already discussed in the previous section
on rewriting. The remaining forms are reviewed in the sequel. We also
refer to~\cite{LV00,LV01} where we report on actual efforts to encode
generic traversal strategies as generic functions.

Let us check the requirement for generic traversal, that is, the
ability to descend into terms. Clearly, algebraic datatypes model sets
of typed terms in functional programming. Extensional polymorphism,
intensional polymorphism and polytypism have in common that they offer
some form of generic function definition based on structural pattern
matching on types.  These forms can be used to encode traversal. In
the cases of extensional and intensional polymorphism, type-based
induction involves cases for basic datatypes, products and
functions. The mere structure of algebraic datatypes implies that a
case for sums is also needed.  In fact, polytypic programming
considers algebraic datatypes as sums of products, and adds a
corresponding pattern for type induction.  It would be straightforward
to extend extensional and intensional polymorphism accordingly.

The idea of strategy extension implies that generic strategies are
aware of many sorts, say \emph{systems} of \emph{named} algebraic
datatypes in the sense of functional programming. However, all the
aforementioned forms of polymorphism are geared towards
\emph{structural} induction on types, that is, they do not 
involve a notion of checking the coincidence of two (names of) types
as it is required for strategy extension. This crucial difference is
discussed in \cite{Glew99}. This shortcoming has been addressed in
recent work on polytypic programming to some extent, namely different
proposals for Generic Haskell include support for some form of
type-specific cases (cf.\ ad-hoc definitions in~\cite{Hinze99-HW}) in
an otherwise structural induction on types.

In fact, Generic Haskell appears to offer the most complete feature
list for an encoding of rewriting strategies because generic term
traversal and specific type cases are offered by the language design
of Generic Haskell. However, we cannot reconstruct \mycalc\ in this
language setup for the following reasons. Firstly, polytypic functions
are not first-class citizens. In particular, one cannot pass a
polytypic function as an argument to another polytypic
function. First-class functions are needed to model traversal
combinators, traversal schemes or other parameterised
strategies. Secondly, type case is based on polytypic function
definition as a top-level form of declaration. This restricts the
separation and composition of type-specific vs.\ generic
functionality. More generally, polytypic programming does not support
combinator style of generic programming whereas strategic programming
relies on combinator style.


\section{Conclusion}
\label{S:concl}

\paragraph*{Typed generic traversal strategies}

In the present paper, we developed a typed calculus \mycalc\ for term
rewriting strategies. The main contribution of the paper is that 
\emph{generic traversal} is covered. The idea of generic traversal
combinators is already present in previous work on strategic
rewriting, however, only in untyped settings. It turned out that
existing combinators for \emph{intentionally} type-preserving
traversal could be easily typed. However, the typical approach to
type-unifying traversal is hopelessly untyped (cf.\ generic term
destruction and construction \`a la $\cdot\#\cdot$ in Stratego,
``=..'' in Prolog). To resolve this problem, we proposed designated
traversal combinators for type-unifying traversal. The key idea
underlying our type system is the \mycalc-like style of type-safe
extension of many-sorted strategies. This approach allows us to
combine many-sorted and generic functionality in a very flexible
manner without confusing different kinds of strategy composition (cf.\
\extend{\cdot}{\cdot}\ vs.\ \lchoice{\cdot}{\cdot}). The type system
separates many-sorted and generic strategies in a way that the
precision of the underlying many-sorted type system is preserved.

\paragraph*{Simple generic programming}

At a design level, our declared goal was to obtain a simple,
self-contained model of typed generic traversal on the grounds of
basically many-sorted, first-order term rewriting. In fact, the type
system of \mycalc\ is simple, and the complete calculus is
straightforward to implement. To explain what we mean by ``simple type
system'' and ``straightforward implementation'', we mention that the
development of the Prolog prototype, which we discussed earlier, took
two days. Contrast that with other approaches to generic programming
such as PolyP and Generic Haskell which usually require(d) several man
years of design and (prototype) implementation. Generic term traversal
based on the designed suite of traversal combinators is very potent
but it certainly does not cover the full range of generic programming
(cf.\ kind-indexed polytypic definitions, generic anamorphisms, and
others). Also, the overall setting of \mycalc, especially the
restriction to a basically first-order, many-sorted setting, rules out
several powerful programming idioms, e.g., higher-order functions.
Nevertheless, the prime justification for the restricted approach is
the well-defined application domain covered by the chosen
expressiveness, namely program transformation and analysis for large
language syntaxes (cf.~\cite{VBT98,LV00,Visser01-WRS,BKV01,LV01}).

\paragraph*{Functional strategies}

In our ongoing work, we transpose strategic term rewriting to the
functional programming paradigm (cf.\ the Haskell-based generic
programming bundle Strafunski; see
{\small\url{http://www.cs.vu.nl/Strafunski/}}). In~\cite{LV01}, we
motivated and characterised a corresponding notion of functional
strategies, and we provided a corresponding combinator library for
generic functions.  This approach complements existing approaches to
generic functional programming in that it supports \emph{first-class}
generic functions which can traverse into terms of systems of
algebraic datatypes while mixing uniform and type-specific behaviour.
In fact, we investigate different models to support strategies in
functional programming. One model which we also discuss in~\cite{LV01}
is based on the formula ``strategies as functions on a universal
representation type'' as discussed for ELAN in Section~\ref{SS:ELAN}.

\paragraph*{Future work}

Besides the notion of functional strategies, we are also interested in
the further development of the strategic rewriting paradigm in
general. We indicate an open-ended list of challenges for future work:
\begin{itemize}
\item Typed-based optimisation of traversals.
\item Typeful treatment of impure extensions of Stratego.
\item Fusion-like principles for traversal strategies~\cite{JV01}.
\item Systematic derivation of one-step traversal combinators.
\item Interaction of constraint mechanisms and traversal strategies.
\item Application of strategic programming to document processing.
\item Coverage of generic datatype-changing transformations~\cite{LL01}.
\item More precise types as for success and failure behaviour~\cite{MoreauPhD}.
\item More precise types as for kinds of involved polymorphic behaviour.
\item Coverage of generic term construction in the sense of anamorphisms.
\item Comparison of attribute grammar approaches and strategic programming.
\item Comparison of strategic programming and adaptive programming~\cite{LP97}.
\end{itemize}


\newpage

\bibliographystyle{alpha}
\bibliography{report}

\newcommand{\etalchar}[1]{$^{#1}$}
\begin{thebibliography}{CDMO01}

\bibitem[ABK{\etalchar{+}}01]{CASL01}
E.~Astesiano, M.~Bidoit, H.~Kirchner, B.~Krieg-Br{\"u}ckner, P.D. Mosses,
  D.~Sannella, and A.~Tarlecki.
\newblock {CASL: The Common Algebraic Specification Language}.
\newblock {\em Theoretical Computer Science}, 2001.
\newblock To appear.

\bibitem[ACPP91]{ACPP91}
M.~Abadi, L.~Cardelli, B.~Pierce, and G.~Plotkin.
\newblock {Dynamic Typing in a Statically Typed Language}.
\newblock {\em ACM Transactions on Programming Languages and Systems},
  13(2):237--268, April 1991.

\bibitem[ACPR92]{ACPR92}
M.~Abadi, L.~Cardelli, B.~Pierce, and D.~R{\'{e}}my.
\newblock {Dynamic Typing in Polymorphic Languages}.
\newblock In {\em {Proceedings of the 1992 {ACM} Workshop on {ML} and its
  Applications}}, pages 92--103, San Francisco, June 1992. Association for
  Computing Machinery.

\bibitem[Bar92]{Barendregt92}
H.~Barendregt.
\newblock {Lambda Calculi with Types}.
\newblock In {\em {Handbook of Logic in Computer Science}}, volume~2. Oxford
  University Press, 1992.

\bibitem[BDCd95]{BDL95}
F.~Barbanera, M.~Dezani-Ciancaglini, and U.~de'Liguoro.
\newblock {Intersection and Union Types: Syntax and Semantics}.
\newblock {\em Information and Computation}, 119(2):202--230, June 1995.

\bibitem[BHJ{\etalchar{+}}01]{B+01}
{M.G.J. van den} Brand, J.~Heering, {H. de} Jong, {M. de} Jonge, T.~Kuipers,
  P.~Klint, L.~Moonen, P.~Oliver, J.~Scheerder, J.~Vinju, E.~Visser, and
  J.~Visser.
\newblock The {ASF+SDF Meta-Environment}: a component-based language
  development environment.
\newblock In {\em Compiler Construction 2001 (CC 2001)}, volume 2027 of {\em
  LNCS}. Springer-Verlag, 2001.

\bibitem[BHK89]{BHK89}
J.A. Bergstra, J.~Heering, and P.~Klint.
\newblock {The Algebraic Specification Formalism ASF}.
\newblock In {\em {Algebraic Specification}}, chapter~1, pages 1--66. The ACM
  Press in cooperation with Addison-Wesley, 1989.

\bibitem[BKK96]{BKK96}
P.~Borovansky, C.~Kirchner, and H.~Kirchner.
\newblock {Controlling Rewriting by Rewriting}.
\newblock In Meseguer \cite{RWLW96}.

\bibitem[BKK{\etalchar{+}}98]{BKKMR98}
P.~Borovansk{\'y}, C.~Kirchner, H.~Kirchner, P.-E. Moreau, and C.~Ringeissen.
\newblock {An Overview of ELAN}.
\newblock In Kirchner and Kirchner \cite{WRLA98}.

\bibitem[BKKR01]{BKKR01}
P.~Borovansk\'y, C.~Kirchner, H.~Kirchner, and C.~Ringeissen.
\newblock {Rewriting with strategies in {{\sf ELAN}}: a functional semantics}.
\newblock {\em {International Journal of Foundations of Computer Science}},
  2001.

\bibitem[BKV01]{BKV01}
{M.G.J. van den} Brand, P.~Klint, and J.J. Vinju.
\newblock Term rewriting with traversal functions.
\newblock Technical Report SEN-R0121, CWI, July 2001.

\bibitem[BLRU97]{BLRU97}
{S.~van} Bakel, L.~Liquori, {S.R.~della} Rocca, and P.~Urzyczyn.
\newblock Comparing cubes of typed and type assignment systems.
\newblock {\em Annals of Pure and Applied Logic}, 86(3):267--303, July 1997.

\bibitem[BSV97]{BSV97}
{M.G.J. van den} Brand, M.P.A. Sellink, and C.~Verhoef.
\newblock Generation of components for software renovation factories from
  context-free grammars.
\newblock In I.D. Baxter, A.~Quilici, and C.~Verhoef, editors, {\em Proceedings
  Fourth Working Conference on Reverse Engineering}, pages 144--153, 1997.

\bibitem[BSV00]{BSV00}
{M.G.J. van den} Brand, M.P.A. Sellink, and C.~Verhoef.
\newblock {Generation of Components for Software Renovation Factories from
  Context-free Grammars}.
\newblock {\em Science of Computer Programming}, 36(2--3):209--266, 2000.

\bibitem[CDCV81]{CDCV81}
M.~Coppo, M.~Dezani-Ciancaglini, and B.~Venneri.
\newblock Functional characters of solvable terms.
\newblock {\em {Zeitschrift f\"{u}r Mathematische Logik und Grundlagen der
  Mathematik}}, 27:45--58, 1981.

\bibitem[CDE{\etalchar{+}}99]{Maude99}
M.~Clavel, F.~Dur{\'a}n, S.~Eker, P.~Lincoln, N.~Mart{\'\i}-Oliet, J.~Meseguer,
  and J.F. Quesada.
\newblock {The Maude System---System Description}.
\newblock In P.~Narendran and M.~Rusinowitch, editors, {\em {Proceedings of the
  10th International Conference on Rewriting Techniques and Applications
  (RTA-99)}}, volume 1631 of {\em LNCS}, pages 240--243, Trento, Italy, July
  1999. Springer-Verlag.

\bibitem[CDMO01]{CFM01}
M.~Clavel, F.~Duran, and N.~Marti-Oliet.
\newblock {Polytypic Programming in Maude}.
\newblock In K.~Futatsugi, editor, {\em {ENTCS}}, volume~36. Elsevier Science,
  2001.

\bibitem[CELM96]{CELM96}
M.~Clavel, S.~Eker, P.~Lincoln, and J.~Meseguer.
\newblock {Principles of Maude}.
\newblock In Meseguer \cite{RWLW96}.

\bibitem[CGL95]{CGL95}
G.~Castagna, G.~Ghelli, and G.~Longo.
\newblock A calculus for overloaded functions with subtyping.
\newblock {\em Information and Computation}, 117(1):115--135, February 1995.

\bibitem[CK99]{CK99}
H.~Cirstea and C.~Kirchner.
\newblock {Introduction to the Rewriting Calculus}.
\newblock Rapport de recherche 3818, INRIA, December 1999.

\bibitem[CKL01]{CKL01}
H.~Cirstea, C.~Kirchner, and L.~Liquori.
\newblock {The Rho Cube}.
\newblock In Furio Honsell, editor, {\em Foundations of Software Science and
  Computation Structures}, volume 2030 of {\em LNCS}, pages 168--183, Genova,
  Italy, April 2001.

\bibitem[CW85]{CW85}
L.~Cardelli and P.~Wegner.
\newblock {On Understanding Types, Data Abstraction, and Polymorphism}.
\newblock {\em ACM Computing Surveys}, 17(4):471--522, December 1985.

\bibitem[CWM99]{CWM99}
K.~Crary, S.~Weirich, and G.~Morrisett.
\newblock Intensional polymorphism in type-erasure semantics.
\newblock In {\em Proceedings of the ACM SIGPLAN International Conference on
  Functional Programming (ICFP '98)}, volume 34(1) of {\em ACM SIGPLAN
  Notices}, pages 301--312. ACM, June 1999.

\bibitem[Des88]{Despeyroux88}
T.~Despeyroux.
\newblock {TYPOL: A formalism to implement natural semantics}.
\newblock Technical report~94, INRIA, March 1988.

\bibitem[DRW95]{DRW95}
C.~Dubois, F.~Rouaix, and P.~Weis.
\newblock Extensional polymorphism.
\newblock In {\em Proceedings of the 22th ACM Conference on Principles of
  Programming Languages}, January 1995.

\bibitem[Geu93]{GeuversPhD}
H.~Geuvers.
\newblock {\em {Logics and Type Systems}}.
\newblock PhD thesis, Computer Science Institute, Katholieke Universiteit
  Nijmegen, 1993.

\bibitem[Gir72]{GirardPhD}
J.-Y. Girard.
\newblock {\em {Interpr\'{e}tation fonctionelle et \'{e}limination des coupures
  dans l'arith\'{e}tique d'ordre sup\'{e}rieur}}.
\newblock PhD thesis, Universit\'{e} Paris VII, 1972.

\bibitem[GL01]{WRS01}
B.~Gramlich and S.~Lucas, editors.
\newblock {\em {Proc.\ International Workshop on Reduction Strategies in
  Rewriting and Programming (WRS 2001)}}, volume SPUPV 2359, Utrecht, The
  Netherlands, May 2001. Servicio de Publicaciones - Universidad
  Polit{\'e}cnica de Valencia.

\bibitem[Gle99]{Glew99}
N.~Glew.
\newblock Type dispatch for named hierarchical types.
\newblock In {\em Proceedings of the Fourth {ACM} {SIGPLAN} International
  Conference on Functional Programming ({ICFP}-99)}, volume 34.9 of {\em ACM
  Sigplan Notices}, pages 172--182, N.Y., September ~27--29 1999. ACM Press.

\bibitem[Hin99]{Hinze99-HW}
R.~Hinze.
\newblock A generic programming extension for {Haskell}.
\newblock In E.~Meijer, editor, {\em {Proceedings of the 3rd {Haskell}
  Workshop, {Paris}, {France}}}, September 1999.
\newblock Technical report, Universiteit Utrecht, UU-CS-1999-28.

\bibitem[HM95]{HM95}
R.~Harper and G.~Morrisett.
\newblock Compiling polymorphism using intentional type analysis.
\newblock In {\em Conference Record of POPL '95: 22nd Annual ACM SIGPLAN-SIGACT
  Symposium on Principles of Programming Languages, San Francisco, Calif.},
  pages 130--141, New York, NY, January 1995. ACM.

\bibitem[Jeu00]{WGP00}
J.~Jeuring, editor.
\newblock {\em {Proceedings of WGP'2000, Technical Report, Universiteit
  Utrecht}}, July 2000.

\bibitem[JJ97]{JJ97}
P.~Jansson and J.~Jeuring.
\newblock Poly{P} - a polytypic programming language extension.
\newblock In {\em {POPL} '97: The 24th {ACM SIGPLAN-SIGACT} {S}ymposium on
  {P}rinciples of {P}rogramming {L}anguages}, pages 470--482. ACM Press, 1997.

\bibitem[Jon95]{Jones95}
M.P. Jones.
\newblock {Functional Programming with Overloading and Higher-Order
  Polymorphism}.
\newblock In J.~Jeuring and E.~Meijer, editors, {\em {Advanced Functional
  Programming}}, volume 925 of {\em LNCS}, pages 97--136. Springer-Verlag,
  1995.

\bibitem[JV01]{JV01}
P.~Johann and E.~Visser.
\newblock {Fusing Logic and Control with Local Transformations: An Example
  Optimization}.
\newblock Technical report, Institute of Information and Computing Sciences,
  Universiteit Utrecht, 2001.

\bibitem[Kah87]{Kahn87}
G.~Kahn.
\newblock {Natural Semantics}.
\newblock In {\em {4th Annual Symposium on Theoretical Aspects of Computer
  Science}}, volume 247 of {\em LNCS}, pages 22--39, Passau, Germany,
  19--21~February 1987. Springer-Verlag.

\bibitem[KK98]{WRLA98}
C.~Kirchner and H.~Kirchner, editors.
\newblock {\em {Proceedings of the International Workshop on Rewriting Logic
  and its Applications (WRLA'98)}}, volume~15 of {\em ENTCS},
  Pont-{\`a}-Mousson, France, September 1998. Elsevier Science.

\bibitem[Kli93]{Klint93}
P.~Klint.
\newblock {A meta-environment for generating programming environments}.
\newblock {\em ACM Transactions on Software Engineering and Methodology, 2(2)},
  pages 176--201, 1993.

\bibitem[L{\"a}m01]{Laemmel01-WRS}
R.~L{\"a}mmel.
\newblock {Generic Type-preserving Traversal Strategies}.
\newblock In Gramlich and Lucas \cite{WRS01}.

\bibitem[LL01]{LL01}
R.~L{\"a}mmel and W.~Lohmann.
\newblock {Format Evolution}.
\newblock In {\em {Proc.\ 7th International Conference on Reverse Engineering
  for Information Systems (RETIS 2001)}}, volume 155 of {\em books@ocg.at},
  pages 113--134. OCG, 2001.

\bibitem[LPS97]{LP97}
K.J. Lieberherr and B.~Patt-Shamir.
\newblock {Traversals of Object Structures: Specification and Efficient
  Implementation}.
\newblock Technical Report {NU-CCS-97-15}, College of Computer Science,
  Northeastern University, Boston, MA, July 1997.

\bibitem[LR01]{LR01}
R.~L{\"a}mmel and G.~Riedewald.
\newblock {Prological Language Processing}.
\newblock In {M.G.J. van den} Brand and D.~Parigot, editors, {\em {Proc.\
  LDTA'01}}, volume~44 of {\em ENTCS}. Elsevier Science, April 2001.

\bibitem[LV97]{LV97}
B.~Luttik and E.~Visser.
\newblock Specification of rewriting strategies.
\newblock In M.~P.~A. Sellink, editor, {\em 2nd International Workshop on the
  Theory and Practice of Algebraic Specifications (ASF+SDF'97)}, Electronic
  Workshops in Computing, Berlin, November 1997. Springer-Verlag.

\bibitem[LV00]{LV00}
R.~L{\"a}mmel and J.~Visser.
\newblock {Type-safe Functional Strategies}.
\newblock In {\em Draft proc.\ of SFP'00, St Andrews}, July 2000.

\bibitem[LV01]{LV01}
R.~L{\"a}mmel and J.~Visser.
\newblock {Typed Combinators for Generic Traversal}.
\newblock Technical Report SEN-R0124, Centrum voor Wiskunde en Informatica,
  August 2001.
\newblock 34 pages; also published in Proc.\ of PADL'02, LNCS 2257,
  Springer-Verlag.

\bibitem[LVK00]{LVK00}
R.~L{\"a}mmel, J.~Visser, and J.~Kort.
\newblock {Dealing with Large Bananas}.
\newblock In Jeuring \cite{WGP00}, pages 46--59.

\bibitem[LW01]{LW01}
R.~L{\"a}mmel and G.~Wachsmuth.
\newblock {Transformation of SDF syntax definitions in the ASF+SDF
  Meta-Environment}.
\newblock In M.~van~den Brand and D.~Parigot, editors, {\em {Proceedings of the
  First Workshop on Language Descriptions, Tools and Applications (LDTA'01),
  Genova, Italy, April 7, 2001, Satellite event of ETAPS'2001}}, volume~44 of
  {\em ENTCS}. Elsevier Science, April 2001.

\bibitem[Mee96]{Meertens96}
L.~Meertens.
\newblock {Calculate Polytypically!}
\newblock In H.~Kuchen and S.D. Swierstra, editors, {\em {Int. Symp. on Progr.
  Languages, Implementations, Logics and Programs (PLILP'96)}}, volume 1140 of
  {\em LNCS}, pages 1--16. Springer-Verlag, 1996.

\bibitem[Mes96]{RWLW96}
J.~Meseguer, editor.
\newblock {\em {Proceedings of the 1st International Workshop on Rewriting
  Logic and its Applications, RWLW'96, (Asilomar, Pacific Grove, CA, USA)}},
  volume~4 of {\em ENTCS}, September 1996.

\bibitem[Mil78]{Milner78}
R.~Milner.
\newblock {A Theory of Type Polymorphism in Programming}.
\newblock {\em Journal of Computer and System Sciences}, 17(3):348--375,
  December 1978.

\bibitem[Mor99]{MoreauPhD}
P.-E. Moreau.
\newblock {\em {Compilation de r\`egles de r\'e\'ecriture et de strat\'egies
  non-d\'eterministes}}.
\newblock PhD thesis, {Universit\'e Henri Poincar\'e - Nancy 1}, 1999.

\bibitem[Pau83]{Paulson83}
L.C. Paulson.
\newblock {A Higher-Order Implementation of Rewriting}.
\newblock {\em Science of Computer Programming}, 3(2):119--149, August 1983.

\bibitem[Pet94]{Pettersson94}
M.~Pettersson.
\newblock {RML} -- a new language and implementation for natural semantics.
\newblock In M.~Hermenegildo and J.~Penjam, editors, {\em Proceedings of the
  6th International Symposium on Programming Language Implementation and Logic
  Programming, PLILP'94}, volume 844 of {\em LNCS}, pages 117--131.
  Springer-Verlag, 1994.

\bibitem[Rey74]{Reynolds74}
J.C. Reynolds.
\newblock {Towards a Theory of Type Structures}.
\newblock In {\em {Programming Symposium (Colloque sur la Programmation,
  Paris)}}, volume~19 of {\em LNCS}, pages 408--425. Springer-Verlag, 1974.

\bibitem[Rey83]{Reynolds83}
J.C. Reynolds.
\newblock Types, abstraction and parametric polymorphism.
\newblock In R.~E.~A. Mason, editor, {\em Proc.\ of 9th IFIP World Computer
  Congress, Information Processing '83, Paris, France, 19--23 Sept 1983}, pages
  513--523. North-Holland, Amsterdam, 1983.

\bibitem[Sch94]{Schmidt94}
D.A. Schmidt.
\newblock {\em {The Structure of Typed Programming Languages}}.
\newblock Foundations of Computing Series. MIT Press, 1994.

\bibitem[Spi90]{Spivey90}
M.~Spivey.
\newblock A functional theory of exceptions.
\newblock {\em Science of Computer Programming}, 14:25--42, 1990.

\bibitem[VB98]{VB98}
E.~Visser and Z.~Benaissa.
\newblock {A Core Language for Rewriting}.
\newblock In Kirchner and Kirchner \cite{WRLA98}.

\bibitem[VBT98]{VBT98}
E.~Visser, Z.~Benaissa, and A.~Tolmach.
\newblock {Building Program Optimizers with Rewriting Strategies}.
\newblock In {\em International Conference on Functional Programming (ICFP'98),
  Baltimore, Maryland. ACM SIGPLAN}, pages 13--26, September 1998.

\bibitem[Vis00]{Visser00}
E.~Visser.
\newblock {Language Independent Traversals for Program Transformation}.
\newblock In Jeuring \cite{WGP00}, pages 86--104.

\bibitem[Vis01]{Visser01-WRS}
E.~Visser.
\newblock {A Survey of Strategies in Program Transformation Systems}.
\newblock In Gramlich and Lucas \cite{WRS01}.

\bibitem[Wad89]{Wadler89}
P.~Wadler.
\newblock {Theorems for Free!}
\newblock In {\em {Proceedings 4th Int.\ Conf.\ on Funct.\ Prog.\ Languages and
  Computer Arch., FPCA'89, London, UK, 11--13 Sept 1989}}, pages 347--359. ACM
  Press, 1989.

\bibitem[Wad92]{Wadler92}
P.~Wadler.
\newblock The essence of functional programming.
\newblock In {ACM}, editor, {\em Conference record of the Nineteenth Annual
  {ACM} {SIGPLAN-SIGACT} Symposium on Principles of Programming Languages,
  {Albuquerque, New Mexico}, {January} 19--22, 1992}, pages 1--14. ACM Press,
  1992.

\bibitem[WB89]{WB89}
P.~Wadler and S.~Blott.
\newblock How to make {\it{ad-hoc}} polymorphism less {\it{ad hoc}}.
\newblock In ACM-SIGPLAN ACM-SIGACT, editor, {\em Conference Record of the 16th
  Annual {ACM} Symposium on Principles of Programming Languages ({POPL} '89)},
  pages 60--76, Austin, TX, USA, January 1989. ACM Press.

\end{thebibliography}

\end{document}